\documentclass[11pt]{article}

\usepackage[dvipsnames]{xcolor}
\usepackage{amsmath}
\usepackage{amssymb}
\usepackage{graphicx}
\usepackage{ifthen}
\usepackage{authblk}
\usepackage{endnotes}
\usepackage{geometry}
\usepackage{booktabs}
\usepackage{sectsty}
\usepackage{verbatim}
\usepackage{longtable}
\usepackage{rotating}
\usepackage{multirow}
\usepackage{colortbl}
\usepackage{csquotes}
\usepackage{tcolorbox}
\usepackage{enumitem}
\usepackage{makecell}
\usepackage{float}
\usepackage{placeins}
\usepackage{subcaption}
\usepackage[style=authoryear, natbib=true]{biblatex}
\usepackage[
    colorlinks=true,
    linkcolor=NavyBlue,
    citecolor=NavyBlue,
    urlcolor=NavyBlue
]{hyperref}
\usepackage{listings}
\newcommand{\param}[1]{\texttt{#1}}
\lstdefinestyle{pystyle}{
  language=Python,
  basicstyle=\ttfamily\small,
  keywordstyle=\color{blue!70!black}\bfseries,
  commentstyle=\color{gray!70!black}\itshape,
  stringstyle=\color{orange!80!black},
  numbers=left,
  numberstyle=\tiny\color{gray},
  numbersep=6pt,
  frame=single,
  breaklines=true,
  showstringspaces=false,
  tabsize=4,
  columns=fullflexible
}
\newcommand{\beginsupplement}{%
    \setcounter{section}{0}
    \renewcommand{\thesection}{S\arabic{section}}
    \setcounter{table}{0}
    \renewcommand{\thetable}{S\arabic{table}}
    \setcounter{figure}{0}
    \renewcommand{\thefigure}{S\arabic{figure}}
    \setcounter{equation}{0}
    \renewcommand{\theequation}{S\arabic{equation}}
}

\title{Modelling sexual partnership dynamics and population heterogeneities in agent-based dynamic network models}

\author[1]{Priyanka Nair-Turkich}
\affil[1]{School of Computing and Information Systems, The University of Melbourne, Parkville, Victoria, Australia}
\author[2]{Patricia T. Campbell}
\affil[2]{Department of Infectious Diseases, The University of Melbourne, at the Peter Doherty Institute for Infection and Immunity, Melbourne, VIC, Australia}
\author[1]{Nicholas Geard}

\date{} 

\begin{document}

\maketitle
\begin{refsection}
\begin{abstract}

{Population-level heterogeneities, combined with temporal fluctuations in sexual partnerships, shape the structure of sexual contact networks and can substantially influence the spread of sexually transmitted infections (STIs). Traditional static network models, which assume fixed attributes of partnerships, such as count and duration, may not adequately capture the effects of the formation, dissolution, and concurrency of partnerships on STI transmission. In contrast, agent-based dynamic network models offer a flexible framework for incorporating individual and population-level heterogeneities. We developed an agent-based dynamic network model in which partnership formation and dissolution probabilities, stratified by age, sex, and sexual orientation (including bisexual individuals), govern the formation of monogamous and concurrent partnerships and their dissolution via a duration-dependent hazard. Partnership statistics from the National Survey of Sexual Attitudes and Lifestyles (NATSAL-3) were used as model calibration targets, and Latin Hypercube Sampling (LHS) was used to generate candidate parameter combinations. Parameter estimation was performed by selecting the combination that produced the lowest Mean Squared Error (MSE) between the model outputs and the calibration targets. Our study addresses three questions: (1) how well can the observed characteristics of sexual partnerships in NATSAL-3 be reproduced using an agent-based model; (2) how does concurrency shape the structure of dynamic sexual contact networks; and (3) how do concurrent partnerships affect the dynamics of STI transmission. In this study, we find that interactions between individual characteristics such as age, sex, and sexual orientation, and partnership attributes such as count, duration, and concurrency play a critical role in shaping the population-level sexual contact network and, in turn, the dynamics of STI transmission. This study provides a generalisable framework for modelling bacterial STIs across diverse populations, highlighting the importance of including partnership dynamics and population heterogeneity to understand transmission.}
\end{abstract}


\begin{tcolorbox}[title=Key terms in this paper, colback=gray!5, colframe=black!60, boxrule=0.4pt, sharp corners]
\begin{description}[leftmargin=0pt, labelindent=0pt, labelsep=0pt, style=nextline, font=\bfseries]
\item[Heterogeneity] Variation across individuals in the population in attributes such as age, sex, and sexual orientation
\item[Partnership] A sexual relationship between two individuals independent of the frequency of sex acts within it.
\item[Formation] The event by which a new sexual partnership is created between individuals.
\item[Dissolution] The termination of an existing sexual partnership between individuals.
\item[Concurrency] The ability to hold two or more partnerships that overlap in time, as distinct from monogamous partnerships, in which an individual has only one active partnership at any given time.
\item[Sexual contact network] A representation of population-level sexual activity in which nodes denote individuals and edges denote partnerships between them.
\end{description}
\end{tcolorbox}

\section{Introduction}
\label{sec:introduction}
Sexually Transmitted Infections (STIs) such as herpes, HIV, chlamydia, gonorrhoea, and syphilis can significantly affect reproductive and sexual health, sometimes leading to serious outcomes, including cancer and death. The global incidence of STIs has increased since the onset of the COVID-19 pandemic in 2020 \citep{soriano2023rebound}, with key populations such as men who have sex with men (MSM), people who inject drugs, transgender people, sex workers, people in prisons and other closed settings, and their partners often experiencing a disproportionate burden of STIs \citep{gottileb, WHO_2022}. STI transmission dynamics are shaped by complex interactions between individual behaviour, partnership dynamics such as count and duration, and population heterogeneities such as age, sex, and sexual orientation. A better understanding of these drivers is essential to design effective interventions to control STIs. 

In STI research, networks serve as abstractions of real-world sexual contact patterns and provide a framework for understanding population-level sexual activity, where nodes represent individuals and edges represent sexual partnerships \citep{eames2002modeling}. The structure, or topology, of sexual contact networks — including the number of partners per individual (degree) and the path length — shapes the patterns of transmission of STIs \citep{ghani1998measuring, eames2002modeling}. Static networks represent partnerships as fixed and do not capture the temporal processes through which sexual partnerships form, persist, overlap, and dissolve \citep{fefferman2007disease}. Although static network representations may be a good approximation of certain real-world scenarios, dynamic networks extend the abstraction of static networks by explicitly incorporating changes in partnership structures over time, providing a more realistic framework for representing sexual partnerships and modelling STI transmission \citep{frieswijk2023time}. Computational agent-based models (ABMs) are well suited for the simulation of STI transmission, as they capture individual-level attributes, behaviours, and interactions \citep{zhao2024relationship}. Within this framework, such models capture partnerships, exposures, risk factors, and other variables derived from observed behaviours and events over time to estimate the risk of STI transmission \citep{garnett_introduction_2002}. These models are widely used to inform health policies and interventions, guide prevention efforts, and optimise resource allocation to limit the spread of STIs. ABMs have been used, for example, to examine the role of partnership duration in STI transmission \citep{azizi_using_2021}.

The dynamics of partnerships shaping the transmission of STIs, particularly through casual partnerships, have been investigated by several studies. \citet{azizi_using_2021}, \citet{vajdi_multilayer_2020}, \citet{frieswijk2023time}, and \citet{Tsoumanis_et_al} highlighted the importance of accurately representing the types of partnerships and their duration, particularly short-term partnerships. \citet{vajdi_multilayer_2020} modelled permanent and casual partnerships using a multi-layer temporal model to simulate the spread of STIs. Similarly, \citet{Tsoumanis_et_al} simulated daily sexual activities between casual, one-time and steady partnerships to understand how people form different types of partnerships. \citet{frieswijk2023time} also indicated that casual partners play a central role in the transmission of STIs. These studies demonstrate that the duration and timing of sexual partnerships are critical determinants of the potential for STI transmission. 

Partnership formation, dissolution, and concurrency are key areas of research to understand the dynamics of sexual partnerships. Several studies have modelled the formation and dissolution of concurrent partnerships \citep{rao_partnership_2021, vajdi_multilayer_2020}, highlighting that concurrency accelerates transmission by increasing the number of infectious contacts in a short period of time and by providing a conduit between individuals who would otherwise not be connected. \citet{whittles2019dynamic} developed a dynamic power-law sexual network model of gonorrhoea transmission, in which most individuals had relatively few sexual contacts, while a small subset of individuals had substantially higher numbers of contacts. Their model demonstrates that interventions targeted at highly active individuals could significantly reduce the burden of gonorrhoea. 

Despite significant advances in modelling methodologies, representing individual and partnership attributes — such as diverse sexual orientations and concurrency — within a sexual contact network remains an ongoing challenge. Limited information on the duration, timing, and overlap of partnerships, particularly for concurrent and casual ones, can make it difficult to estimate partnership dynamics from behavioural surveys alone. These limitations make scenario modelling challenging in an already data-poor environment of STI research, particularly in low-resource settings \citep{rao_partnership_2021, hopkins2024importance}. Existing models often focus only on the heterosexual population or key populations, such as MSM, overlooking subpopulations such as bisexual individuals who can potentially serve as a source of bridging partnerships between heterosexual individuals and the MSM population \citep{Tsoumanis_et_al, vajdi_multilayer_2020, weiss_epidemiological_2019, azizi_using_2021}. Additionally, the effect of concurrency on partnerships has not been fully investigated in existing models \citep{Tsoumanis_et_al, vajdi_multilayer_2020}. Addressing the gaps related to bridging partnerships and concurrency requires identifying mechanisms through which these attributes can be included in an agent-based model to reproduce empirically observed sexual partnership characteristics.

We developed a dynamic agent-based model that represents individuals with three different sexual orientations, namely opposite-sex, same-sex, and bisexual, and ages from 16 to 74 years. The model was calibrated to the partnership data from the third National Survey of Sexual Attitudes and Lifestyles (NATSAL-3) \citep{Mercer2013SexualAttitudesNatsal, Clifton2023Natsal3RefTables}. Latin Hypercube Sampling (LHS) \citep{McKay1979ComparisonThreeMethods} was used to explore the input parameter space, combined with a Mean Squared Error (MSE) method to identify the combination of parameters that best reproduced the characteristics of sexual partnerships observed empirically in NATSAL-3. Our study builds on the existing modelling approaches of \citet{Tsoumanis_et_al} and \citet{azizi_using_2021}, to further examine how the underlying network structure and concurrent partnerships influence the transmission dynamics of STIs by using a \textit{Susceptible-Infected-Susceptible (SIS)} transmission model \citep{keeling_rohani} as an illustrative framework. This framework provides a basis for subsequent applications to specific bacterial STIs such as chlamydia, gonorrhoea, and syphilis. 

In this study, we address three research questions: (1) how well can the observed characteristics of sexual partnerships in NATSAL-3 be reproduced using an agent-based model; (2) how does concurrency shape the structure of dynamic sexual contact networks; and (3) how do concurrent partnerships affect the dynamics of STI transmission. In addressing these questions, this study provides a generalisable framework for modelling bacterial STIs across diverse populations, highlighting the importance of partnership dynamics and population heterogeneities for transmission. In this paper, we first describe the model structure, followed by the parameter selection and model calibration processes. We then compare network characteristics with and without concurrency included. Finally, we discuss the implications of these findings for modelling of STI transmission, along with the strengths and limitations of our modelling framework, and outline directions for future work.
\section{Methods} \label{sec:methods}
\subsection{Agent-based dynamic network modelling framework description} \label{subsec:model_description}
\begin{figure}[H]
\centering
\includegraphics[width=1\linewidth]{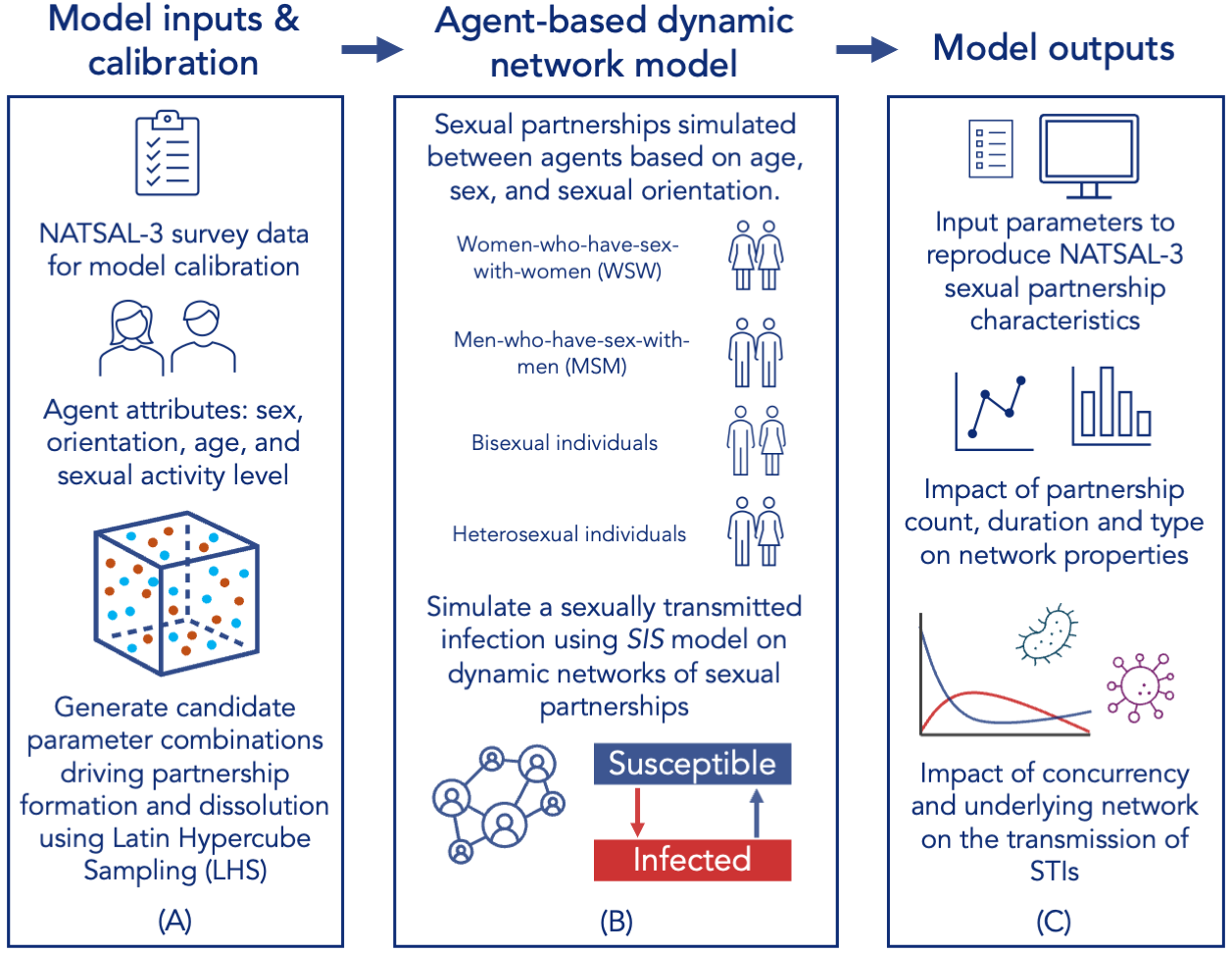}
\caption{Agent-based dynamic network modelling framework overview, from input parameters and model calibration data (A) through model structure (B), to model outputs (C)}
\label{fig:study-overview}
\end{figure}

We model a population using a dynamic sexual contact network in which agents, stratified by age, sex, and sexual orientation, form and dissolve partnerships probabilistically at daily time steps. Agents are characterised by fixed attributes (sex and sexual orientation) and time-varying attributes (age, sexually-active status, and current partnerships), with age, sex, and sexual orientation distributions matched to the NATSAL-3 data. Partnership formation is restricted to compatible pairings of sex and sexual orientations, weighted by age-assortative mixing, and governed by formation and dissolution probabilities that vary by age group, sex, and sexual orientation. Individual heterogeneity in sexual activity is represented by agent-specific multipliers drawn from a negative binomial distribution and applied to both partnership formation and dissolution. A proportion of agents are designated as concurrency-eligible, allowing multiple simultaneous partnerships up to an individual-specific maximum, whereas all other agents are restricted to monogamy. Partnership dissolution follows a duration-dependent hazard, with the daily probability of dissolution decreasing as the duration of the sexual partnership increases, reflecting the empirically observed higher risk of dissolution in the early stages of a partnership \citep{nelson2010age}. Demographic stability is maintained through dynamic ageing of agents, probabilistic sexual debut, and replacement of agents leaving the sexually-active population. Figure~\ref{fig:study-overview} summarises the input data and parameters, design, and output of the agent-based dynamic network model presented in this study. 

\subsection{Agent attributes} 
\label{subsec:agent_attributes}

\subsubsection{Population size} In the agent-based dynamic network model developed in this study, the formation and dissolution of sexual partnerships are simulated for a population of agents $N = 15{,}000$. Each agent $i$ is characterised by attributes that are fixed throughout the simulation (sex $s_i$ and sexual orientation $o_i$) and attributes that evolve over time (age $a_i(t)$, sexually-active status, and current set of partnerships). The distributions of age, sex, and sexual orientation reflect the NATSAL-3 technical reference tables \citep{Mercer2013SexualAttitudesNatsal, Clifton2023Natsal3RefTables}.
\subsubsection{Age distribution and ageing} 
Initial ages are drawn from a discrete uniform distribution between 16 and 74 years (inclusive):
\begin{equation}
a_i(0) \sim \text{DiscreteUniform}(16, 74).
\end{equation}
Ages are grouped into ten--year categories $g = 1, \dots, 6$, where $g=i$ corresponds to $16\text{--}24$, $g=2$ to $25\text{--}34$, $g=3$ to $35\text{--}44$, $g=4$ to $45\text{--}54$, $g=5$ to $55\text{--}64$, and $g=6$ to $65\text{--}74$. To avoid synchronised ageing events, each agent is assigned a uniformly distributed offset $d_i(0)$ representing the number of days elapsed since their last birthday, drawn uniformly from:
\begin{equation}
d_i(0) \sim \text{DiscreteUniform}(0, 364).
\end{equation}
The ageing process operates deterministically at a daily time step. The day counter increases by one each day. When $d_i(t) = 365$, the agent ages by one year and the counter resets:
\begin{equation}
a_i(t+1) = a_i(t) + 1, \quad d_i(t+1) = 0 \quad \text{if } d_i(t) = 365.
\end{equation}
\subsubsection{Sex and sexual orientation} 
\label{subsubsec:sex-orientation} At initialisation, each agent is randomly assigned a biological sex with equal probability, producing a balanced sex ratio in the population:
\begin{equation}\label{eq:sex_split}
\Pr(s_i = \text{Male}) = \Pr(s_i = \text{Female}) = 0.5.
\end{equation}
The binary sex classification reflects the structure of available empirical data from sexual behaviour surveys such as NATSAL-3, which collect data using male/female categories. The NATSAL-3 survey captures data related to the number of partners of individuals who form same-sex, opposite-sex, and same- and opposite-sex partnerships. The survey does not explicitly capture the sexual behaviour of bisexual individuals; therefore, an additional demographic stratum of bisexual individuals was constructed based on the responses given to the question related to sexual attraction in the survey. The proportion of survey respondents who selected the following responses have been classified as bisexual individuals in our model:
\begin{enumerate}
\item \textit{More often to (females/males) and at least once to (male/female)}
\item \textit{Approximately equally often to (females/males) and to (males/females)}
\item \textit{More often to (males/females) and at least once to (female/male)}
\end{enumerate}
The sexual orientation of an agent $o_i$ is conditional on the assignment of biological sex $s_i$. The conditional probabilities reflect NATSAL-3 findings that females report approximately twice the rate of non-heterosexual attraction compared to males. 

For males:
\begin{equation} \label{eq:male_sex_assignment}
\Pr(o_i \mid s_i = \text{Male}) =
\begin{cases}
0.90 & o_i = \text{Opposite-sex} \\
0.05 & o_i = \text{Same-sex} \\
0.05 & o_i = \text{Bisexual}
\end{cases}
\end{equation}
For females:
\begin{equation} \label{eq:female_sex_assignment}
\Pr(o_i \mid s_i = \text{Female}) =
\begin{cases}
0.80 & o_i = \text{Opposite-sex} \\
0.10 & o_i = \text{Same-sex} \\
0.10 & o_i = \text{Bisexual}
\end{cases}
\end{equation}
\subsubsection{Sexual debut} 
\label{subsubsec:sexual_debut}
Each agent has a sexually-active status that determines eligibility to form partnerships. Agents aged 21 and over are treated as having already made their sexual debut. At initialisation, the sexually-active status of agents aged 16--20 is determined by sampling against the cumulative debut probability up to their assigned age. The probability distribution for this parameter reflects a simplified assumption such that the modelled median age at first sexual contact is approximately 16 years. This is broadly consistent with the estimates from previous NATSAL surveys \citep{lewis2017heterosexual, wellings2001sexual}. The cumulative annual probabilities of sexual debut are given in the Supplementary material.
\subsection{Partnership types}
\subsubsection{Concurrency-eligible individuals}
\label{subsubsec:concurrency}
Concurrent (overlapping) partnerships are an important determinant of the transmission of STIs through the connected components they create within a network \citep{Mercer2018TimingSexualPartnerships}. A fixed proportion $\theta_{\mathrm{conc}}$ of agents are designated as concurrency-eligible at initialisation, while the remainder are monogamous. Each concurrency-eligible agent $i$ is assigned an individual upper bound $K_i$ on the number of simultaneous partnerships, drawn from a Poisson distribution with rate parameter $\lambda$ and lower bound of two:
\begin{equation}
K_i = \max\!\left(2,\ \text{Poisson}(\lambda)\right).
\end{equation}
Concurrency is treated as a fixed individual attribute rather than a dynamic one and $\lambda$ is constant across all combinations of age, sex, and sexual orientation. Replacement agents introduced through population turnover are designated as concurrency-eligible with probability equal to the current population-level concurrency proportion $\theta_{\mathrm{conc}}$. 
\subsubsection{Non-concurrency-eligible individuals}
\label{subsubsec:monogamy}
Agents not designated concurrency-eligible are strictly monogamous; upon dissolution of their sole active partnership, they return to the pool of agents eligible to form a new partnership. This constraint is enforced throughout the simulation, including for replacement agents introduced through population turnover who are assigned the strictly-monogamous status with the probability of $1 - \theta_{\mathrm{conc}}$.
\subsection{Simulation of sexual partnerships}
\label{subsec:sexual_partnerships}
\subsubsection{Partnership formation}
\label{subsubsec:partnership_formation}
\textbf{Baseline probabilities and multipliers:} Daily formation probabilities are organised in a three-dimensional matrix stratified by the combination of age group and sex-sexual orientation ($6 \times 6 = 36$ strata), with the reference stratum $p^{\text{form}}_{\text{base}}$ defined for opposite-sex females aged 25--34. Probabilities for other strata are obtained via a multiplicative adjustment given by:

1) an age multiplier;
\begin{equation}
\mu^{\text{age}}_{g} =
\begin{cases}
\beta & g = 1 \\
1 & g = 2 \\
\exp\!\left(-\kappa \cdot (g - 2)\right) & g = 3, 4, \ldots, 6,
\end{cases}
\end{equation}
parameterised by a \textit{youth boost} $\beta$ for ages 16--24 and an exponential \textit{age decay} rate $\kappa$ for agents over 34 years of age, with the age band 25--34 as the baseline ($\mu^{\text{age}}_{25\text{--}34}=1$), and

2) a sex-sexual orientation multiplier $\mu^{\text{ori}}_{s,o}$, calibrated so that stratum-specific formation probabilities reproduce the relative mean partnership rates observed by sex and sexual orientation in NATSAL-3, relative to the opposite-sex female baseline stratum:
\begin{equation}
\label{eq:formation_strat}
p^{\text{form}}_{s,o,g} = p^{\text{form}}_{\text{base}} \cdot \mu^{\text{ori}}_{s,o} \cdot \mu^{\text{age}}_{g}.
\end{equation}
The \textit{youth boost} mechanism is based on empirical evidence that the youngest age group reported the highest mean number of sexual partnerships among all age groups, which then decreases smoothly with age \citep{Clifton2023Natsal3RefTables}. Full details along with a numerical example of assigning probabilities based on age, sex, and sexual orientation, are given in the Supplementary material.

\textbf{Formation steps:} At each time step, a pool of eligible sexually-active agents is constructed who are below their allowed partnership maximum (1 for monogamous agents, $K_i \geq 2$ for concurrency-eligible agents (see Section 2.6). Eligible agents are processed in random order and independently attempt formation at their individually-adjusted demographic-stratum-specific probability (see Section 2.13). Candidate partners must satisfy the compatibility of sex and sexual orientation outlined in Section 2.3. Among compatible candidates, partner selection uses Gaussian age-assortative weighting, ($w_{ij}$), given in equation \eqref{eq:age_assortativity}, so that partners of similar age are more likely to be selected. Each agent may form at most one new partnership per time step.
\begin{equation}
\label{eq:age_assortativity}
w_{ij} \propto \exp\!\left(-\frac{(a_j - a_i)^2}{2\sigma^2}\right), \qquad \sigma = 4.0,
\end{equation}

\subsubsection{Partnership dissolution}
\label{subsubsec:partnership_dissolution}
The daily probabilities of partnership dissolution $p^{\text{diss}}_{\text{base}, s, o, g}$ follow the same stratified structure as formation (see Section 2.11). To reflect the empirical observation that the risk of partnership dissolution is highest at the beginning and decreases with its duration \citep{nelson2010age}, the baseline rate is adjusted by a Weibull-like hazard function of the duration of partnerships $d$,
\begin{equation}
\label{eq:dissolution_adj}
p^{\text{diss}}_{\text{adj}}(d) = p^{\text{diss}}_{\text{base}, s, o, g} \cdot f(d),
\end{equation}
\begin{equation}
\label{eq:hazard}
f(d) = \left(1 + \frac{d}{\alpha}\right)^{-\gamma}, \qquad \alpha = 1500, \ \gamma = 2,
\end{equation}
such that dissolution risk decays smoothly with partnership age and approaches near-zero for long-duration partnerships, consistent with the persistence of stable long-term partnerships. The same hazard parameters apply to internal partnerships (where both partners are active) and external partnerships (one partner removed at age 75), and require the duration of the partnership to be at least one day ($d >= 1$) to be eligible for dissolution. The step-by-step evaluation procedure is given in the Supplementary material.
\subsubsection{Individual-level heterogeneity}
\label{subsubsec:heterogeneity}
To incorporate heterogeneity in sexual activity among agents sharing the same age, sex, and sexual orientation, each agent is assigned an independent multiplicative effect on the formation and dissolution probabilities of partnerships during initialisation. Each agent draws an integer $X_i \sim \text{NegBin}(r, p)$, with mean $\mathbb{E}[X_i] = r(1-p)/p$, converted into a multiplier
\begin{equation}
\label{eq:nb_multiplier}
\eta_i = 1 + \frac{X_i}{\mathbb{E}[X_i]} \geq 1,
\end{equation}
applied to the agent's demographic-stratum-specific baseline formation and dissolution probabilities,
\begin{equation}
p^{\text{form}}_{i} = \eta^{\text{form}}_{i} \cdot p^{\text{form}}_{s_i, o_i, g_i(t)}, \qquad p^{\text{diss}}_{i} = \eta^{\text{diss}}_{i} \cdot p^{\text{diss}}_{\text{base}, s_i, o_i, g_i(t)},
\end{equation}
and fixed for the duration of the simulation. Since $\eta_i \geq 1$ by construction and $\eta^{\text{form}}_{i}$ and $\eta^{\text{diss}}_{i}$ are drawn independently, multipliers inflate the probabilities of formation and dissolution at the start of the simulation. The resulting probabilities are bounded to $[10^{-4}, 0.99]$ to avoid unrealistic values. This mechanism yields a right-skewed, heavy-tailed distribution of activity levels, with a small proportion of agents accounting for disproportionately high sexual activity. An example of applying $\eta^{\text{form}}_{i}$ and $\eta^{\text{diss}}_{i}$ multipliers, and the resulting partnership formation and dissolution probabilities, is given in the Supplementary material.
\subsubsection{Population turnover}
Agents whose age exceeds 74 years are flagged for removal from the sexually-active pool through the continuous ageing process, and are removed on reaching age 75. To maintain a constant population size, each removed agent is replaced by a new 16-year-old agent, assigned:
\begin{enumerate}
\item a unique agent identifier, one greater than the previous maximum;
\item an initial age of 16 years;
\item sex and sexual orientation, sampled from equations ~\eqref{eq:sex_split}, ~\eqref{eq:male_sex_assignment} and~\eqref{eq:female_sex_assignment};
\item an empty partnership set;
\item a uniformly distributed birthday offset $d_i \sim \text{DiscreteUniform}(0, 364)$;
\item independent formation and dissolution heterogeneity multipliers (see Section 2.13).
\end{enumerate}
When an agent over the age of 74 is removed, surviving partners retain the partnership record, reclassified as an \textit{external partnership}, and subject to the same dissolution hazard as internal partnerships (see Section 2.12), reflecting the assumption that the departure of the exiting partner from the sexually active pool does not end the partnership. 

\subsubsection{Temporal dynamics and partnership records}
\label{subsubsec:temporal_recording_partnerships}
\begin{figure}[H]
\centering
\includegraphics[width=1\linewidth, height=15cm, keepaspectratio]{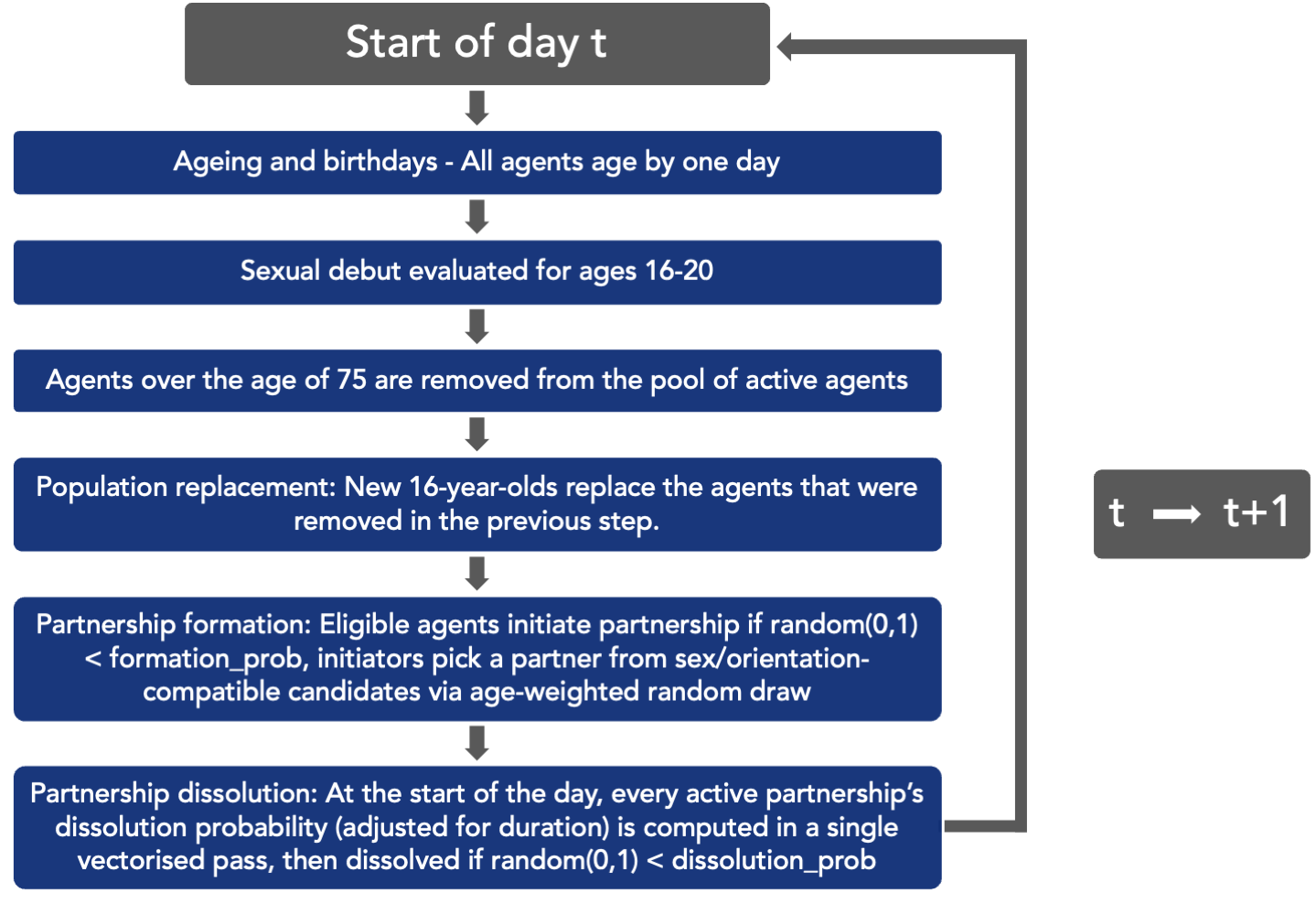}
\caption{Daily process sequence in the model}
\label{fig:daily_timestep}
\end{figure}

The simulation operates in discrete daily time steps. At each daily time step, processes are applied sequentially — ageing and birthday updates, sexual debut, agent removal, population replacement, partnership formation, and partnership dissolution. Figure~\ref{fig:daily_timestep} summarises the sequence of operations in each step with the details of implementation provided in the subsequent sections and the Supplementary material.

Every partnership generated during the simulation is recorded, with fields covering partner demographics (age, sex, and sexual orientation) and partnership attributes (formation and dissolution time, duration, and censoring status). The full list of recorded parameters is provided in the Supplementary material.

\subsection{Model calibration using Latin Hypercube Sampling (LHS) and parameter estimation via Mean Squared Error (MSE)} 
\label{subsec:lhs}
Latin Hypercube Sampling (LHS) is a class of Monte Carlo sampling methods introduced by \citet{McKay1979ComparisonThreeMethods}. LHS provides an efficient stratified sampling strategy for exploring the input parameter space \citep{Marino2008MethodologyGlobalUncertainty, McKay1979ComparisonThreeMethods} by ensuring that each variable distribution is well represented while using fewer samples than the random sampling method. For a parameter space of dimension $k$ and sample size $n$, each parameter distribution is divided into $n$ intervals of equal probability, and then $n$ model replicates are simulated using combinations of parameter values \citep{Marino2008MethodologyGlobalUncertainty}. 
\subsubsection{LHS experimental design}
\label{subsubsec:lhs_experimental_design}
LHS allows systematic exploration of plausible partner behaviours under heterogeneity by age group, sex, and sexual orientation, supporting the calibration of the model against empirical patterns observed in NATSAL-3. The LHS design for this study comprises 60{,}000 samples with the same random seed, generated under two scenarios: a no-concurrency scenario ($\theta_{\mathrm{conc}} = 0$), in which all agents are strictly monogamous, and a 15\%-concurrency scenario ($\theta_{\mathrm{conc}} = 0.15$), in which 15\% of agents are eligible to form concurrent partnerships. The 60{,}000 parameter sets for each scenario are identical due to the same seed, and any differences in model calibration are primarily attributed to the inclusion of concurrency. The sampled parameters fall into three groups: baseline formation and dissolution probabilities for the reference stratum (opposite-sex females aged 25--34); sex-sexual orientation multipliers for the remaining five combinations of sex and sexual orientations; and age-structure parameters, comprising a \textit{youth boost} $\beta$ for the 16--24 age group and an exponential \textit{age decay} rate $\kappa$ for older age groups. The full parameter ranges are provided in \hyperref[app:model_parameters]{Appendix A}.

Each combination is simulated using the dynamic network generator with $N = 15{,}000$ agents for 1{,}875 time steps. For each simulation generated with a set of parameters, we recorded the following.
\begin{itemize}
\item concurrency proportion and seed values;
\item baseline formation and dissolution probabilities for the opposite-sex females of the 25--34 age group;
\item the multipliers for each demographic stratum;
\item the post-multiplier formation and dissolution probabilities for each demographic stratum;
\item the mean, median, maximum, and standard deviation of partnership counts stratified by age, sex, and sexual orientation.

\end{itemize}

\subsubsection{Model calibration and evaluation}
\label{subsubsec:model_calibration}
\textbf{Calibration targets:} The calibration targets for this study are derived from NATSAL-3, which provides empirical data on the patterns of partnerships between age, sex, and sexual orientation. Calibration targets for mean partnership count for opposite-sex and same-sex individuals were obtained from the NATSAL-3 reference tables 28 and 32, respectively \citep{Clifton2023Natsal3RefTables}. For bisexual agents, calibration targets were derived from table 36, which reports partnership counts involving both sexes. We assume that these responses predominantly reflect bisexual-identifying individuals, while acknowledging that the responses may also include individuals with the opposite-sex or same-sex sexual orientation. The mean partnership count targets for all combinations of age group, sex, and sexual orientation are presented in Table~\ref{tab:calibration_targets}. 
\begin{table}[H]
\centering
\caption{Calibration targets for the mean number of sexual partners in the past five years by sex, age group, and sexual orientation. Values marked with * were treated as outliers and excluded from model calibration.}
\label{tab:calibration_targets}
\begin{tabular}{llcccccc}
\hline
Sexual orientation & Sex & 16--24 & 25--34 & 35--44 & 45--54 & 55--64 & 65--74 \\
\hline

\multirow{2}{*}{Opposite-sex}
& Males  & 4.8 & 3.8 & 2.2 & 1.9 & 1.6 & 1.0 \\
& Females & 3.7 & 2.3 & 2.1 & 1.2 & 0.8 & 0.6 \\

\multirow{2}{*}{Same-sex}
& Males  & 5.4 & 9.6\textsuperscript{*} & 13.2\textsuperscript{*} & 4.2 & 2.4 & 0.3 \\
& Females & 2.3 & 1.1 & 0.9 & 0.6 & 0.4 & 0.2 \\

\multirow{2}{*}{Bisexual}
& Males  & 4.9 & 4.3 & 2.8 & 2.2 & 1.7 & 1.0 \\
& Females & 3.9 & 2.4 & 2.1 & 1.3 & 0.9 & 0.6 \\

\hline
\end{tabular}
\end{table}

\textbf{Mean Squared Error (MSE):} The Mean Squared Error (MSE) quantifies the difference between modelled and target mean partnership counts for each sexual orientation. For each sexual orientation, the error is computed for 12 cells corresponding to the product of sex (2 categories) and age group (6 categories). For a given set of parameters $k$ and sexual orientation $o$, the MSE is calculated as:
\begin{equation}
\mathrm{MSE}_{o}^{(k)}
=
\frac{1}{12}
\sum_{s \in S}
\sum_{g \in G}
\left(
\hat{y}_{o,s,g}^{(k)} - y_{o,s,g}
\right)^2
\end{equation}
where:
\begin{itemize}
\item $\mathrm{MSE}_{o}^{(k)}$ is the mean squared error for a given sexual orientation $o$ for the parameter set $k$,
\item $\hat{y}_{o,s,g}^{(k)}$ denotes the modelled mean partnership count for the parameter set $k$,
\item $y_{o,s,g}$ denotes the corresponding empirical target value,
\item $s \in S$ indexes two sex categories and
\item $g \in G$ indexes six age groups.
\end{itemize}

The global loss function is defined as the mean of the errors specific to the three sexual orientations:
\begin{equation}
\mathrm{Global\ Loss}^{(k)}
=
\frac{
\mathrm{MSE}_{\text{opposite-sex}}^{(k)}
+
\mathrm{MSE}_{\text{same-sex}}^{(k)}
+
\mathrm{MSE}_{\text{bisexual}}^{(k)}
}{3}
\end{equation}

\textbf{Model calibration:} The mean partnership count targets for same-sex males aged 25--34 and 35--44 years were treated as outliers and excluded from calibration in both scenarios, as they were inconsistent with the monotonic age-related decline in partnership counts observed for other combinations of sex and sexual orientation. \footnote{These mean partnership counts had large standard deviations relative to their means (SD~=~ 17.2 for a mean of 9.6 in the 25--34 group; SD~=~ 51.7 for a mean of 13.2 in the 35--44 group), compared to lower variability in adjacent age groups (e.g., SD~=~ 11.2 for a mean of 4.2 in the 45--54 group) \citep{Clifton2023Natsal3RefTables}}. This pattern indicates that the elevated means are driven by a small number of highly active individuals within these subgroups, rather than reflecting a pattern of typical partnership counts. 

The simulation results for the sets of best-fitting parameter sets (id:2777 for no-concurrency and id:557 for 15\%-concurrency scenarios) for each sexual orientation ranked by the global MSE are presented in Figures~\ref{fig:scatter_comparison_opposite-sex},~\ref{fig:scatter_comparison_same-sex} and~\ref{fig:scatter_comparison_bisexual} for males and females of all age groups. 

\begin{enumerate}
\item \textbf{Opposite-sex partnerships.}
To estimate the best set of input parameters for the model, we compare the model outputs against the NATSAL-3 calibration targets for the opposite-sex stratum for all age groups and sexes (Figure~\ref{fig:scatter_comparison_opposite-sex}). We observe that the fit for males in the youngest age group is better for the no-concurrency scenario compared to the 15\%-concurrency scenario. We also observe that the introduction of concurrency increases the upper bound of the SD $\pm$ of the partnership count by $\sim$40\% for 16--24 year old males. 

\begin{figure}[H]
\centering
\begin{subfigure}{0.49\linewidth}
  \centering
  \includegraphics[width=\linewidth, height=12cm, keepaspectratio]{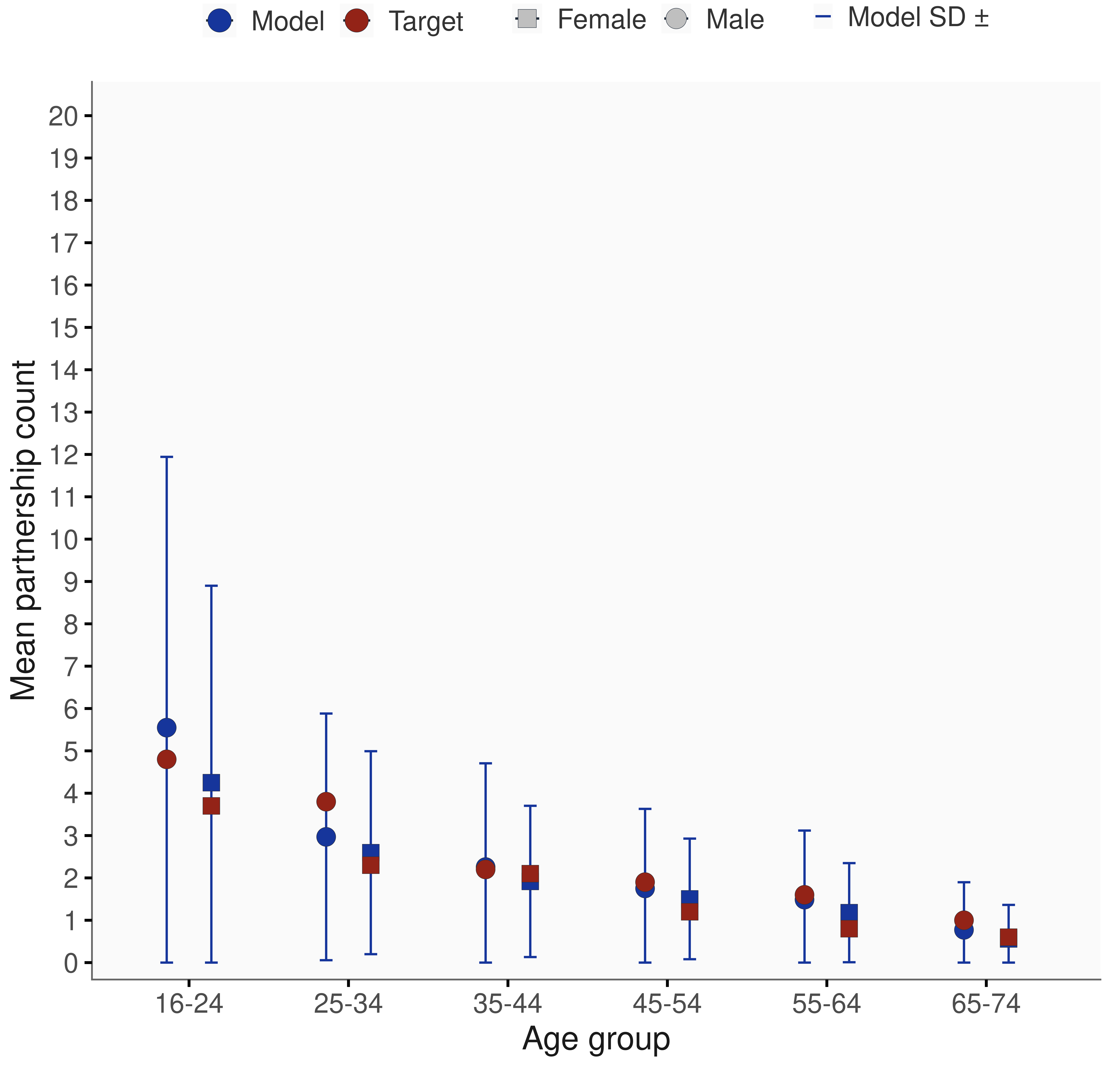}
  \caption{No concurrency ($\theta_{\mathrm{conc}} = 0$), best-fit id:2777}
\end{subfigure}
\hfill
\begin{subfigure}{0.49\linewidth}
  \centering
  \includegraphics[width=\linewidth, height=12cm, keepaspectratio]{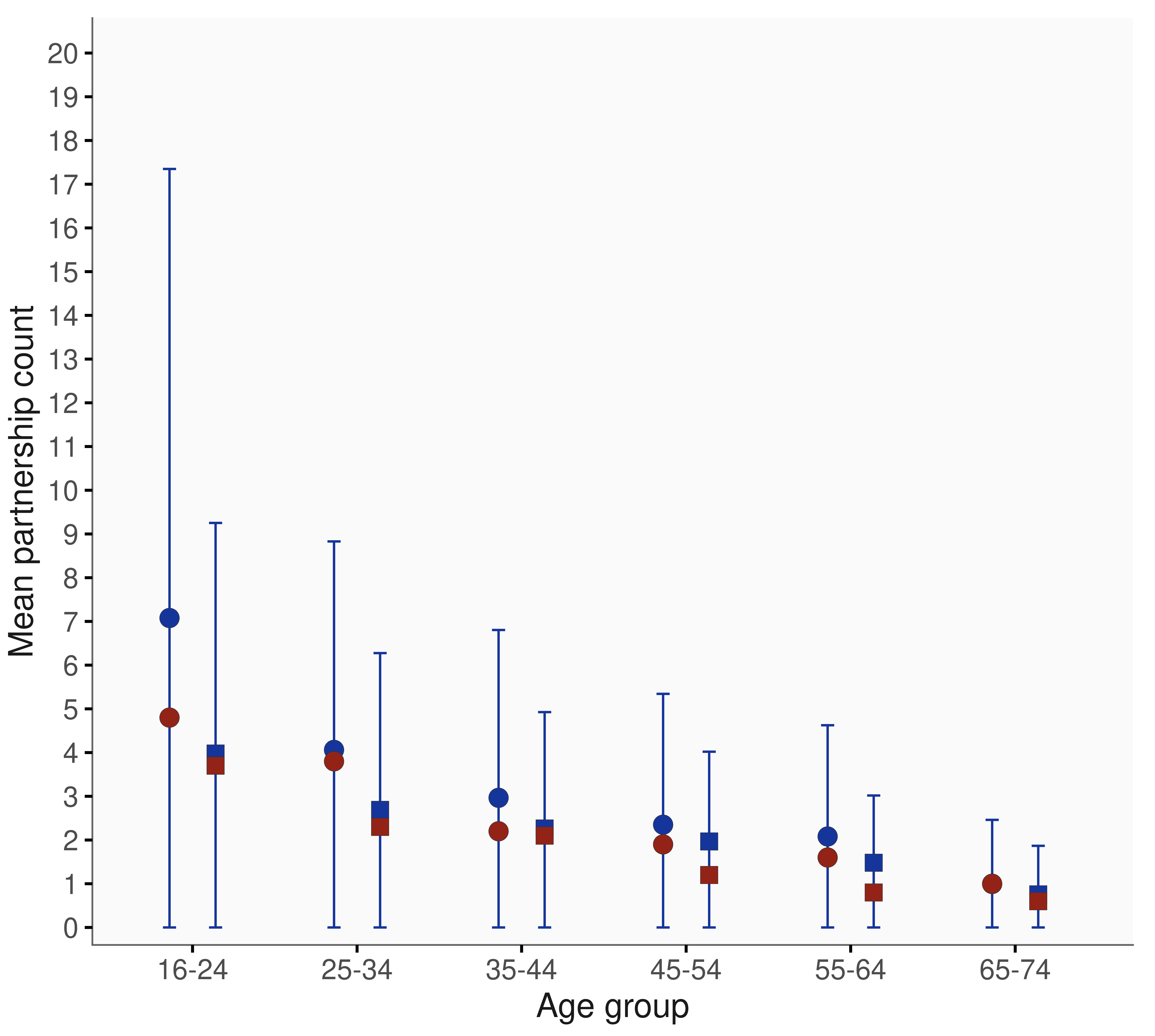}
  \caption{15\%-concurrency ($\theta_{\mathrm{conc}} = 0.15$), best-fit id:557}
\end{subfigure}
\caption{Mean partnerships by age group, opposite-sex stratum. The fit is reasonably good across both scenarios, with the model reproducing the monotonic age-related decline in the mean partnership count. }
\label{fig:scatter_comparison_opposite-sex}
\end{figure}

\item \textbf{Same-sex partnerships.}
We observe that the mean partner counts are higher than for the opposite-sex stratum, particularly among males in the calibration results for the same-sex stratum (Figure~\ref{fig:scatter_comparison_same-sex}). We also observe that the model fits better for females of all age groups compared to males. 

\begin{figure}[H]
\centering
\begin{subfigure}{0.49\linewidth}
\centering
\includegraphics[width=\linewidth, height=9cm, keepaspectratio]{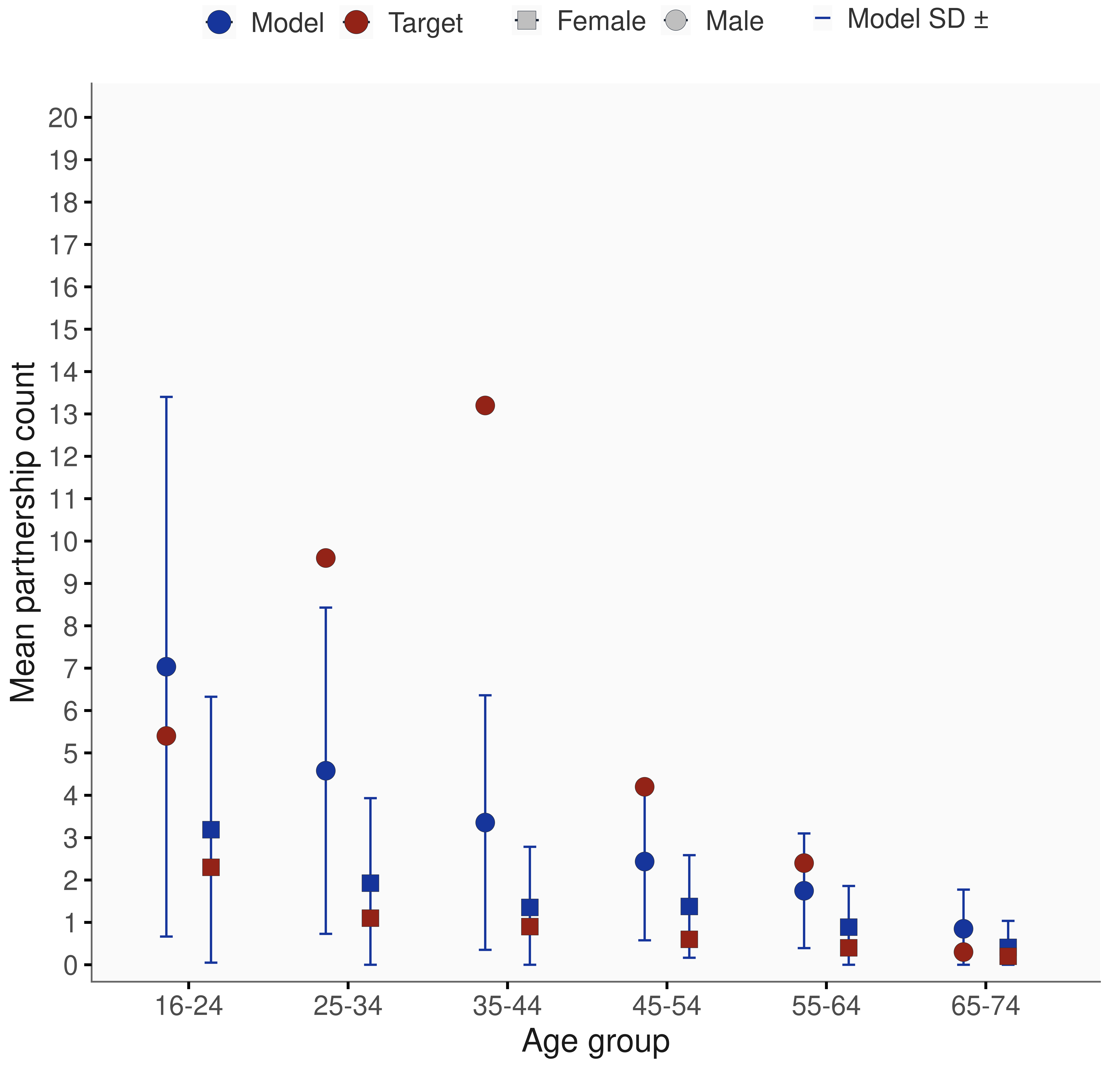}
\caption{No concurrency ($\theta_{\mathrm{conc}} = 0$), best-fit id:2777}
\end{subfigure}
\hfill
\begin{subfigure}{0.49\linewidth}
\centering
\includegraphics[width=\linewidth, height=9cm, keepaspectratio]{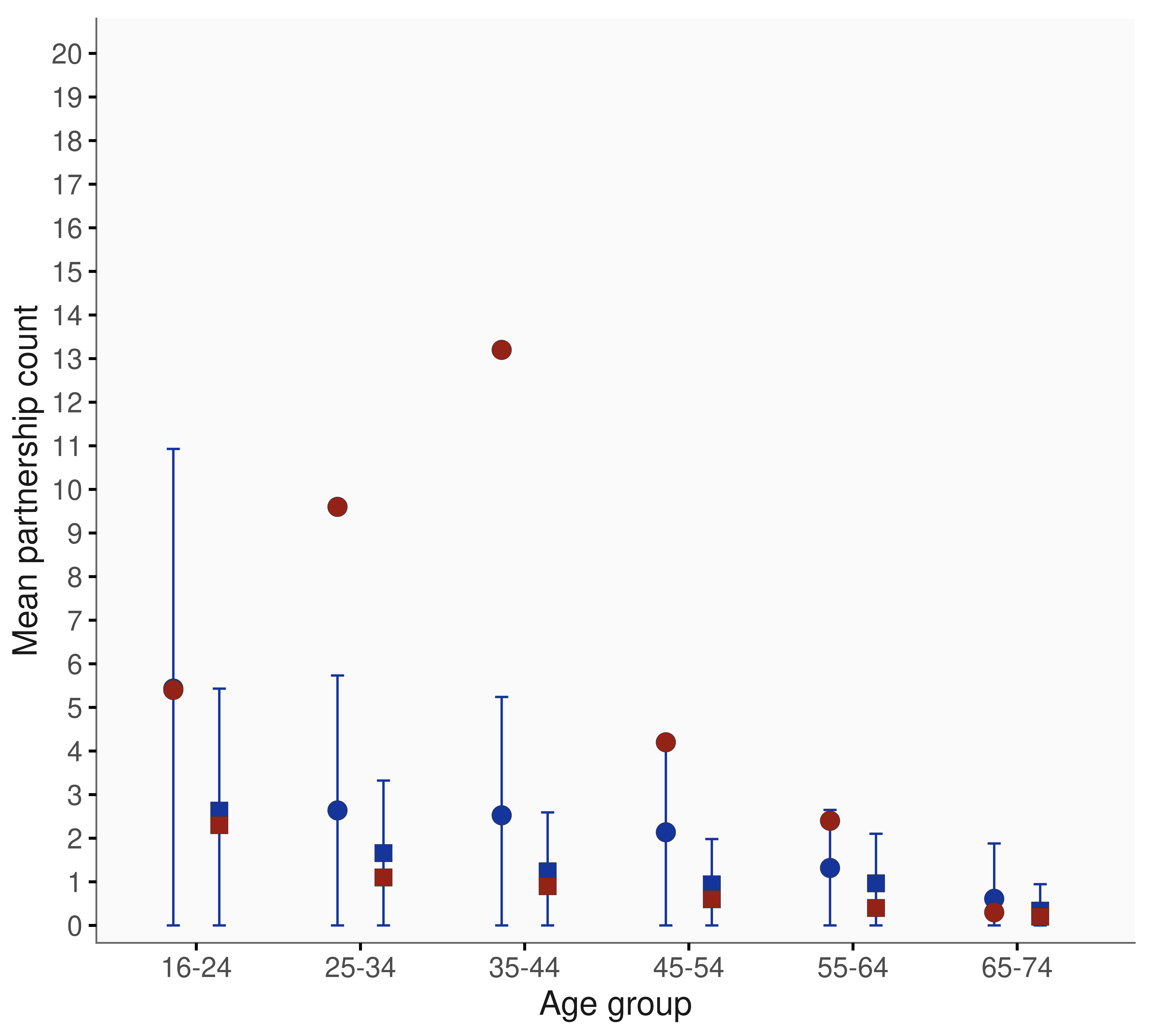}
\caption{15\%-concurrency ($\theta_{\mathrm{conc}} = 0.15$), best-fit id:557}
\end{subfigure}
\caption{Mean partnerships by age group, same-sex stratum. The mean partnership count for the 25--34 (9.6) and 35--44 (13.2) age groups, fall outside the standard deviation bounds of the model output for both scenarios.}
\label{fig:scatter_comparison_same-sex}
\end{figure}
\item \textbf{Bisexual partnerships.}
We observe that the fit for females is better in the older age groups for both scenarios compared to males in the calibration results for bisexual agents (Figure~\ref{fig:scatter_comparison_bisexual}). We also observe that the fit for males is best achieved for the 16--24 age group in the no-concurrency scenario, followed by the older age groups in the 15-\% concurrency scenario. 
\begin{figure}[H]
\centering
\begin{subfigure}{0.49\linewidth}
\centering
\includegraphics[width=\linewidth, height=9cm, keepaspectratio]{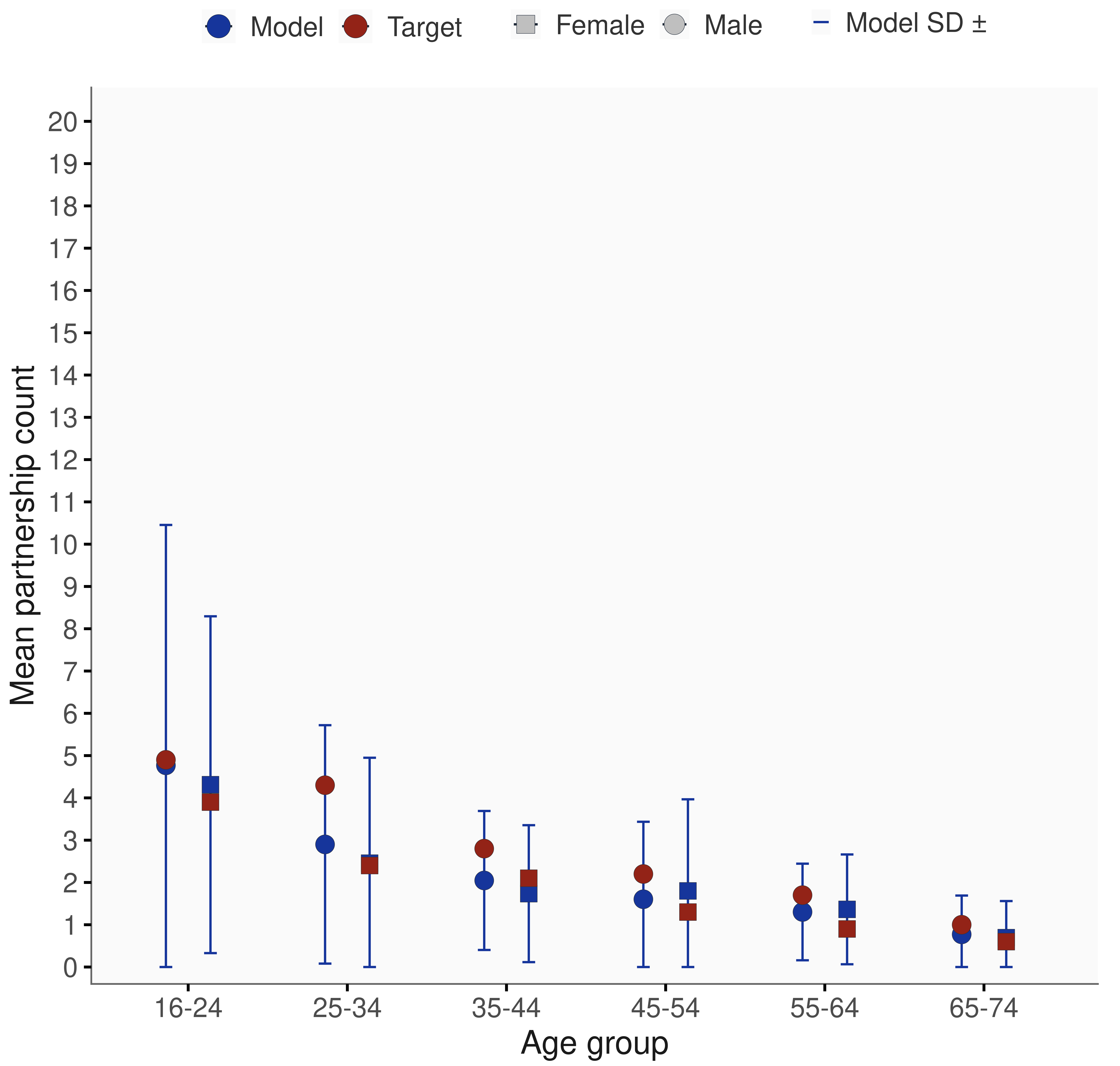}
\caption{No concurrency ($\theta_{\mathrm{conc}} = 0$), best-fit id:2777}
\end{subfigure}
\hfill
\begin{subfigure}{0.49\linewidth}
\centering
\includegraphics[width=\linewidth, height=9cm, keepaspectratio]{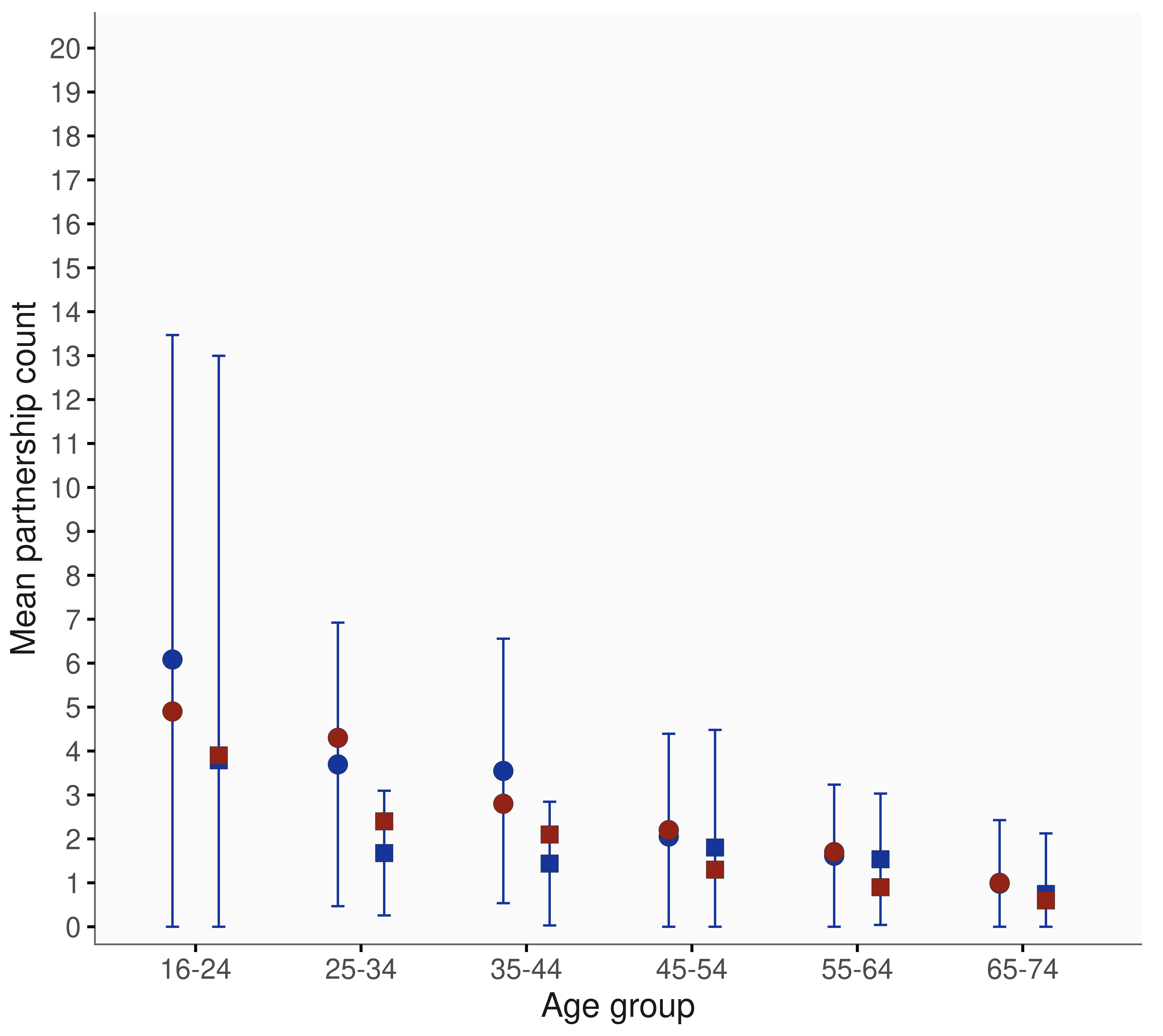}
\caption{15\%-concurrency ($\theta_{\mathrm{conc}} = 0.15$), best-fit id:557}
\end{subfigure}
\caption{Mean partnerships by age group, bisexual stratum. The mean value of the target of the partnership count is intermediate between the strata of opposite-sex and the same-sex, and the model achieves the best fit of the three sexual orientation groups for both scenarios.}
\label{fig:scatter_comparison_bisexual}
\end{figure}
\end{enumerate}
\subsubsection{Model input parameters}
\label{subsubsec:model_inputs}
The combinations of parameters that produced the lowest MSE between the calibration targets and the model outputs were selected for the no-concurrency and the 15\%-concurrency scenarios. Table~\ref{tab:bestfit_parameters} in the \hyperref[app:model_parameters]{Appendix A} presents the best-fitting parameters for each stratum that govern the formation and dissolution mechanisms in the model. These include baseline formation and dissolution probabilities, sex- and sexual orientation-specific multipliers for both formation and dissolution processes for both scenarios of concurrency. The age-dependent structure is captured through the \textit{youth boost} parameters and the exponential \textit{age decay} terms for both formation and dissolution. 

Table~\ref{tab:fixed_simulation_params} in \hyperref[app:model_parameters]{Appendix A} summarises the fixed simulation parameters that remain constant for multiple simulation runs. These include population size, temporal resolution, total simulation duration, number of replicates, and concurrency structure. Additionally, the parameters driving individual-level heterogeneity in sexual behaviour such as the negative binomial dispersion and probability terms, as well as duration-dependent decay parameters, are listed in the Supplementary material.

\subsection{\textit{Susceptible-Infectious-Susceptible} (SIS) Transmission Model}
\label{subsec:SIS_model_description}
We use a Susceptible-Infectious-Susceptible (SIS) framework of infection model to illustrate how the partnership model can be applied for epidemiological study of bacterial STIs. The SIS model partitions the population into two states: \textbf{Susceptible}, referring to agents who are able to be infected if exposed to a pathogen, and \textbf{Infectious}, referring to agents who are currently infected and capable of infecting susceptible partners. Individuals move from \textit{S} to \textit{I} upon infection and then from \textit{I} to \textit{S} upon recovery. For many bacterial STIs, infected individuals recover without acquiring long-term immunity and return immediately to the susceptible state, leaving them at risk of reinfection \citep{keeling_rohani, kretzschmar2017pair}. 
\begin{figure}[H]
\centering
\includegraphics[width=0.8\textwidth, height=4cm]{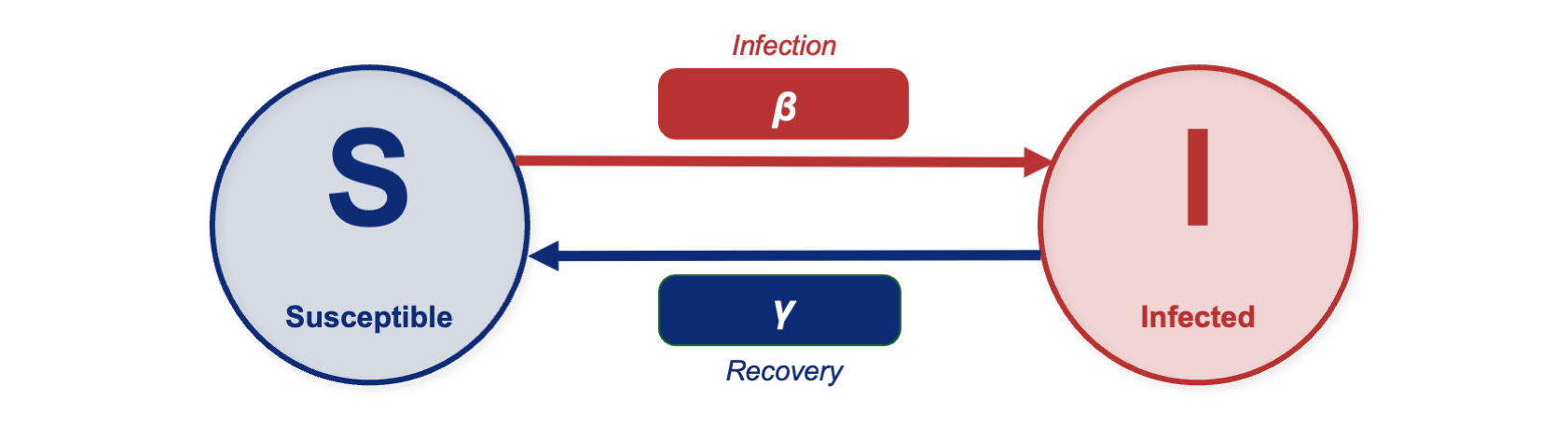}
\caption{SIS model for bacterial STI transmission}
\label{fig:sis_model}
\end{figure} 
Figure~\ref{fig:sis_model} shows the flow diagram for the SIS model, illustrating the transitions between epidemiological states. Solid arrows indicate progression from susceptible (\textit{S}) to infectious (\textit{I}) governed by transmission probability \(\beta\), and from infectious (\textit{I}) back to susceptible (\textit{S}), governed by recovery probability \(\gamma\). Transmission and recovery mechanisms are implemented as the probability per time step instead of the rates, and are adapted from the classical definition of transmission and recovery rates from \citet{keeling_rohani}. $\beta$ denotes the probability that an infectious agent transmits the pathogen to a given susceptible partner in a single active partnership in one time step. $\gamma$ denotes the probability that an infectious agent clears the infection and returns to the susceptible state in a given time step. 

The SIS model in this study simulates the transmission of a hypothetical bacterial STI with synthetic parameters listed in Table~\ref{tab:disease-parameters} in \hyperref[app:model_parameters]{Appendix A}. Disease simulations are run on a randomly selected single partnership network with 100 independent disease seeding and transmission runs simulated on this same underlying network. The reported variation across the 100 disease runs, therefore, reflects the stochasticity in disease seeding and transmission alone, not the variation in the underlying partnership network structure.
\section{Model outputs}\label{sec:model_outputs}
\subsection{Partnership count, duration and activity distributions} \label{subsec:partnership_count_duration}
\begin{figure}[H]
\centering
\begin{subfigure}[b]{\linewidth}
\centering
\includegraphics[width=0.75\linewidth, height=6cm]{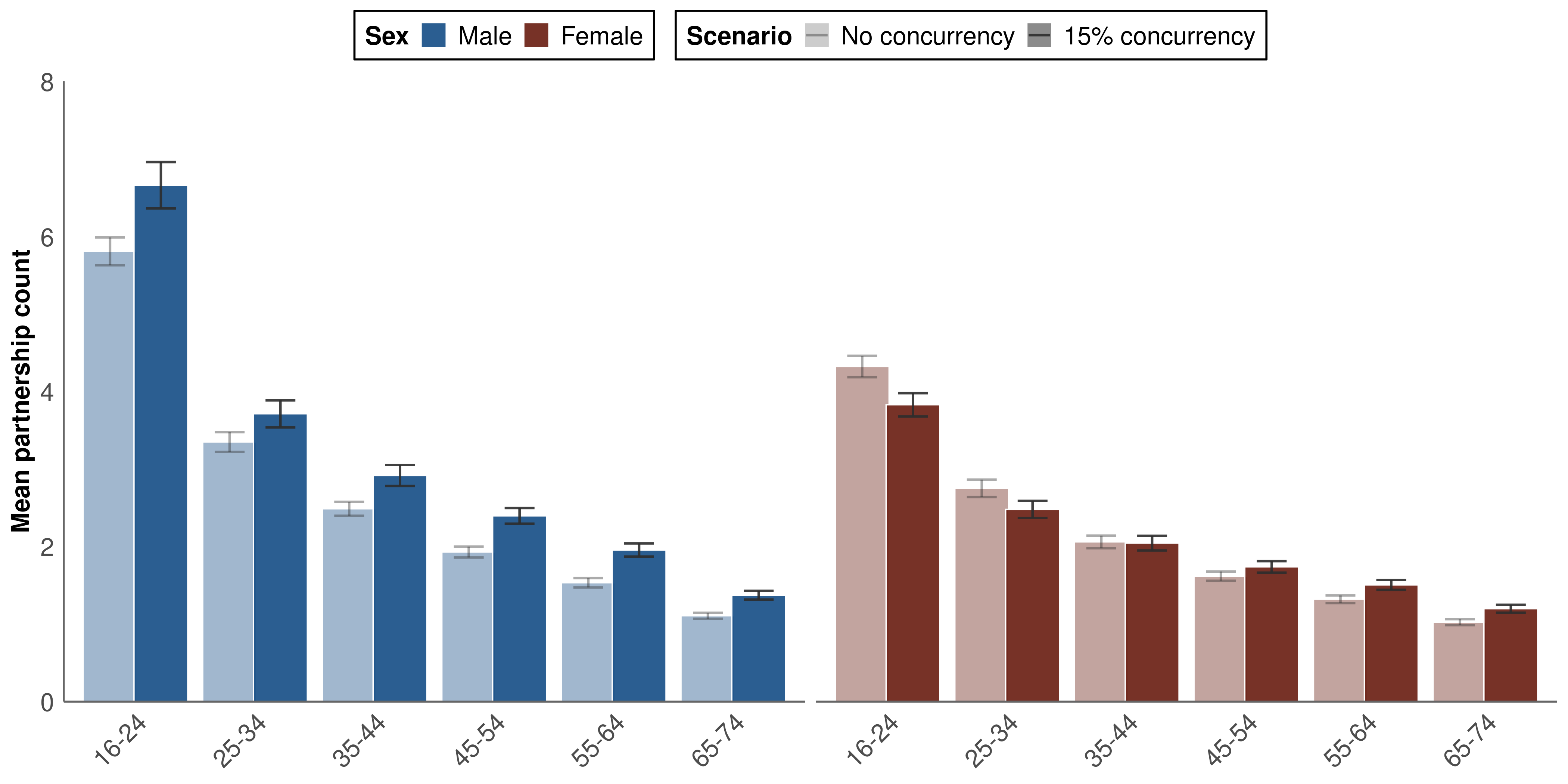}
\caption{Mean partnership count}
\label{fig:opposite_sex_count}
\end{subfigure}
\vspace{6pt}
\begin{subfigure}[b]{\linewidth}
\centering
\includegraphics[width=0.75\linewidth, height=6cm]{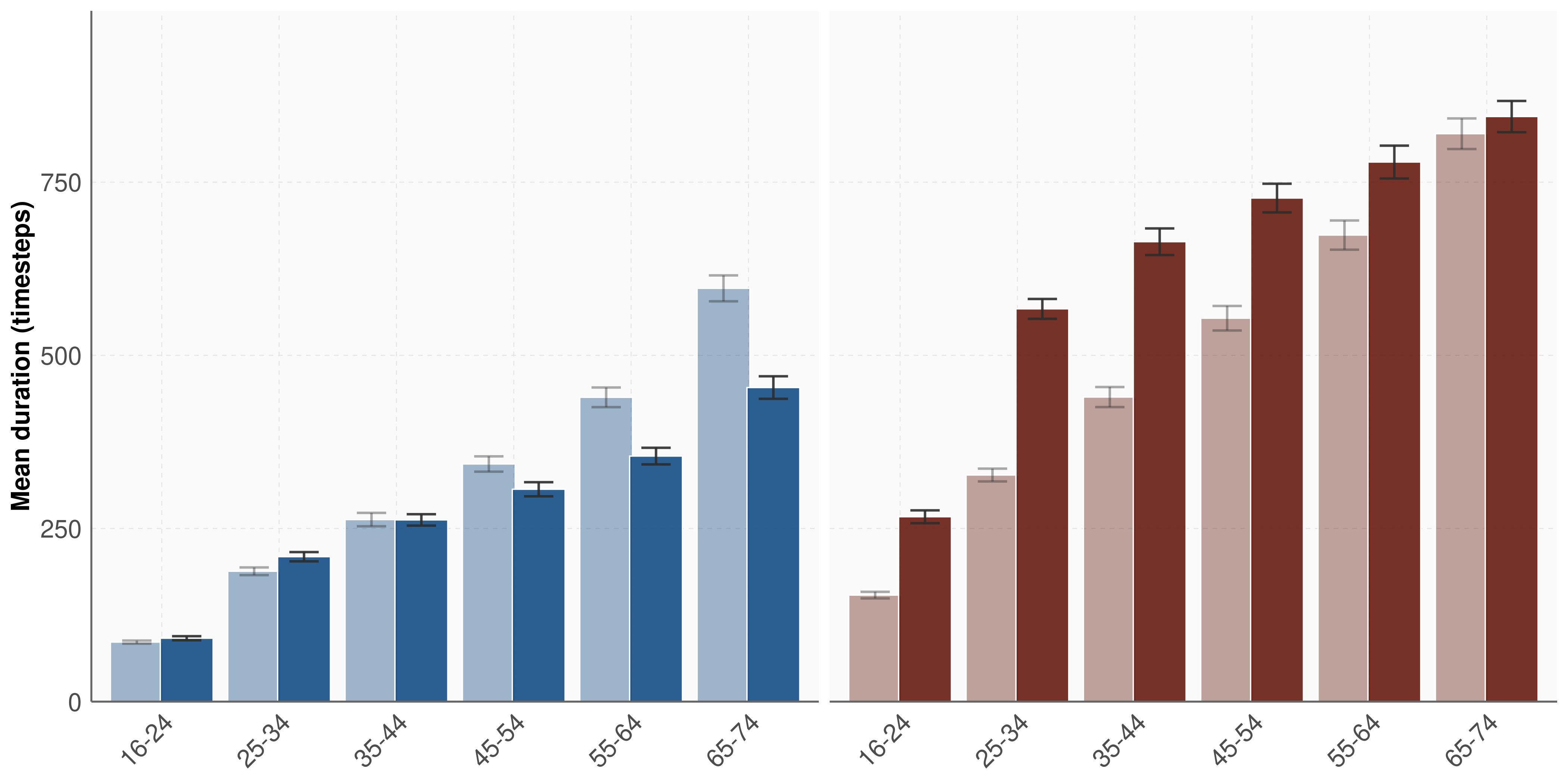}
\caption{Mean partnership duration}
\label{fig:opposite_sex_duration}
\end{subfigure}
\caption{Mean partnership count and duration for opposite-sex sexual orientation across 100 simulations, by age group and sex. Lighter bars:$\theta_{\mathrm{conc}} = 0$; darker bars: $\theta_{\mathrm{conc}} = 0.15$. Error bars: $\pm$1~SD across simulations.}
\label{fig:opposite_sex_count_duration}
\end{figure}

All model outputs presented in this section were simulated using the best fitting parameters for the no-concurr-ency and 15\%-concurrency scenarios provided in Table~\ref{tab:bestfit_parameters} and the fixed parameters described in Table~\ref{tab:fixed_simulation_params} in \hyperref[app:model_parameters]{Appendix A}. For each of the three strata of sexual orientations (opposite-sex, same-sex, and bisexual), we have summarised four key aspects of simulated partnership behaviour, each shown as a two-panel figure comparing: (i) mean partnership count and mean partnership duration by age group and sex, and (ii) the percentage of single agents (0 partners) and the percentage of agents with high-activity (10 or more partners) by age group and sex. All results are averaged over 100 simulations. The results shown below are for the opposite-sex stratum, with the plots for the same-sex and bisexual strata given in \hyperref[app:additional_figures]{Appendix B}.
The introduction of 15\% concurrency increases the mean count of partners for the opposite-sex oriented agents in all age groups for males, with the highest increase in the 16--24 age band (Figure~\ref{fig:opposite_sex_count_duration}). For females, the mean partnership count slightly decreases for the 16--24 and 25--34 age groups. The mean duration of partnerships increases for females in the younger age groups and decreases for males in the older age groups with the inclusion concurrency. This can be attributed to partnerships that dissolve more frequently when agents maintain simultaneous partnerships in combination with a higher probability of dissolution among younger age groups.

\begin{figure}[H]
\centering
\begin{subfigure}[b]{\linewidth}
\centering
\includegraphics[width=0.75\linewidth, height=6cm]{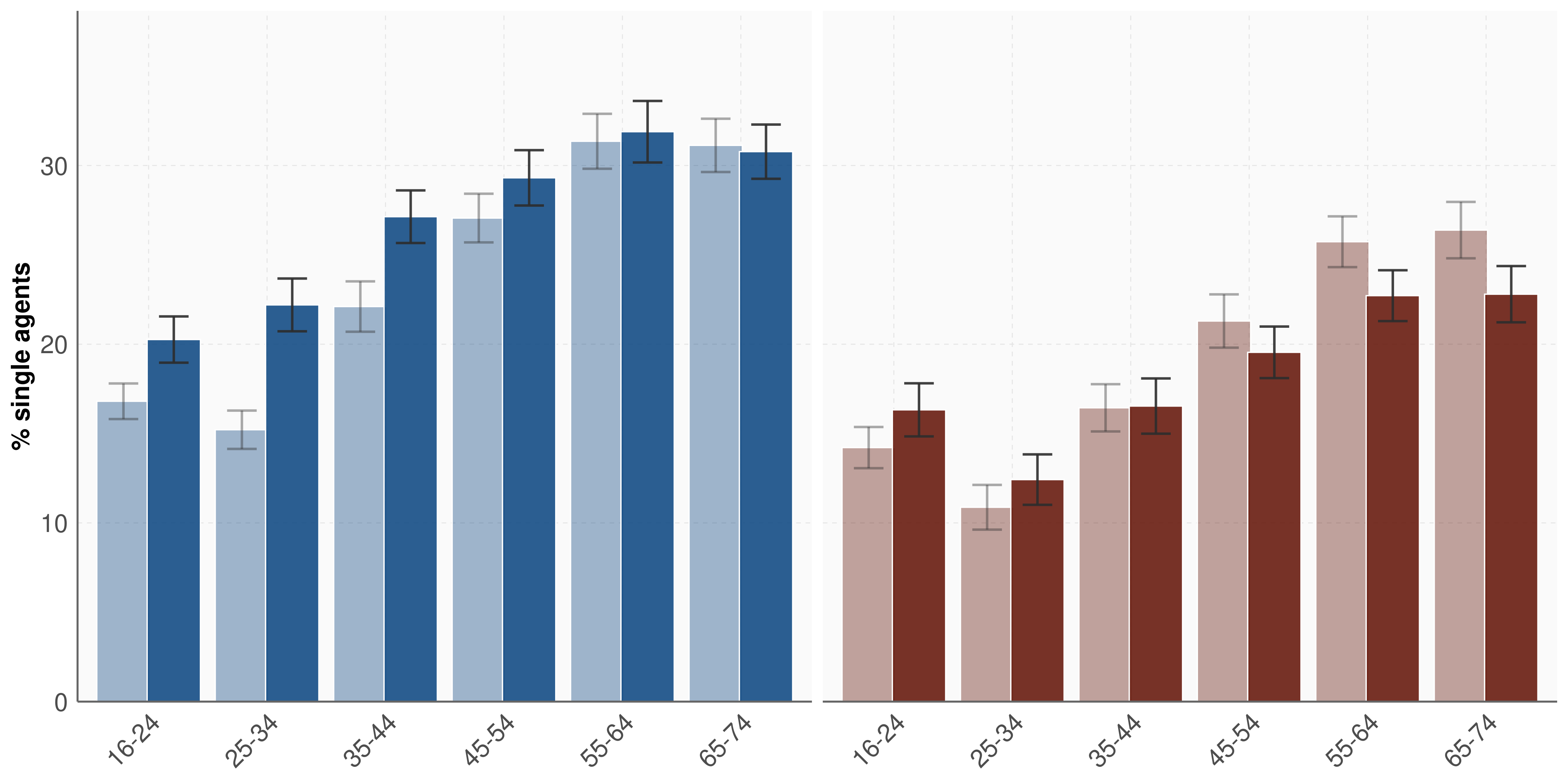}
\caption{\% single agents (0 partners)}
\label{fig:opposite_sex_single}
\end{subfigure}
\vspace{6pt}
\begin{subfigure}[b]{\linewidth}
\centering
\includegraphics[width=0.75\linewidth, height=6cm]{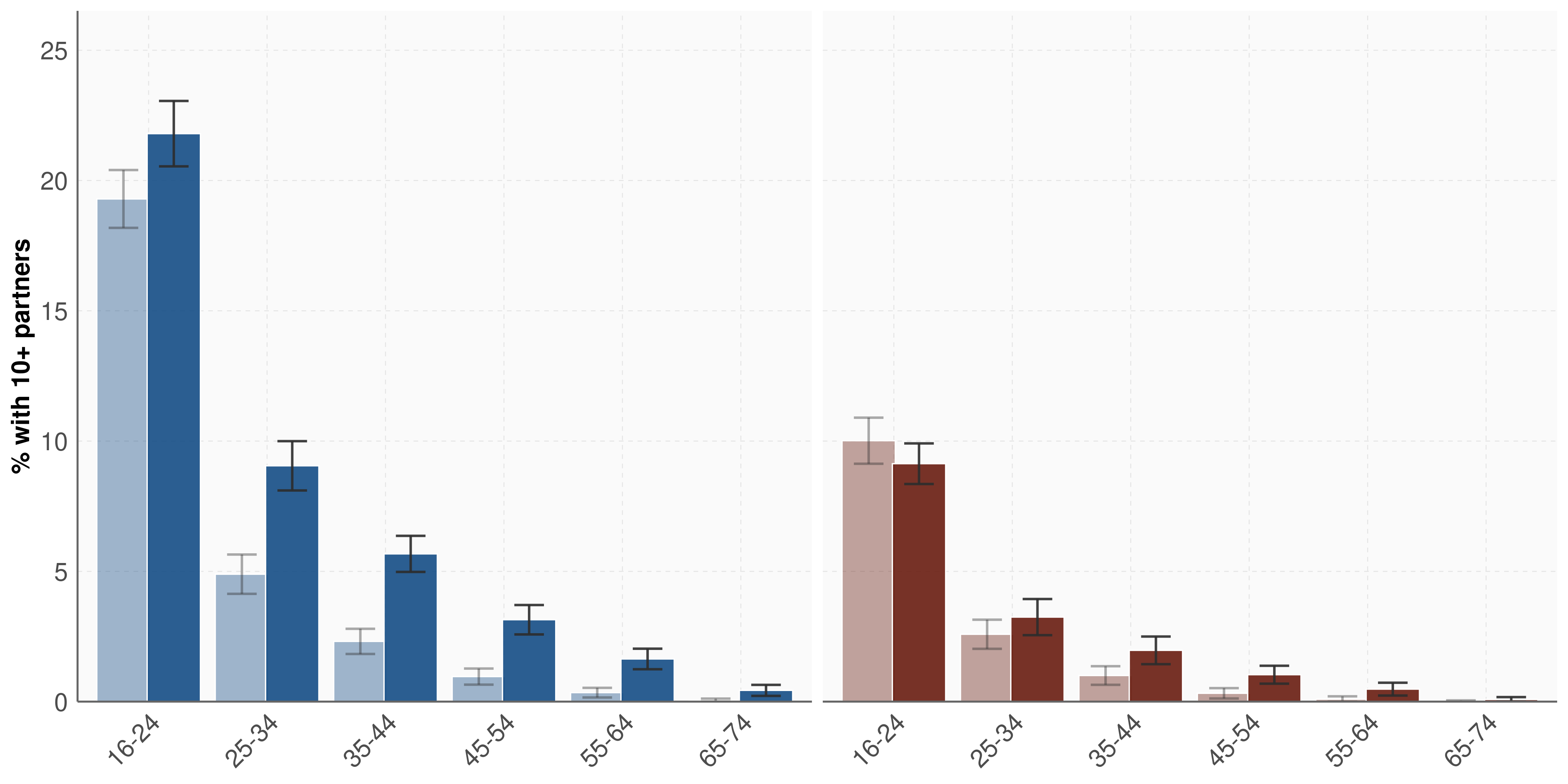}
\caption{\% agents with 10+ partners}
\label{fig:opposite_sex_high}
\end{subfigure}
\caption{Percentage of single agents and percentage of agents with ten or more partners for opposite-sex sexual orientation across 100 simulations, by age group and sex. }
\label{fig:opposite_sex_single_high}
\end{figure}

The percentage of single agents increases with the introduction of concurrency for males in most age groups, for example, from approximately 15\% to 22\% for the age group 25--34 (Figure~\ref{fig:opposite_sex_single_high}). The inclusion of concurrency decreased the proportion of single agents in older age groups among females. The percentage of agents with ten or more partners is highest in the 16--24 age group for both sexes and increases with the introduction of concurrency for males, from approximately 19\% to 22\% for the 16--24 age group, and from approximately 5\% to 9\% for the 25--34 age group, while the increase for females is much smaller and the proportion declines rapidly with age in both scenarios.

\subsection{Mapping partnerships on dynamic networks} 
\label{subsec:networks}
Using partnership records, we constructed dynamic networks to capture partnership turnover that emerges in our modelled population, including short-term partnerships, serial monogamy, and concurrency. The partnership network is represented as an undirected graph $G(t) = (V(t),\ E(t))$, where the node set $V(t)$ comprises all agents active at the time step $t$ and the edge set $E(t)$ contains all partnerships active at $t$. A partnership between agents $i$ and $j$ is present in $E(t)$ if and only if its formation time $t_{\text{start}}$ and dissolution time $t_{\text{end}}$ satisfy $t_{\text{start}} \leq t \leq t_{\text{end}}$. Agents without an active partnership at $t$ are retained as isolated nodes in $G(t)$, preserving the complete population in the network representation. 
\subsubsection{Degree distributions over time}
\label{subsubsec:degree_distribution}
Table~\ref{tab:network_summary} demonstrates that introducing 15\% concurrency altered the structure of the partnership network in ways that increase the potential for transmission of STIs due to greater connectivity. Although the median degree remained close to 1 in both scenarios, the mean degree increased from $0.80 \pm 0.07$ to $1.07 \pm 0.13$ because a subset of individuals formed concurrent partnerships. The concurrent individuals had a mean degree of $3.28 \pm 0.23$, while the monogamous individuals maintained a degree of one in both scenarios, indicating that the increase in network connectivity was concentrated between the concurrent individuals rather than spreading throughout the population. The mean length of the shortest path decreased from 8.08 to 6.92 and the median length of the shortest path decreased from 8 to 7, indicating that individuals were connected through fewer intermediary partnerships.
\begin{table}[htbp]
\centering
\caption{Partnership network summary statistics by concurrency scenario,
averaged across 1000 simulation runs (mean $\pm$ SD).}
\label{tab:network_summary}
\begin{tabular}{lrr}
\toprule
\textbf{Metric} & \textbf{no-concurrency} & \textbf{15\% concurrency} \\
\midrule
\multicolumn{3}{l}{\textit{Degree distribution}} \\
\quad Mean degree & $0.80 \pm 0.07$ & $1.07 \pm 0.13$ \\
\quad Median degree & $0.99 \pm 0.09$ & $1.00 \pm 0.07$ \\
\quad Max degree & $1.00 \pm 0.00$ & $9.56 \pm 0.86$ \\
\addlinespace
\multicolumn{3}{l}{\textit{Degree and counts by concurrency status}} \\
\quad Mean degree, concurrent agents & --- & $3.28 \pm 0.23$ \\
\quad $n$ concurrent agents & $0$ & $1{,}650 \pm 286$ \\
\quad Mean degree, monogamous agents & $1.00 \pm 0.00$ & $1.00 \pm 0.00$ \\
\quad $n$ monogamous agents & $12{,}000 \pm 1{,}060$ & $10{,}600 \pm 758$ \\
\addlinespace
\multicolumn{3}{l}{\textit{Network structure}} \\
\quad Shortest path, mean & $8.08$ & $6.92$ \\
\quad Shortest path, median & $8$ & $7$ \\
\bottomrule
\end{tabular}
\end{table}

\subsubsection{Ego network snapshots}
\label{subsubsec:ego_network_snapshots}

\begin{figure}[H]
\includegraphics[width=1\linewidth, height=18.5cm]{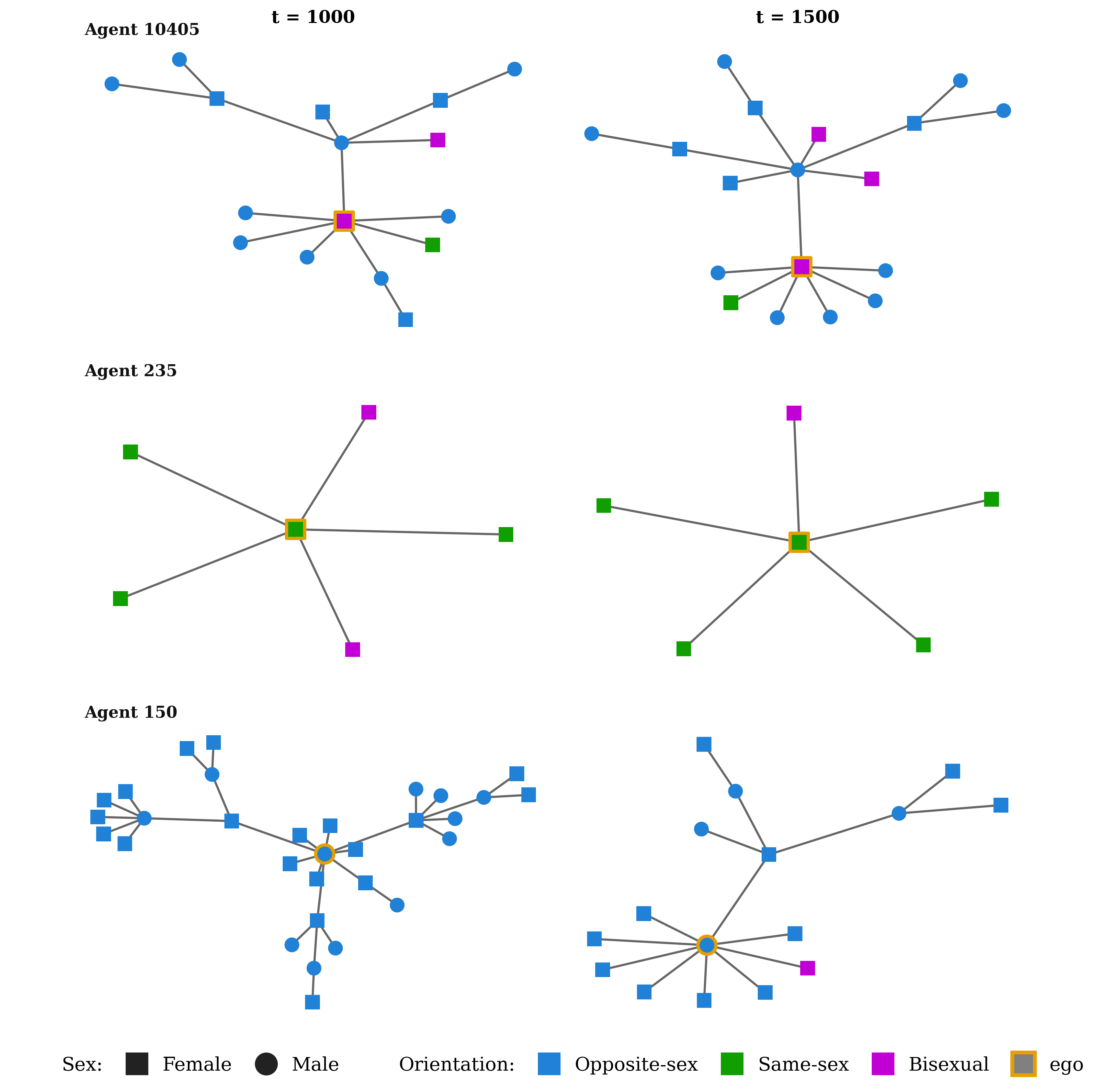}
\caption{Snapshot network (sexual orientation-coloured) for three selected agents at
$t \in \{ 1000, 1500\}$, for 15\%-concurrency scenario. The nodes were selected explicitly for the combination of sex and sexual orientation to demonstrate the role of bridging partnerships}
\label{fig:ego_networks_comparison_15pcconc}
\end{figure}

Under the 15\%-concurrency scenario, agents accumulate substantially larger ego networks over time (Figure ~\ref{fig:ego_networks_comparison_15pcconc}), with multi-hop chains emerging as concurrent partners' own partners become reachable within the 3-hop neighbourhood. Agent 10405, a bisexual female node, is a densely connected agent at both snapshots, embedded throughout in a large network of multiple dense clusters joined by bisexual and same-sex bridging nodes. The bisexual (magenta) and same-sex (green) nodes appearing in the ego networks in the 15\%-concurrency scenario act as bridges connecting clusters of otherwise opposite-sex (blue) partnerships. Agent 235, the same-sex female node, stays stable from time step $t=1000$ to $t=1500$. Agent 150, a male of opposite-sex orientation, depicts the change in network density: a dense cluster at $t=1000$ becomes relatively sparse at $t=1500$. The bisexual bridging behaviour observed in the networks of Agents 235 and 10405 has direct epidemiological relevance: the 15\%-concurrency networks generate large, mixed sexual orientation connected components through which an infection could potentially reach many agents indirectly, including through the boundaries of sexual orientations and through concurrent partnerships. An additional figure of the ego network for the no-concurrency scenario is given in \hyperref[app:additional_figures]{Appendix B} where we see only single chains of connections due to the absence of concurrency.

\begin{figure}[H]
\includegraphics[width=1\linewidth, height=12cm]{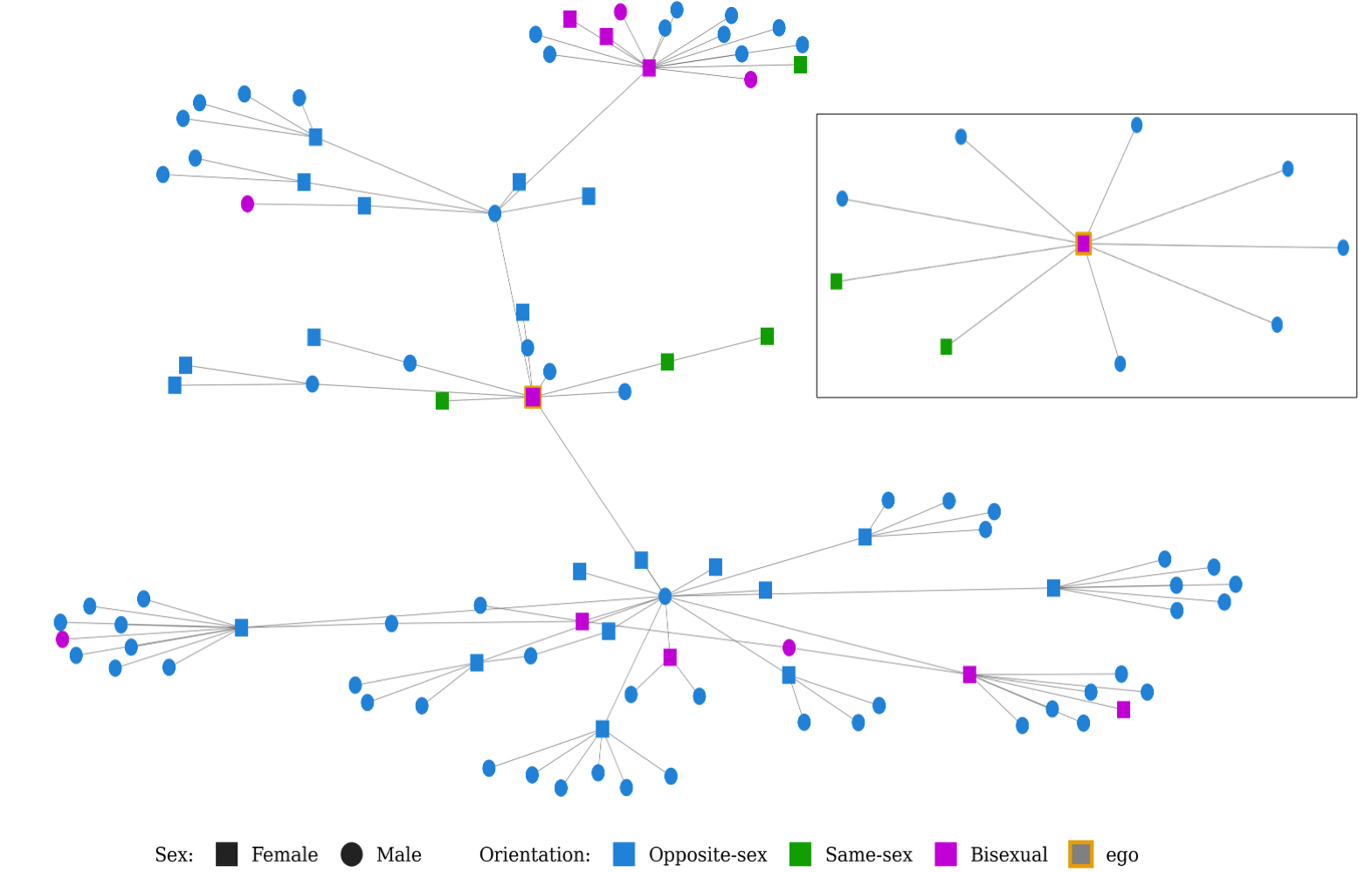}
\caption{3-hop sexual partnership network of a bisexual individual (Agent 10405) aggregated over 5 years of simulation. Inset: same ego's 1-hop network only. Agent 10405 is the sole link between the same-sex (green) and opposite-sex (blue) partnership clusters illustrating how bisexual individuals can structurally bridge otherwise-segregated sexual networks.}
\label{fig:15pcconc_ego_network_aggregated}
\end{figure}

We further look at the accumulated partnerships of Agent 10405 to understand how concurrency enables the formation of connected components during the simulation period. We observe a highly connected component build around Agent 10405, which is the ego node (Figure ~\ref{fig:15pcconc_ego_network_aggregated}). The bisexual female ego node connects to multiple opposite-sex and same-sex partners, several of whom are hubs within their own component, illustrating how concurrency lets one component connect across different sexual orientation groups. The inset box in Figure ~\ref{fig:15pcconc_ego_network_aggregated} is a smaller, separate snapshot of the immediate partners (single-hop) of Agent 10405 accumulated over the course of the simulation. 

\subsection{SIS transmission model} 
\label{subsec:sis_model}
\begin{figure}[H]
\centering
\includegraphics[width=\linewidth, height=6.5cm]{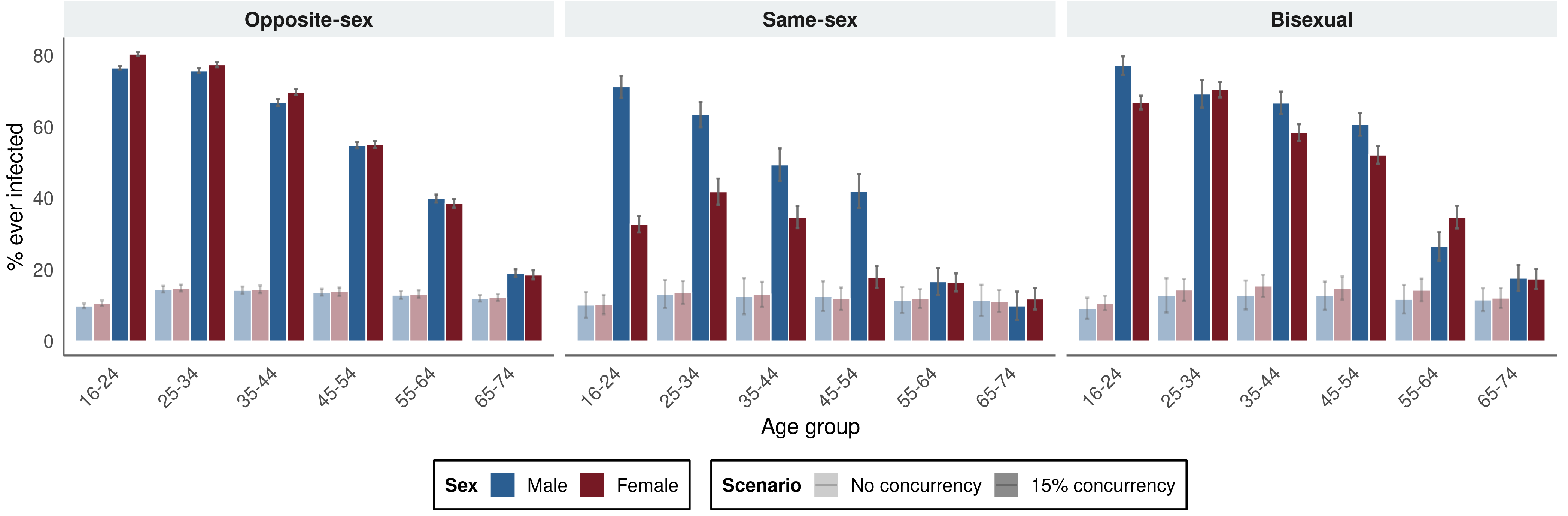}
\caption{Mean percentage of agents ever infected ($\beta = 0.20$ and $\gamma = 0.10$) by age group, sexual orientation, and sex under no-concurrency ($\theta_{\mathrm{conc}} = 0$, transparent bars) and 15\%-concurrency ($\theta_{\mathrm{conc}} = 0.15$, opaque bars) scenarios, averaged across 100 independent disease simulation runs.}
\label{fig:prevalence_age}
\end{figure}

\begin{figure}[H]
\centering
\begin{subfigure}[b]{\linewidth}
\centering
\includegraphics[width=1\linewidth, height=6cm]{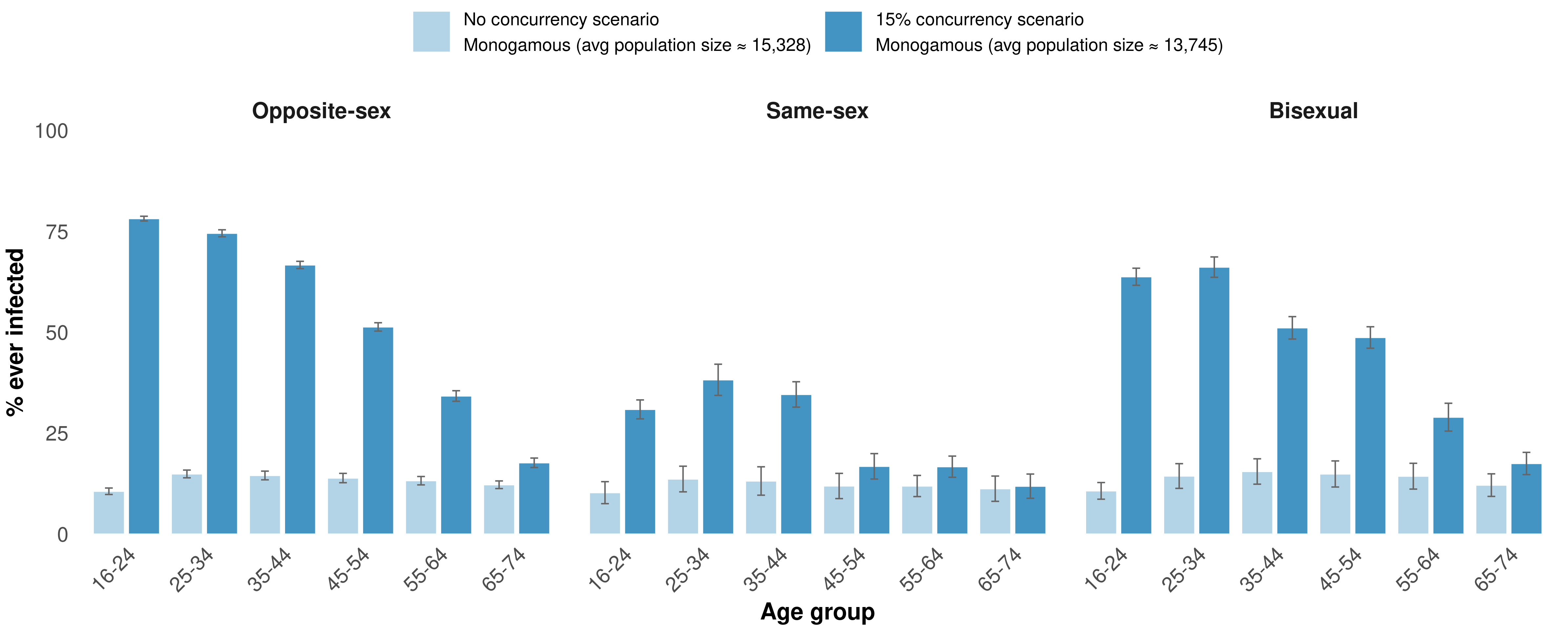}
\caption{Female agents}
\label{fig:spillover_female}
\end{subfigure}

\vspace{0.5cm}

\begin{subfigure}[b]{\linewidth}
\centering
\includegraphics[width=1\linewidth, height=5cm]{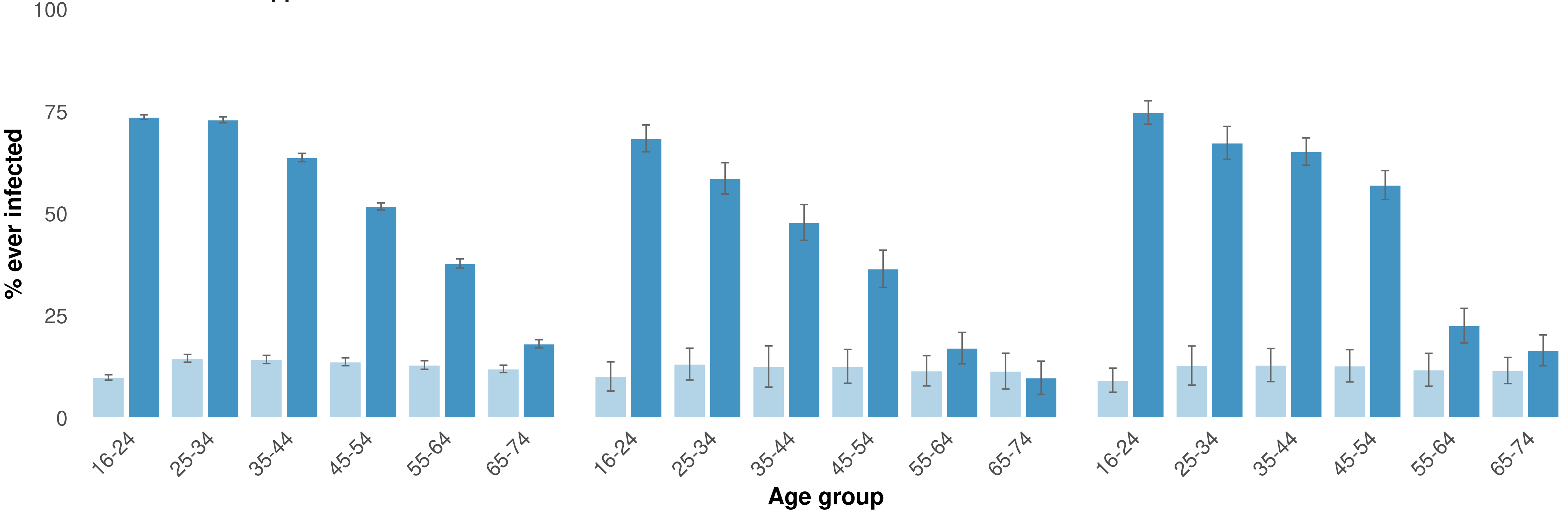}
\caption{Male agents (legend as in panel (a))}
\label{fig:spillover_male}
\end{subfigure}
\caption{Prevalence of infection ($\beta = 0.20$ and $\gamma = 0.10$) by sexual orientation and age group among \textit{Monogamous} agents within the no-concurrency and 15\%-concurrency scenarios, for female (a) and male (b) agents}
\label{fig:spillover}
\end{figure}

We estimate the mean prevalence of infection by age group, sexual orientation, and sex, by running 100 independent \textit{SIS} transmission model, with the same disease parameters but varying seeds on the same network in each concurrency scenario (Figure~\ref{fig:prevalence_age}). In the no-concurrency scenario, the prevalence is low and does not vary noticeably between all sex and sexual orientation groups (9--15\%). The inclusion of concurrency produces a substantial increase in prevalence overall, together with a notable effect related to age, where prevalence is highest among the youngest agents (16--24) and declines steadily with age. This effect is strongest for opposite-sex and bisexual agents, who reach 77--80\% ever-infected at 16--24 before falling to 18--24\% by 65--74. Same-sex agents show consistently lower prevalence under concurrency (37--61\%) and a non-monotonic age pattern, most likely reflecting the comparatively smaller size of the same-sex partnership network and correspondingly greater stochastic variability across runs.
To further examine the impact of concurrency, we classify agents into two main categories: \textit{Monogamous} and \textit{Polygamous}. In the no-concurrency scenario, all agents in partnerships are classified as \textit{Monogamous}. In the 15\%-concurrency scenario, agents that have ever been in a concurrent partnership are classified as \textit{Polygamous}, and those who are never in concurrent partnerships are classified as \textit{Monogamous}. \textit{Monogamous} agents in the 15\%-concurrency scenario include agents who are designated as concurrency-eligible but never form one over the course of the simulation. We compare the prevalence of infection for \textit{Monogamous} agents in the no-concurrency scenario with the \textit{Monogamous} subset within the 15\%-concurrency scenario (Figure~\ref{fig:spillover}). Although the behaviour of these agents is identical across both scenarios, their infection risk increases by roughly $\sim$6--8 $\times$ under 15\%-concurrency for the opposite-sex and bisexual oriented agents in the youngest age groups, falling to roughly $\sim$3-4 $\times$ by middle age and by $\sim$1.5 $\times$ by the oldest age group. For agents of the same-sex orientation, there is a marked difference in prevalence rates between males and females, particularly in the 16--24 and 25--34 age groups. These results demonstrate that an individual's monogamous status does not fully protect them from STI risk introduced by concurrency elsewhere in the sexual partnership network.

\begin{figure}[H]
\centering
\begin{subfigure}[b]{\linewidth}
\centering
\includegraphics[width=1\linewidth, height=6cm]{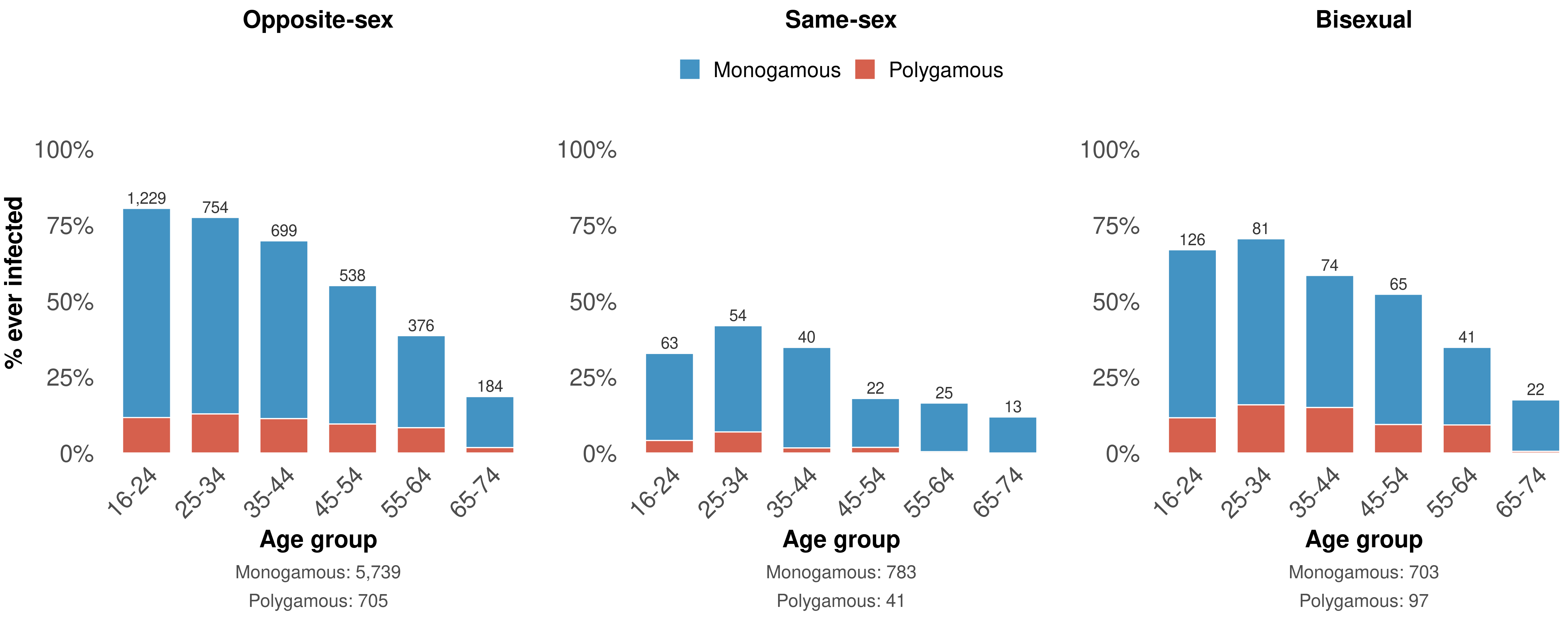}
\caption{Female agents}
\label{fig:prevalence_15pc_female}
\end{subfigure}

\vspace{0.5cm}

\begin{subfigure}[b]{\linewidth}
\centering
\includegraphics[width=1\linewidth, height=5cm]{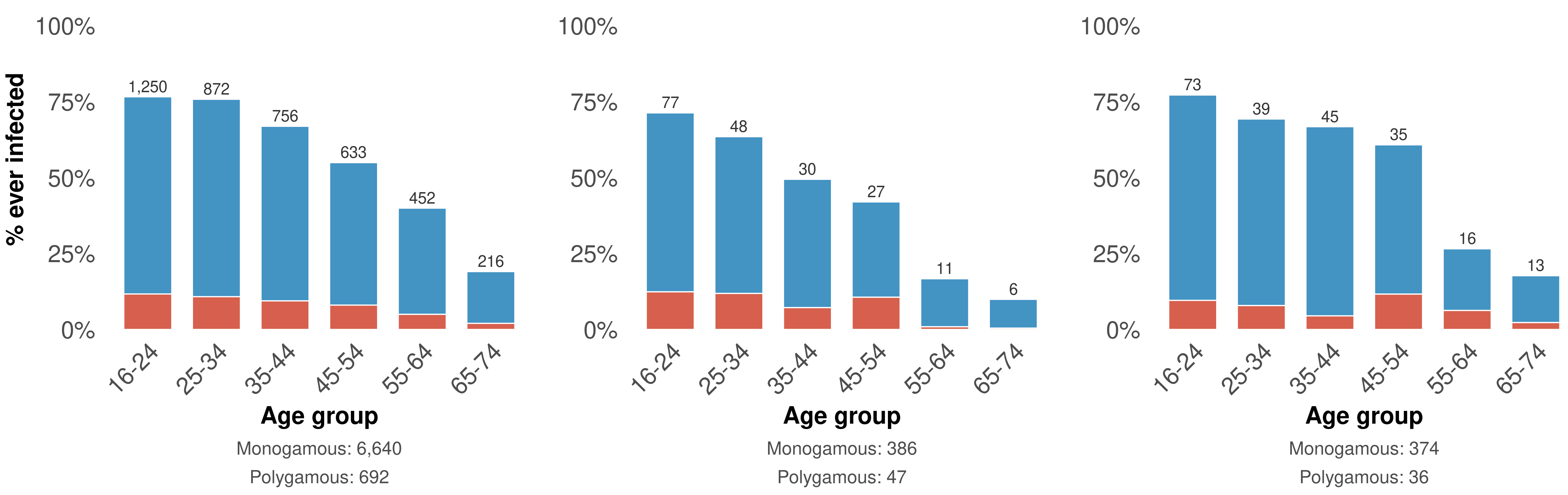}
\caption{Male agents (legend as in panel (a))}
\label{fig:prevalence_15pc_male}
\end{subfigure}
\caption{Prevalence of infection ($\beta = 0.20$ and $\gamma = 0.10$) by age group and sexual orientation, comparing \textit{Monogamous} and \textit{Polygamous} agents in the 15\%-concurrency scenario, for female (a) and male (b) agents. The total counts of infected agents by concurrency status is given under the X-axis and for the respective age groups on top of the bars.}
\label{fig:prevalence_15pc}
\end{figure}

We examine the impact of concurrency on the prevalence of infection by deconstructing the 15\%-concurrency scenario by individual concurrency status 
(Figure~\ref{fig:prevalence_15pc}). Although they are a small proportion of the simulated population, \textit{Polygamous} agents are $\sim$10-12\% of the total share of infected agents in the younger age groups for opposite-sex and bisexual females and males of all orientations. This suggests that a substantial proportion of \textit{Polygamous} agents in the younger age groups for opposite-sex and bisexual females, and males of all orientations get infected in the 15\%-concurrency scenario. For all three orientations, the overall prevalence of infection decreases with age. For the opposite-sex oriented agents, the distribution of infected agents by age group and concurrency status is nearly identical, which is expected. Despite the population of same-sex oriented females being twice as much as same-sex oriented males, we see nearly identical number of infected \textit{Polygamous} agents. The proportion of infected \textit{Monogamous} agents is twice as higher among same-sex oriented males compared to females, particularly in the 16--24 and 25-34 age groups.

\begin{figure}[H]
\centering
\includegraphics[width=1\linewidth, height=6.5cm]{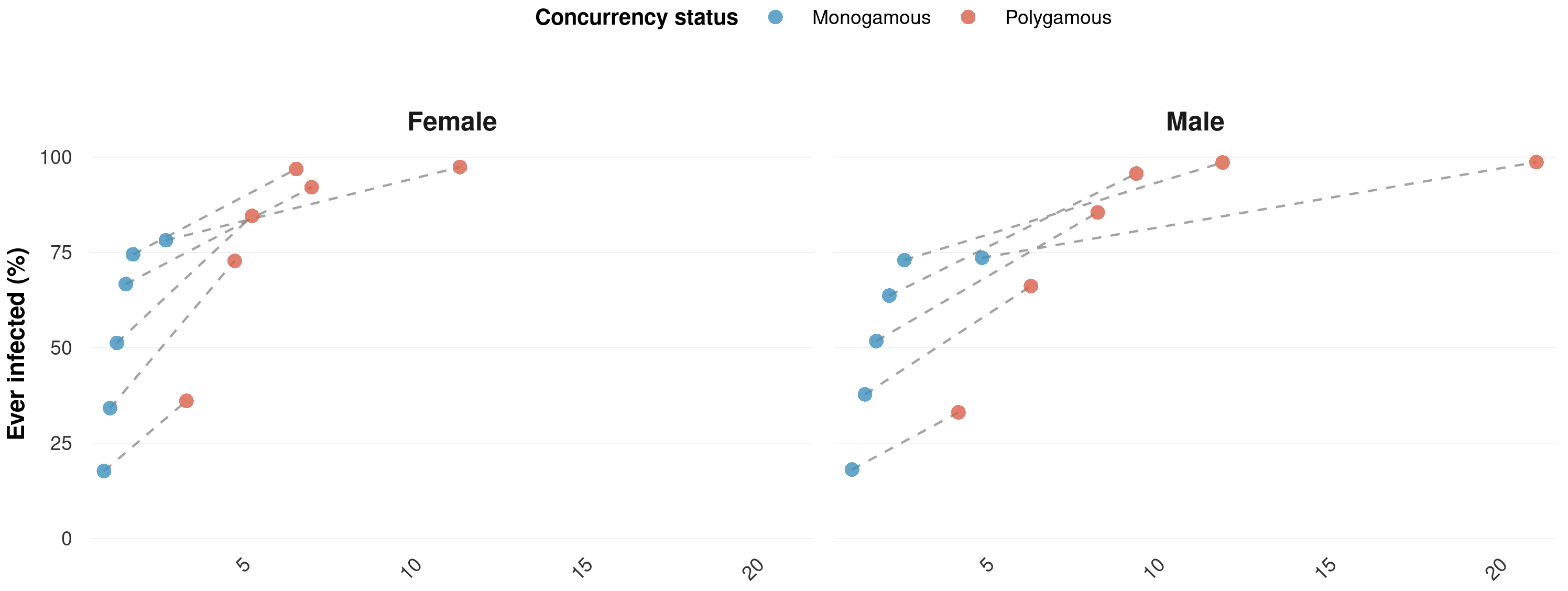}
\caption{Relationship between mean partnership count and mean \% of agents ever infected, across sex and age-group demographic categories in the opposite-sex orientation, split by concurrency status. Each point represents a combination of mean count of partnership and mean \% of infected agents corresponding to a particular age-group within this orientation, for males and females. Dashed lines connect the paired Monogamous/Polygamous values for the same age group.}
\label{fig:scatterplot_opposite}
\end{figure}

To answer the broader research question focused on the impact of partnership dynamics and STI transmission, we explore the relationship between infection prevalence and partnership count. Across both sexes, the mean count of partnership and the prevalence of infection rise monotonically for the opposite-sex oriented agents (Figure~\ref{fig:scatterplot_opposite}). In every age group, \textit{Polygamous} agents have a higher infection rate and partner count than their \textit{Monogamous} counterparts of the same age. The mean partnership counts are $\sim$1–-3 partners for \textit{Monogamous} agents and $\sim$3–-12 partners for \textit{Polygamous} agents, with prevalence increasing from $\sim$18\% at the low end to $\sim$97–99\% at the high end. This suggests that the inclusion of concurrency increases the risk of infection by enabling a higher cumulative accumulation of partnerships. The plots illustrating the relationship between infection prevalence and partnership count among bisexual and same-sex agents along with the full simulation results used to generate the scatter plots are provided in \hyperref[app:additional_figures]{Appendix B} 

\section{Discussion}
\label{sec:discussion}
This study addresses research questions related to modelling partnership dynamics and population heterogeneities in agent-based dynamic network models. We developed an agent-based dynamic network modelling framework that integrates the dynamics of sexual partnerships with population heterogeneities to generate realistic sexual contact networks. By combining probabilistic partnership formation and dissolution mechanisms across different population heterogeneities such as age, sex, and sexual orientation, the model captures the temporal evolution of sexual partnerships in a way that static network representations cannot. The calibration of the partnership model against the mean partnership counts reported in NATSAL-3 using LHS produced parameter sets that reproduced empirical patterns in most demographic subgroups. This addresses the first research question by demonstrating how an agent-based model can be parameterised to include mechanisms of sexual partnership formation and dissolution to reproduce the -age, -sex, and sexual orientation-specific partnership distributions observed in large population surveys.

A novel contribution of this study is the explicit representation of bisexual individuals as a distinct demographic stratum capable of forming partnerships between sex categories. Existing models of sexual contact networks have focussed on heterosexual populations or men who have sex with men, which do not fully capture the role of bridging partnerships in shaping population-level transmission dynamics \citep{Tsoumanis_et_al, vajdi_multilayer_2020}. By including sexual orientation-compatible partnerships between opposite-sex, same-sex, and bisexual agents, our framework provides a more inclusive representation of the sexual contact network. The calibration results indicate that bisexual partnership counts can be reproduced with parameter values intermediate between those for opposite- and same-sex groups, consistent with the empirical data reported in NATSAL-3.

The inclusion of individual-level heterogeneity through negative-binomial multipliers, combined with demographic specific baseline probabilities, generates the heavy-tailed distribution of partnership counts. The proportion of agents who formed ten or more partnerships over a simulation period of $\sim$5 years varied substantially between demographic strata, with the same-sex oriented males in the 16--24 age group reaching approximately 47\%, compared to much lower proportions in opposite-sex groups. This pattern emerges from the interaction of three mechanisms in the model: sexual orientation-specific multipliers, a \textit{youth boost} applied to the youngest age band, and individual-level activity multipliers. Together, these mechanisms replicate population-level sexual behaviour amongst individuals of different attributes, as well as between individuals who share those attributes. 

The use of a duration-dependent Weibull-like hazard for partnership dissolution allowed the model to represent the empirical observation that the risk of dissolution is highest closer to the time of inception of the partnership and decreases over time \citep{nelson2010age} with the mean duration of partnerships increasing with age in all orientations. Female agents consistently exhibited longer mean partnership durations than males, and opposite-sex partnerships were the most durable, followed by same-sex and bisexual partnerships. These patterns are consistent with our broader understanding of partnership dynamics \citep{nelson2010age}. 

Snapshot ego network analysis demonstrated a structural consequence of concurrency that is not captured by aggregate statistics alone. In the 15\%-concurrency scenario, the selected agents accumulated large connected components over time. The presence of bisexual and same-sex nodes within these components illustrates how concurrency-eligible agents can act as bridges between subpopulations of different sexual orientations. These structural findings answer the second research question that concurrency reshapes network connectivity itself as a concurrent partnership at one point in time with more than one individual, potentially increasing the risk of transmission of STIs. This network-level effect is consistent with theoretical accounts of concurrency as a driver of epidemic potential \citep{Mercer2018TimingSexualPartnerships}.

SIS disease simulations provided a proof-of-concept application of the framework to epidemiological investigation. The overall prevalence of infection was driven by the inclusion of concurrent partnerships. Within most strata, the introduction of 15\%-concurrency demonstrated a noticeable change in cumulative prevalence, suggesting that with 15\% of agents allowed to form concurrent partnerships, the additional transmission pathways have an effect on the overall burden of infection. These findings answer the third research question that concurrency not only increases individual risk but also reshapes network connectivity itself, as transmission chains operate through a population-level mechanism, increasing infection risk to monogamous agents. These results are intended as an illustrative demonstration of how the partnership framework can be coupled with a transmission model; a full epidemiological analysis with disease parameters will be undertaken in the future.

We addressed several gaps in existing sexual network models by developing an agent-based simulation that explicitly incorporates individual and partnership attributes often omitted from previously published modelling studies. By including a subpopulation of bisexual individuals, this framework facilitates the exploration of the role of bridging partnerships in the transmission of STIs within heterogeneous populations. This approach enables simulation of the full sexual contact network in a population rather than specific subpopulations. The model also introduces concurrency in sexual partnerships to understand the role of overlapping partnerships in the underlying structure of the sexual network. To address challenges related to modelling the duration of sexual partnerships, our model implemented dissolution through a duration-dependent hazard function, where the probability of partnership dissolution decreases as the length of the partnership increases. This reflects empirically observed patterns of relatively higher risk of dissolution in the early stages of a partnership.

This modelling framework also demonstrates that the elevated infection risk due to concurrency was not limited to agents who were themselves concurrent. We observed that monogamous agents faced a markedly higher risk of infection simply as a consequence of belonging to a population in which concurrency was present elsewhere, with the effect strongest among younger age-groups. This effect demonstrates that individual monogamy is not, by itself, sufficient to protect an agent from concurrency-driven transmission risk when that risk is generated by the structure of the wider sexual network rather than by an agent's own partnership behaviour. This result reiterates the need for better understanding the role of concurrent partnerships as a structural risk factor at the population-level rather than as an individual behavioural one.

The study has several limitations and opportunities for further work. The exclusion of the same-sex male 25--34 and 35--44 sub-groups from model calibration substantially improved global fit but highlighted the ongoing challenge in calibrating models against survey-derived targets. The reported partner counts for these sub-groups were driven by a small subset of highly active individuals rather than a pattern in typical partnership counts. Although this exclusion preserved a monotonic, age-consistent calibration target across strata, it also means the current model does not explicitly capture the heavy-tailed, high-activity behaviour observed within this subgroup. Our approach assumed that the NATSAL-3 estimates represent the partnership patterns of the modelled population, but the differences in sampling time frames and reporting biases may limit the application to other populations. In our modelling framework, the calibration mechanism against available data sources is built into the LHS pipeline. The agent-based modelling framework developed in this study is designed to be generalisable across populations and can be coupled with dynamic transmission models to simulate the spread of bacterial STIs 

The model also makes several simplifying assumptions about the population and partnership dynamics. Partnership formation and dissolution propensities depend only on the demographic stratum and individual multipliers and are modelled as memory less processes, except for the inclusion of duration that plays a role in partnership dissolution. Finally, the binary sex classification reflects the structure of NATSAL-3 but does not capture the full diversity of gender identities relevant to STI epidemiology. The representation of bisexual individuals in this model is a novel contribution, but there is a need for better data to improve the representation of bridging partnership dynamics. Each of these simplifications driving model design choices represents a need for further work to appropriately reflect the real-world complexities of sexual partnerships. 

The SIS disease model used in this study was deliberately simplified, with synthetic transmission parameters, to serve as a proof-of-concept demonstration rather than an epidemiological model. The disease results should not be interpreted as predictions of any particular STI burden. Future work should couple the partnership framework with pathogen-specific transmission models calibrated against surveillance data for chlamydia, gonorrhoea, or syphilis, and investigate the sensitivity of disease outcomes to variation in concurrency prevalence, partnership turnover, and bridging across the various subpopulations defined by their sexual orientations. Each of these extensions represents an application of the flexible and generalisable framework developed here.

Our findings support the view that agent-based dynamic network models offer a flexible and empirically grounded framework for representing the sexual contact networks through which STIs spread. By jointly modelling partnership dynamics and population heterogeneities, the framework captures features such as partnership count and duration, and bridging across orientations that are difficult to represent in static or compartmental approaches. These features are likely to be particularly important for understanding the transmission of bacterial STIs such as chlamydia, gonorrhoea, and syphilis, where partnership turnover and bridging partnerships play a central role in shaping prevalence. This framework can be coupled with an STI transmission model to better understand the impact of partnership features on the dynamics of STI transmission.

\section{Acknowledgements}

We thank the following individuals for their methodological and conceptual contributions, provided through discussions and feedback that helped refine the modelling framework:

\begin{itemize}
    \item \textbf{Peter Doherty Institute for Infection and Immunity}: A/Prof David Price, Dr Kylie Ainslie
    \item \textbf{School of Computing and Information Systems, University of Melbourne}: Dr Thomas Harris, Dr Martin Cyster
    \item \textbf{Melbourne School of Population and Global Health, University of Melbourne}: Dr Rob Moss, Prof Jane Hocking, A/Prof Fabian Kong
    \item \textbf{Public Health Service of Amsterdam (GGD Amsterdam)}: Dr Janneke Heijne
    \item \textbf{University of Maastricht}: A/Prof Nicole Dukers-Muijrers, Dr Zo\"{i}e Alexiou
\end{itemize}

All computational simulations were performed using resources supported by the University of Melbourne's Research Computing Services (RCS).

\section{Author Contributions: CRediT taxonomy} \textbf{P.N.T}: Conceptualization, Methodology, Software, Formal analysis, Investigation, Visualization, Writing -- Original Draft, Writing -- Review \& Editing, Project administration. \textbf{N.G}: Conceptualization, Investigation, Methodology, Supervision, Writing -- Review \& Editing. \textbf{P.T.C}: Conceptualization, Investigation, Methodology, Supervision, Writing -- Review \& Editing.

\section{Funding}
\textbf{P.N.T} is supported by the Australian Government Research Training Program (RTP) Scholarship. The funding body had no role in study design, data collection, data analysis and interpretation, or in the preparation and submission of the article. 

\section{Conflict of Interest}
The authors declare no conflict of interest.

\section {Data Availability}
This study uses publicly available data from the third National Survey of Sexual Attitudes and Lifestyles (NATSAL-3). The authors acknowledge the NATSAL-3 research team and all survey participants who contributed to the collection of these data. 

\section{Code Availability}
The full source code is available at: \url{https://github.com/pnt-id-models/partnersim-dynet-lhs} (Partnership model), \url{https://github.com/pnt-id-models/partnersim-dynet-lhs} (LHS Analysis), and \url{https://github.com/pnt-id-models/partnersim-dynet-sti} (\textit{SIS} transmission model)
\section{Ethics Approval}
Not applicable for this study, as we have used publicly available data.

\section{Participant Consent}
Not applicable for this study, as we have used publicly available data. 


\section{Appendix A: Model Parameters}
\label{app:model_parameters}

This appendix details the parameter ranges explored during LHS calibration in Sections 2.19-2.21, the best-fitting parameter values obtained under the no-concurrency and 15\%-concurrency scenarios, the fixed simulation parameters (Section 2.22), and the SIS transmission model parameters used for all model runs (Section 2.24-2.26).
\subsection{Sex-sexual orientation multipliers}
The sex-specific sexual orientation multipliers $\mu^{\text{ori}}_{s,o}$ for the no-concurrency ($\pi_{\text{conc}} = 0$) and 15\%-concurrency ($\pi_{\text{conc}} = 0.15$) scenarios scale the probability of an agent in demographic stratum $(s, o)$ relative to the reference category. The female opposite-sex multiplier stays fixed; the remaining five combinations of sex and sexual orientation are sampled independently for both formation and dissolution. Table~\ref{tab:sex_orientation_multipliers} lists the lower and upper bounds for the five parameters. 
\begin{table}[h]
\centering
\caption{Sampled ranges for sex-specific sexual orientation multipliers, applied to both formation and dissolution probabilities. The female opposite-sex multiplier is fixed at 1.0 as the reference category.}
\begin{tabular}{lcc}
\toprule
Multiplier & Lower bound & Upper bound \\
\midrule
Female same-sex & 0.5 & 6.0 \\
Female bisexual & 0.5 & 6.0 \\
Male opposite-sex & 0.5 & 6.0 \\
Male same-sex & 0.5 & 6.0 \\
Male bisexual & 0.5 & 6.0 \\
\bottomrule
\end{tabular}

\label{tab:sex_orientation_multipliers}
\end{table}
\subsection{Age multipliers}
The age multipliers $\mu^{\text{age}}_{g}$ for the no-concurrency and 15\%-concurrency scenarios for the six age bands are determined by two sampled parameters per behaviour (formation, dissolution): a \textit{youth boost} $\beta$ applied to the 16--24 band, and an exponential \textit{age decay} rate $\kappa$ that determines the decline in $\mu^{\text{age}}_{g}$ from the age group 25--34 onwards. Table~\ref{tab:age_multipliers} lists the lower and upper bounds for the \textit{youth boost} and \textit{age decay} rate. 

\begin{table}[h]
\centering
\caption{Sampled ranges for the \textit{youth boost} and \textit{age decay} parameters governing the age structure of formation and dissolution probabilities.}
\begin{tabular}{lcc}
\toprule
Parameter & Lower bound & Upper bound \\
\midrule
\textit{Youth boost} ($\beta$), formation & 2.0 & 4.0 \\
\textit{Youth boost} ($\beta$), dissolution & 2.0 & 4.0 \\
\textit{Age decay} ($\kappa$), formation & 0.1 & 1.0 \\
\textit{Age decay} ($\kappa$), dissolution & 0.1 & 1.0 \\
\bottomrule
\end{tabular}
\label{tab:age_multipliers}
\end{table}
\subsection{Best-fitting parameters}
Table~\ref{tab:bestfit_parameters} presents the best-fitting parameters for each stratum without concurrency and with concurrency, with the female opposite-sex stratum fixed as the reference stratum (scale = 1.000). The baseline and age-structure parameters are broadly similar between scenarios; the largest divergence occurs for the same-sex male dissolution scale and the same-sex female formation scale, where the no-concurrency scenario produced systematically lower formation and dissolution scales relative to the 15\%-concurrency (ratios $<$1 indicate lower values at $\theta_{\mathrm{conc}}=0$). 

\begin{table}[H]
\centering
\small
\renewcommand{\arraystretch}{0.88}
\caption{Best-fit parameter values under no-concurrency ($\theta_{\mathrm{conc}} = 0$, id:2777) and 15\%-concurrency ($\theta_{\mathrm{conc}} = 0.15$, id:557) scenarios. Ratio = value at $\theta_{\mathrm{conc}}=0$ divided by value at $\theta_{\mathrm{conc}}=0.15$.}
\label{tab:bestfit_parameters}
\begin{tabular}{lccc}
\toprule
\textbf{Parameter} & $\theta_{\mathrm{conc}} = 0$ & $\theta_{\mathrm{conc}} = 0.15$ & \textbf{Ratio} \\
\midrule
\multicolumn{4}{l}{\textit{Baseline (Female opposite-sex reference 25--34 age group)}} \\
\quad Formation probability & 0.003 & 0.002 & 1.50 \\
\quad Dissolution probability  & 0.002 & 0.001 & 2.00 \\
\midrule
\multicolumn{4}{l}{\textit{Youth boost multiplier for 16--24 age group across all sexes and orientations}} \\
\quad Formation youth boost  & 2.159 & 2.376 & 0.90 \\
\quad Dissolution youth boost   & 2.083 & 2.162 & 0.96 \\
\midrule
\multicolumn{4}{l}{\textit{Age decay parameters for ages 25 and above for all sexes and orientations}} \\
\quad Formation age decay & 0.300 & 0.372 & 0.80 \\
\quad Dissolution age decay  & 0.286 & 0.155 & 1.84 \\
\midrule
\multicolumn{4}{l}{\textit{Multipliers for opposite-sex orientation}} \\
\quad Male formation scale   & 4.778 & 5.730 & 0.83 \\
\quad Male dissolution scale & 1.934 & 3.589 & 0.54\\
\quad Female formation scale & 1.000 & 1.000 & 1.00 \\
\quad Female dissolution scale  & 1.000 & 1.000 & 1.00 \\
\midrule
\multicolumn{4}{l}{\textit{Multipliers for bisexual orientation}} \\
\quad Male formation scale   & 1.197 & 2.030 & 0.59 \\
\quad Male dissolution scale & 1.559 & 4.705 & 0.33 \\
\quad Female formation scale & 2.155 & 5.168 & 0.42\\
\quad Female dissolution scale  & 1.024 & 0.556 & 1.84 \\
\midrule
\multicolumn{4}{l}{\textit{Multipliers for same-sex orientation}} \\
\quad Male formation scale   & 1.569 & 2.532 & 0.62 \\
\quad Male dissolution scale & 5.261 & 2.312 & 2.28\\
\quad Female formation scale & 1.804 & 0.595 & 3.03 \\
\quad Female dissolution scale  & 0.964 & 1.422 & 0.68\\
\midrule
\multicolumn{4}{l}{\textit{Model fit (MSE)}} \\
\quad Global MSE   & 0.466 & 0.540 \\
\quad Opposite-sex MSE & 0.167 & 0.634 \\
\quad Same-sex MSE & 0.907 & 0.652 \\
\quad Bisexual MSE & 0.324 & 0.334 \\
\bottomrule
\end{tabular}
\end{table}

\subsection{Fixed simulation parameters}
Table~\ref{tab:fixed_simulation_params} summarises the fixed simulation parameters that remain constant for multiple simulation runs. These include population size, temporal resolution, total simulation duration, number of replicates, and concurrency structure. Additionally, the parameters driving individual-level heterogeneity in sexual behaviour such as the negative binomial dispersion and probability terms, as well as duration-dependent decay parameters, are listed where applicable. 
\begin{table}[H]
\centering
\caption{Fixed simulation parameters used across multiple simulation runs.}
\label{tab:fixed_simulation_params}
\begin{tabular}{ll}
\hline
\textbf{Parameter} & \textbf{Value} \\
\hline

Number of agents ($N$) & 15,000 \\
Number of replicates & 100 \\
Simulation duration & 1,875 time steps \\
Time step resolution & 1 day \\
Concurrency proportion & 0.00, 0.15 \\
Concurrency parameter ($\lambda$) & 2 \\

\hline
Negative binomial dispersion ($r$) & 0.5 \\
Negative binomial probability ($p$) & 0.5 \\

Duration-dependent decay ($\alpha$) & 1500 \\
Duration-dependent decay ($\gamma$) & 2 \\

\hline
\end{tabular}
\end{table}
\subsection{Disease transmission parameters}
This SIS model in this study simulates a hypothetical bacterial STI infection with synthetic parameters listed in Table~\ref{tab:disease-parameters}

\FloatBarrier

\begin{table}[H]
\centering
\caption{Disease transmission parameters}
\label{tab:disease-parameters}
\begin{tabular}{lc}
\hline
\textbf{Parameter} & \textbf{Value} \\
\hline
Total population size & 15000 \\
Total number of disease replicates & 100 \\
Initial proportion of infected agents & 0.10 \\
Infection probability per partnership per time step ($\beta$) & 0.20 \\
Recovery probability per time step ($\gamma$) & 0.10 \\
Time step to start seeding of initial infections & 51 \\
Maximum number of time steps simulated & 1825  \\
\hline
\end{tabular}
\end{table}

\section{Appendix B: Additional Figures and Tables}
\label{app:additional_figures}

This appendix presents supplementary figures supporting the partnership and network results in the main text. We have reported the effect of concurrency on partnership count, duration, and activity-level distributions for the same-sex and bisexual strata, respectively, complementing the summary results discussed in the main text for the opposite-sex stratum. This appendix also includes network plots visualised for the no-concurrency scenario. 

\subsection{Same-sex partnership dynamics under concurrency}

\begin{figure}[H]
\centering
\begin{subfigure}[b]{\linewidth}
\centering
\includegraphics[width=0.8\linewidth, height=6cm]{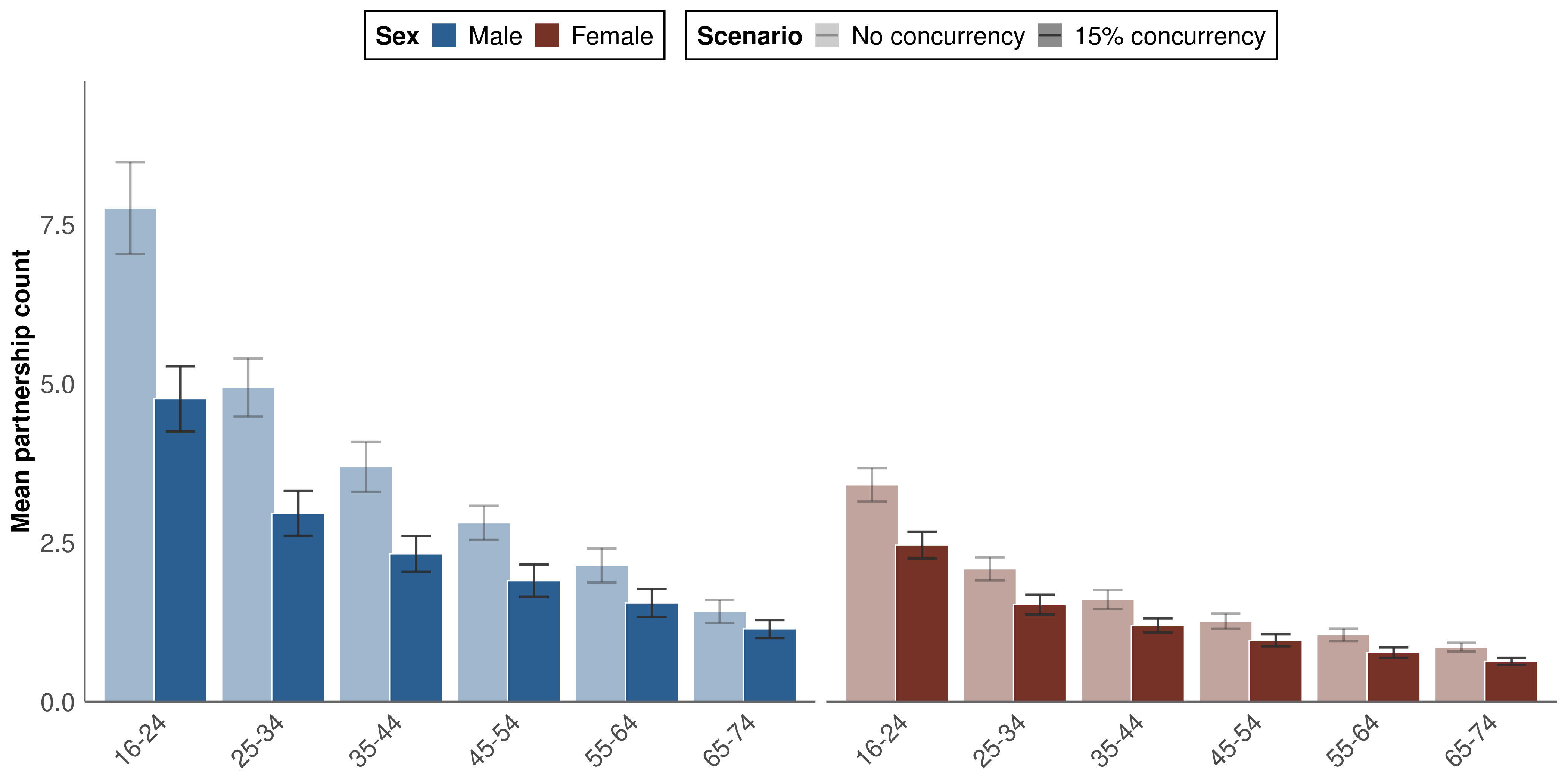}
\caption{Mean partnership count}
\label{fig:same_sex_count}
\end{subfigure}
\vspace{6pt}
\begin{subfigure}[b]{\linewidth}
\centering
\includegraphics[width=0.8\linewidth, height=6cm]{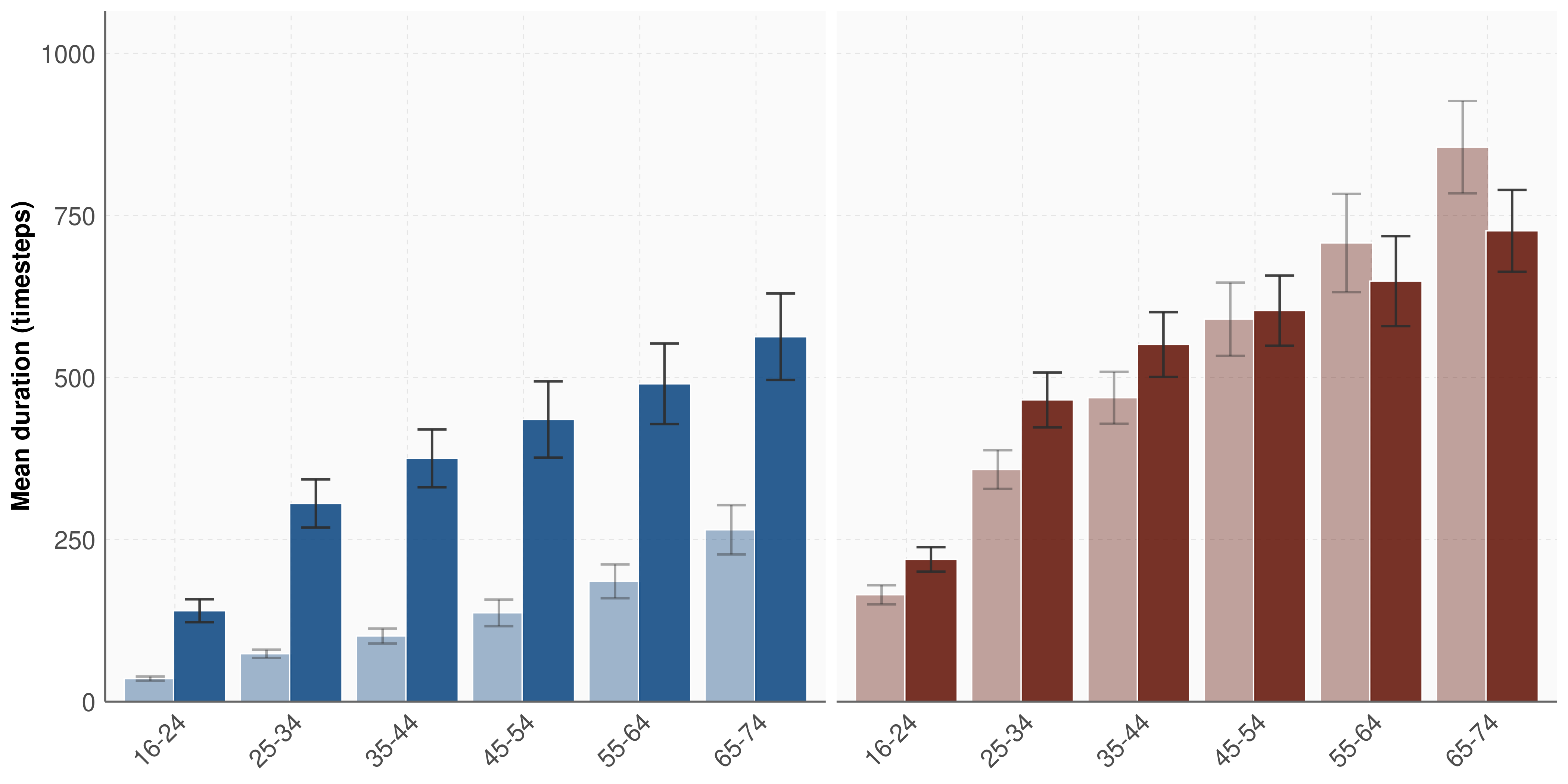}
\caption{Mean partnership duration}
\label{fig:same_sex_duration}
\end{subfigure}
\caption{Mean partnership count and duration for same-sex sexual orientation across 100 simulations}
\label{fig:same_sex_count_duration}
\end{figure}

In Figure~\ref{fig:same_sex_count_duration}, the introduction of 15\% concurrency produces a notable decrease in mean partnership count among same-sex males in the 16--24 and 25--34 age groups. The addition of concurrency also increases the mean duration among same-sex males, particularly among younger age groups. Among females, the addition of concurrency does not markedly impact the partnership count and only slightly increases the mean partnership duration in younger age groups with a slight decrease in the older age groups.

\begin{figure}[H]
\centering
\begin{subfigure}[b]{\linewidth}
\centering
\includegraphics[width=0.8\linewidth, height=6cm]{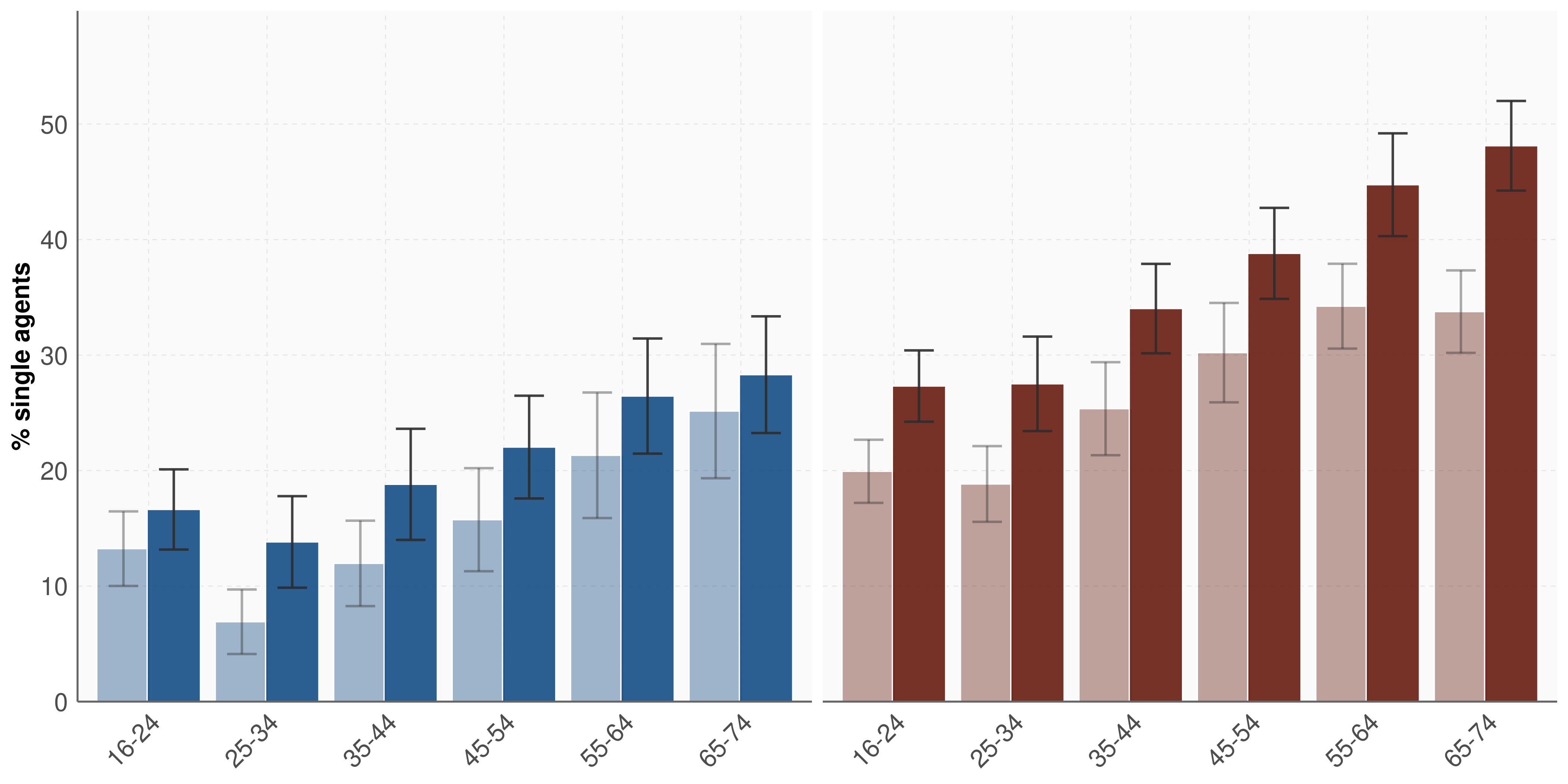}
\caption{\% single agents (0 partners)}
\label{fig:same_sex_single}
\end{subfigure}
\vspace{2pt}
\begin{subfigure}[b]{\linewidth}
\centering
\includegraphics[width=0.8\linewidth, height=6cm]{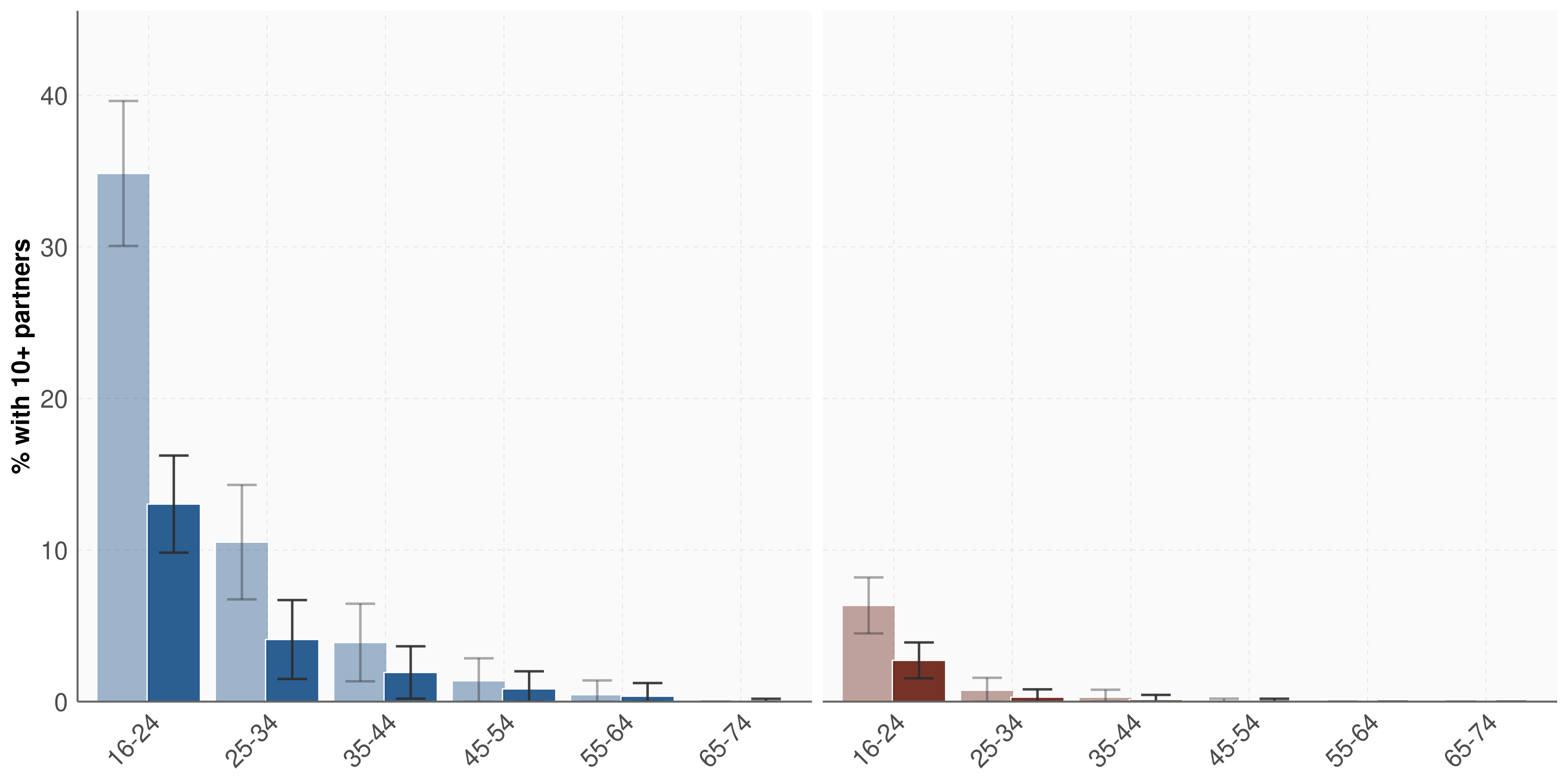}
\caption{\% agents with 10+ partners}
\label{fig:same_sex_high}
\end{subfigure}
\caption{Percentage of single agents and percentage of agents with ten or more partners for same-sex sexual orientation across 100 simulations, by age group and sex}
\label{fig:same_sex_single_high}
\end{figure}

In Figure~\ref{fig:same_sex_single_high}, we observed the largest effect of concurrency on single-agent percentages of any stratum: for same-sex females, the proportion of single individuals increases steadily with age and is markedly higher in the older age groups, approximately 33\% to 48\% for the 65--74 age group, while the increase for males is smaller. The percentage of agents with ten or more partners is notably lower in the scenario with concurrency compared to without. For example, among same-sex males, the percentage of single agents decreases from approximately 35\% to 13\% for the 16--24 age group, indicating that concurrency distributes same-sex male partnerships across a broader group of agents. 

\subsection{Bisexual partnership dynamics under concurrency}
\begin{figure}[H]
\centering
\begin{subfigure}[b]{\linewidth}
\centering
\includegraphics[width=0.8\linewidth, height=6cm]{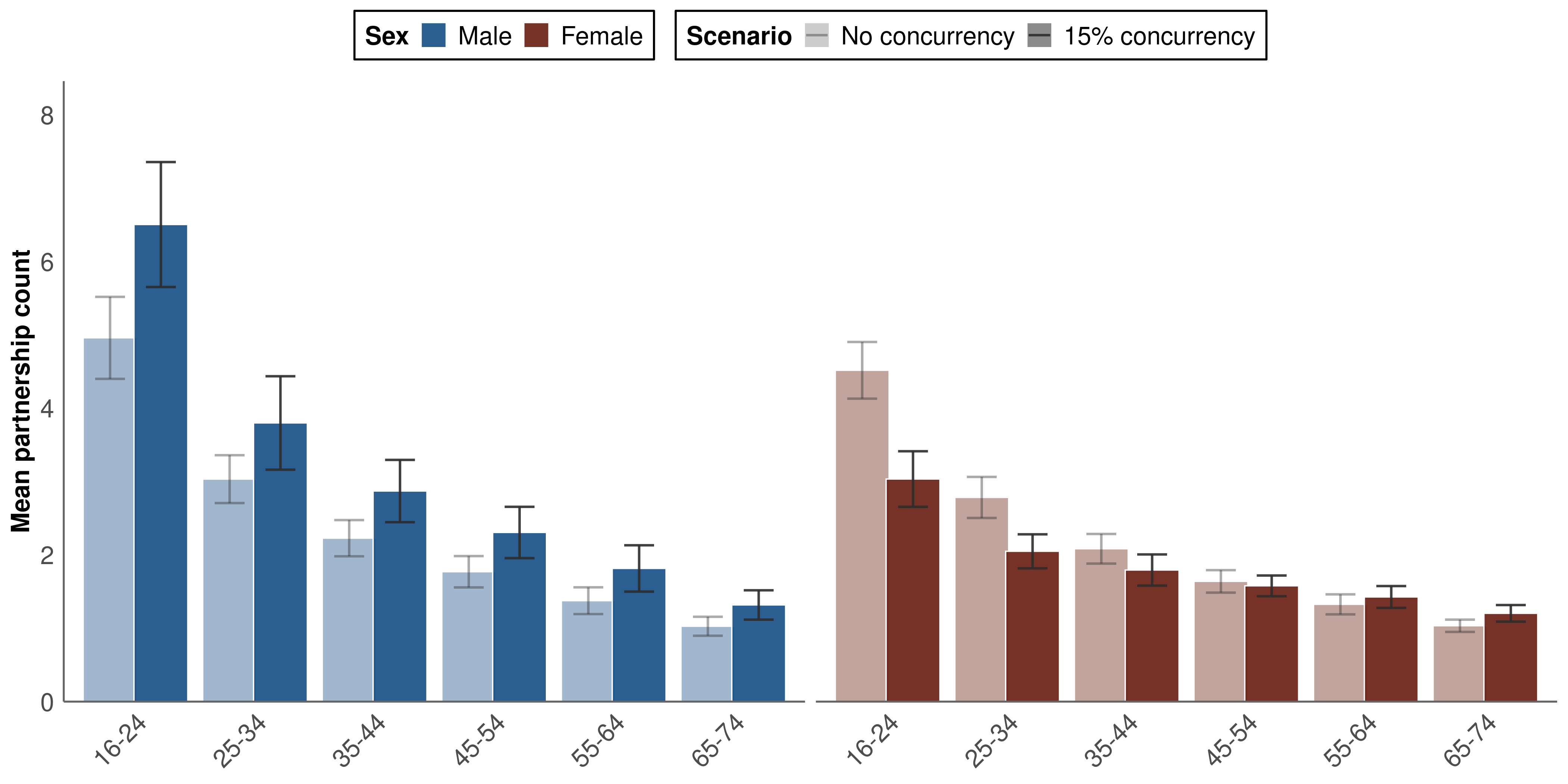}
\caption{Mean partnership count}
\label{fig:bisexual_count}
\end{subfigure}
\vspace{6pt}
\begin{subfigure}[b]{\linewidth}
\centering
\includegraphics[width=0.8\linewidth, height=6cm]{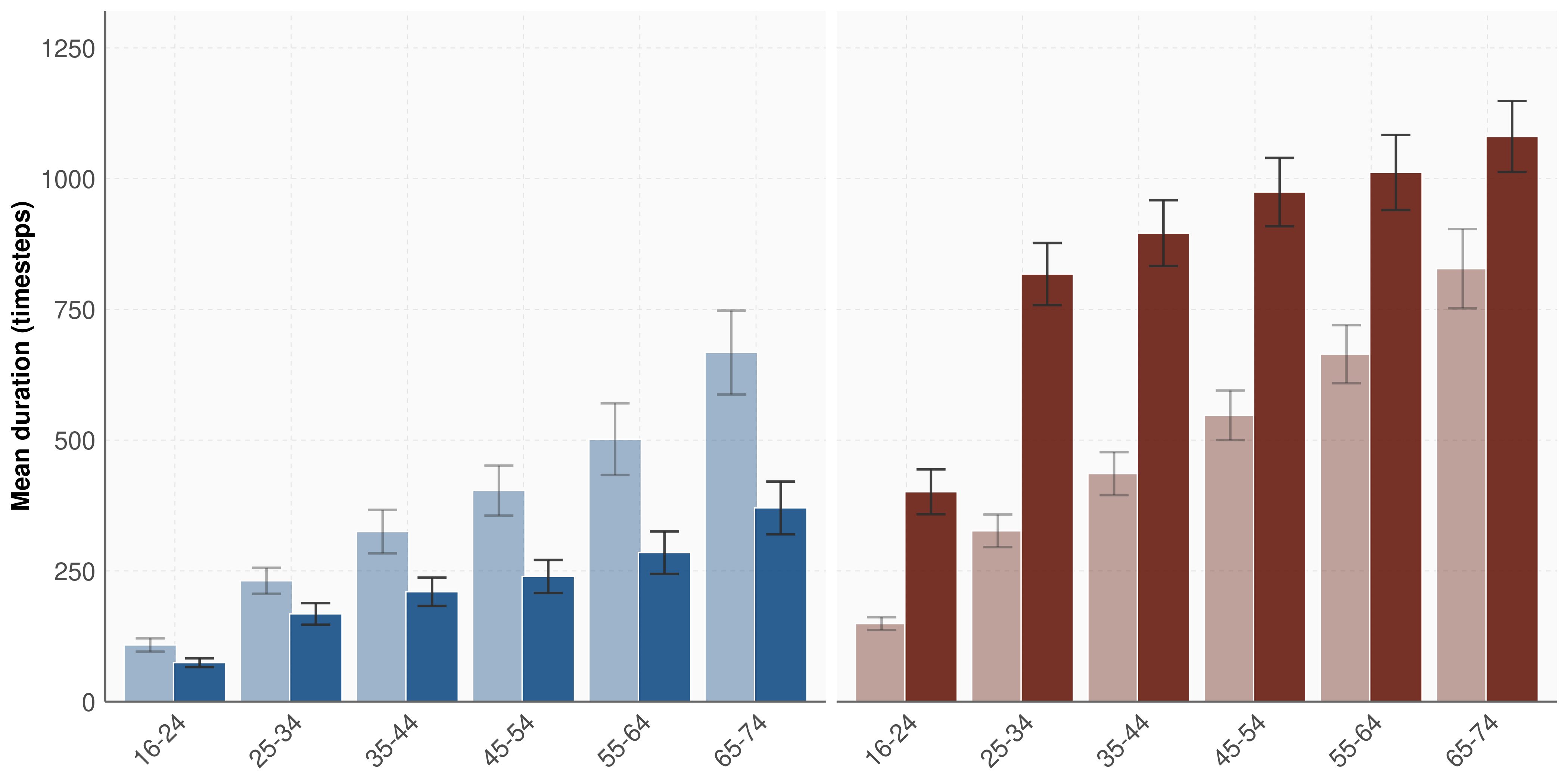}
\caption{Mean partnership duration}
\label{fig:bisexual_duration}
\end{subfigure}
\caption{Mean partnership count and duration for bisexual sexual orientation across 100 simulations, by age group and sex}
\label{fig:bisexual_count_duration}
\end{figure}

In Figure~\ref{fig:bisexual_count_duration}, the introduction of 15\% concurrency produces a notable increase in the mean partnership count for younger males but a slight decrease for younger females. The mean partnership count remains slightly higher in the 15\%-concurrency scenario for males from age group 35--44 onwards, and stays nearly equal across both concurrency scenarios from age group 35--44 onwards for females. The mean duration of partnerships increases with age for both scenarios, and concurrency produces a marked increase in the mean duration of partnership in all age groups for females, but a notable decrease for males.

\begin{figure}[H]
\centering
\begin{subfigure}[b]{\linewidth}
\centering
\includegraphics[width=0.8\linewidth, height=6cm]{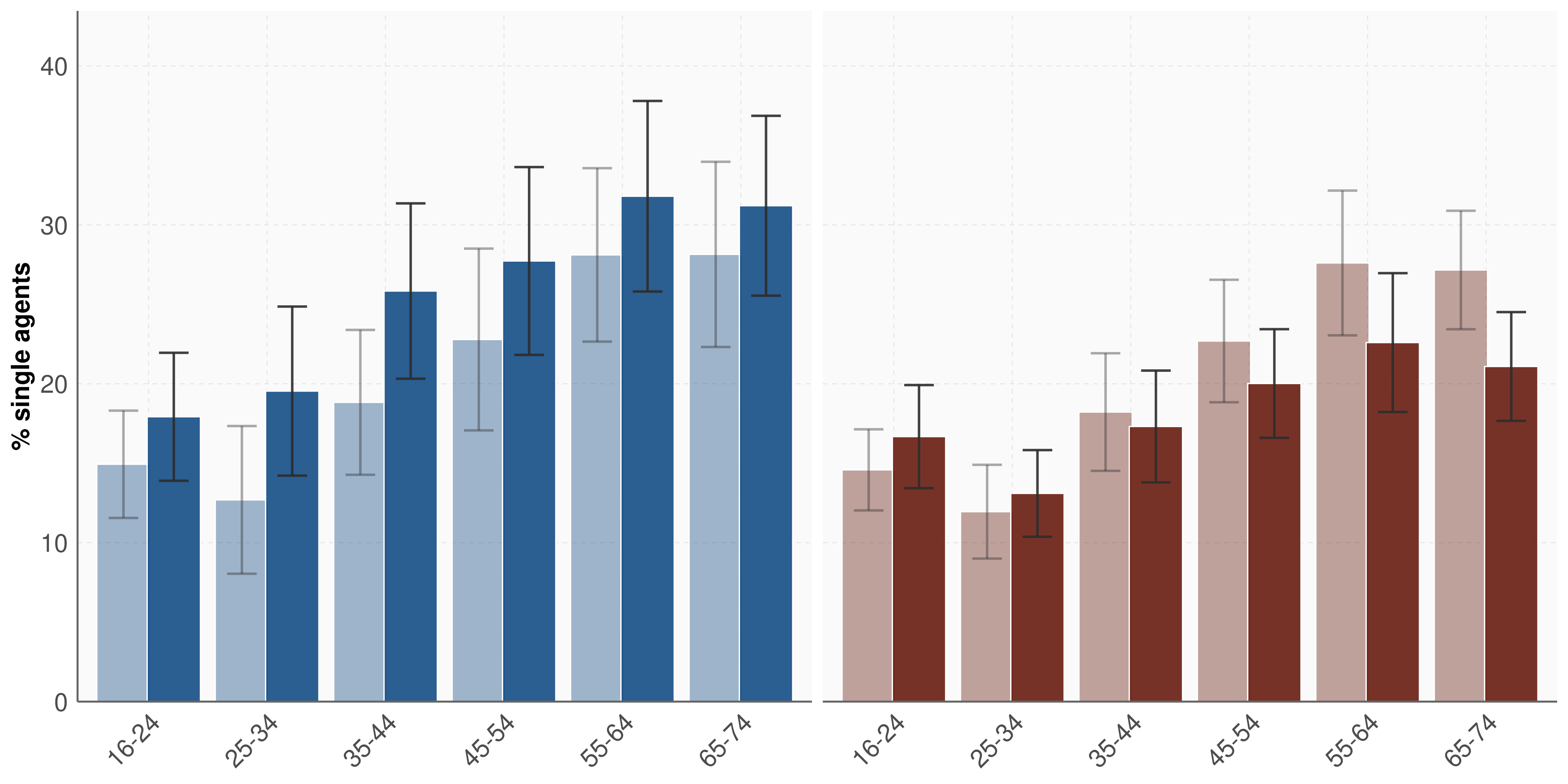}
\caption{\% single agents (0 partners)}
\label{fig:bisexual_single}
\end{subfigure}
\vspace{6pt}
\begin{subfigure}[b]{\linewidth}
\centering
\includegraphics[width=0.8\linewidth, height=6cm]{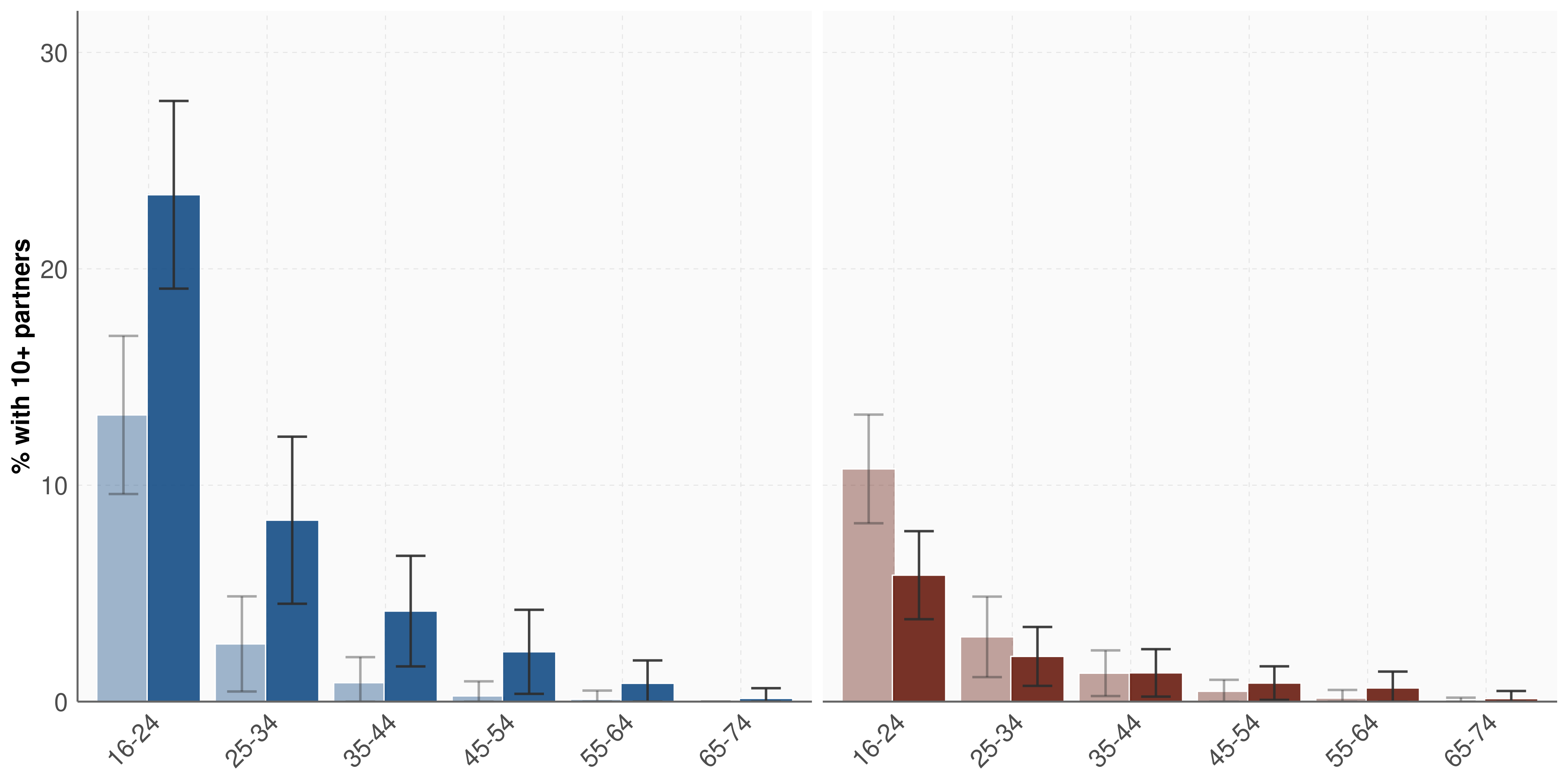}
\caption{\% agents with 10+ partners}
\label{fig:bisexual_high}
\end{subfigure}
\caption{Percentage of single agents and percentage of agents with ten or more partners for bisexual sexual orientation across 100 simulations, by age group and sex. Lighter bars: $\theta_{\mathrm{conc}} = 0$; darker bars: $\theta_{\mathrm{conc}} = 0.15$. Error bars: $\pm$1~SD across simulations.}
\label{fig:bisexual_single_high}
\end{figure}

In Figure~\ref{fig:bisexual_single_high}, we observed that concurrency increases the percentage of single agents for males for all age groups. For example, the percentage of single agents increases from approximately 12\% to 20\% for the 25--34 age group. For females, the percentage of single agents decreases in the 15\%-concurrency in the older age groups, for example, from approximately 27\% to 22\% for the 55--64 age group. The percentage of agents with ten or more partners increases markedly in the 15\%-concurrency for males, for example, from approximately 14\% to 24\% for the 16--24 age group, but decreases for females approximately 11\% to 6\% for the 16--24 age group. 

Across all three sexual orientations, the 16--24 age group consistently shows the highest proportion of high-activity agents. The effect of concurrency on high-activity behaviour varies between males and females within the same-sex and bisexual groups: it increases the proportion of high-activity agents in males, but decreases in females. The wider error bars in the same-sex and bisexual plots reflect greater stochastic variability across simulations in these relatively smaller subpopulations. 

\subsection{Ego networks under strict monogamy}

\begin{figure}[H]
\centering
\includegraphics[width=1\linewidth, height=16cm]{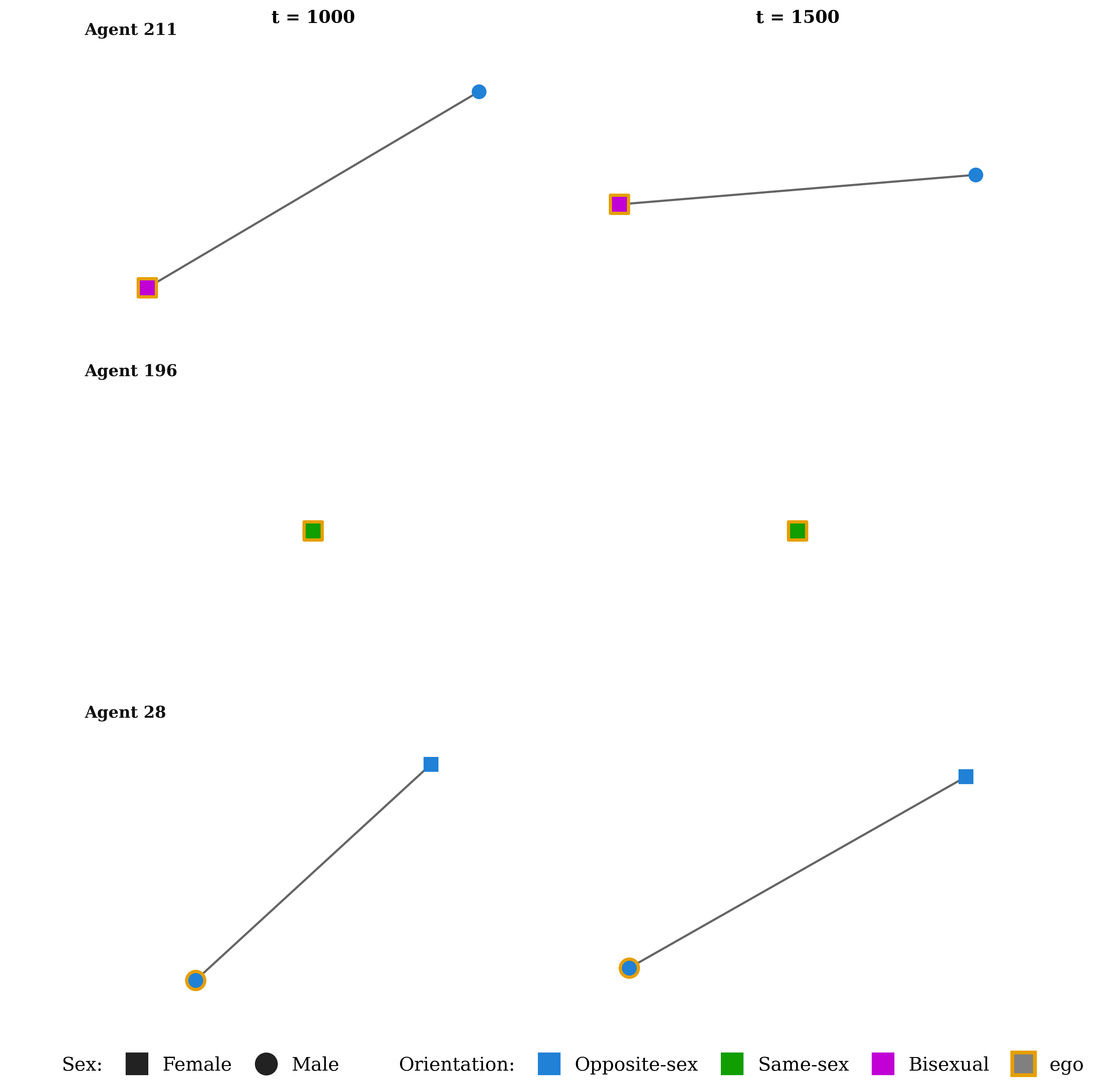}
\caption{Snapshot network (sexual orientation-coloured) for six selected agents at
$t \in \{ 1000, 1500\}$, under no-concurrency.}
\label{fig:ego_networks_comparison_0pcconc}
\end{figure}

Under the no-concurrency scenario in (Figure~\ref{fig:ego_networks_comparison_0pcconc}), all three selected agents remain in partnerships of at most one simultaneous partner throughout the simulation, producing ego networks that never exceed two nodes and one edge at any snapshot. No higher-order (2- or 3-hop) structure can be formed.

\subsection{Partnership count and infection prevalence}

\begin{figure}[H]
\centering
\includegraphics[width=1\linewidth, height=8cm]{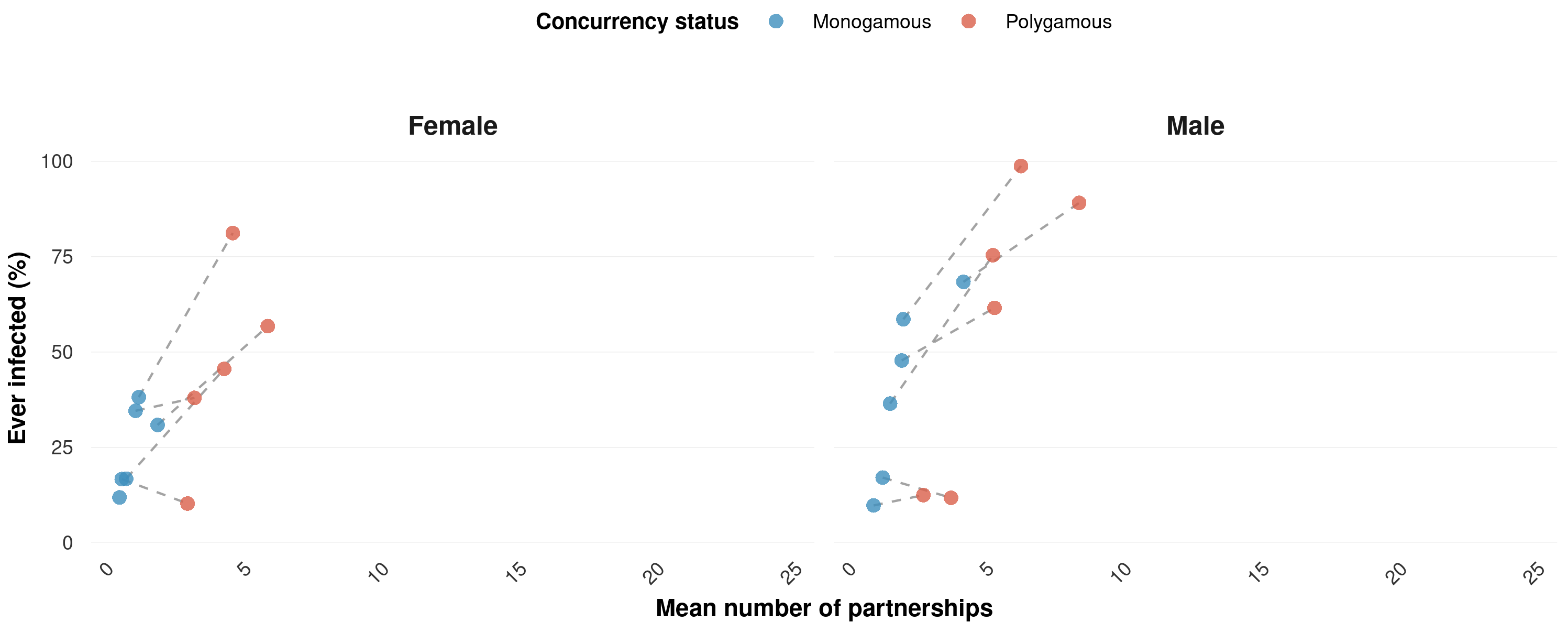}
\caption{Relationship between mean partnership count and mean \% of agents ever infected, across sex and age-group demographic categories in the bisexual orientation, split by concurrency status. Each point represents a combination of mean count of partnership and mean \% of infected agents corresponding to a particular age-group within this orientation, for males and females. Dashed lines connect the paired Monogamous/Polygamous values for the same age group.}
\label{fig:scatter_prevalence_partnership_same}
\end{figure}

Across both sexes, the mean number of partnerships and the prevalence of infections rise monotonically for agents of the same-sex over a narrower range (Figure~\ref{fig:scatter_prevalence_partnership_same}) compared to agents of opposite-sex and bisexuality. In every age group, \textit{Polygamous} agents have a higher infection rate and partner count than their \textit{Monogamous} counterparts of the same age group. Partnership counts are below 2 for nearly all \textit{Monogamous} agents and $\sim$3–10 for all \textit{Polygamous} agents. The \textit{Polygamous} same-sex oriented males reach both higher partnership counts and higher infection prevalence ($\sim$99\% with $\sim$6 mean partners) than same-sex females (peaking at $\sim$81\% with a similar partnership count), consistent with results showing a higher share of infected agents among \textit{Monogamous} same-sex oriented males in the main article. 

\begin{figure}[H]
\centering
\includegraphics[width=1\linewidth, height=8cm]{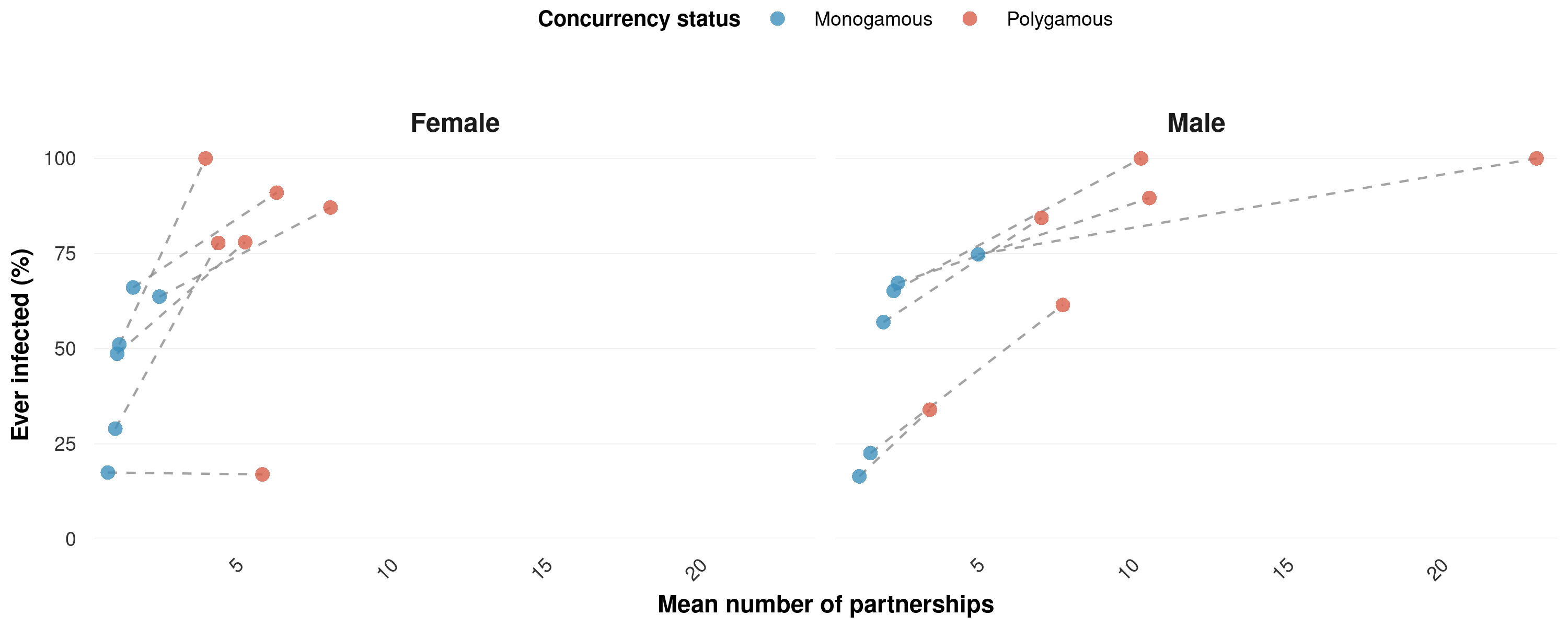}
\caption{Relationship between mean partnership count and mean \% of agents ever infected, across sex and age-group demographic categories in the bisexual orientation, split by concurrency status. Each point represents a combination of mean count of partnership and mean \% of infected agents corresponding to a particular age-group within this orientation, for males and females. Dashed lines connect the paired Monogamous/Polygamous values for the same age group.}
\label{fig:scatter_prevalence_partnership_bisexual}
\end{figure}

Across both sexes, mean partnership count and infection prevalence rise monotonically for same-sex agents over a wide range (Figure~\ref{fig:scatter_prevalence_partnership_bisexual}). In nearly every age group, \textit{Polygamous} agents have a higher infection rate as well as partner count than their \textit{Monogamous} counterparts of the same age group. Among \textit{Polygamous} males, whose mean partnership counts extend up to $\sim$22 compared to a maximum of $\sim$8 for bisexual females, with the prevalence reaching nearly 100\% for both sexes. Partnership counts stay below 2 for nearly all \textit{Monogamous} agents and $\sim$5–10 for nearly all male \textit{Polygamous} agents and between $\sim$3–8 for all female \textit{Polygamous} agents.

In Table~\ref{app_tab:full_sim_results}, we can see the detailed breakdown of the mean count of partnerships and the mean prevalence of infection for each stratum of the population. Infection prevalence and mean count of partnerships decline monotonically with age for nearly all strata. The data from Table~\ref{app_tab:full_sim_results} were used to plot Figure~\ref{fig:scatter_prevalence_partnership_bisexual}, Figure~\ref{fig:scatter_prevalence_partnership_same} and Figure~\ref{fig:scatterplot_opposite} in the main article. 

\begin{longtable}[t]{lllllrrl}
\caption{\label{app_tab:full_sim_results} Full simulation results for disease transmission model by sex (S; M = male, F = female), sexual orientation (O; Opp = opposite-sex, Same = same-sex, Bi = bisexual), age group (A), and partnership concurrency status (Mono = monogamous, Poly = polygamous). Results are reported overall and by sex, sex-orientation, sex x orientation x age group categories. N agents denotes the mean number of agents in each stratum, mean partnership count is the mean of number of partnerships per agent, and \% ever infected is reported as the mean percentage infected across simulations $\pm$ SD}\\
\toprule
Level & S & O & A & \shortstack[c]{Concurrency\\status} & \shortstack[c]{N agents} & \shortstack[c]{Mean\\partnership\\count} & \shortstack[c]{\% ever\\infected}\\
\midrule
Overall & NA & NA & NA & Mono & 13745 & 2.03 & 52.2 $\pm$ 0.3\\
 & NA & NA & NA & Poly & 1618 & 9.00 & 84.5 $\pm$ 0.6\\
By S & M & NA & NA & Mono & 6995 & 2.47 & 53.3 $\pm$ 0.4\\
 & M & NA & NA & Poly & 775 & 11.45 & 84.5 $\pm$ 0.8\\
 & F & NA & NA & Mono & 6750 & 1.58 & 51.1 $\pm$ 0.4\\
 & F & NA & NA & Poly & 843 & 6.74 & 84.5 $\pm$ 0.7\\
By S x O & M & Opp & NA & Mono & 6280 & 2.47 & 54.0 $\pm$ 0.4\\
 & M & Opp & NA & Poly & 692 & 11.84 & 85.6 $\pm$ 0.7\\
 & M & Same & NA & Mono & 361 & 2.19 & 42.5 $\pm$ 1.8\\
 & M & Same & NA & Poly & 47 & 6.09 & 72.8 $\pm$ 4.3\\
 & M & Bi & NA & Mono & 354 & 2.62 & 51.4 $\pm$ 1.4\\
 & M & Bi & NA & Poly & 36 & 10.92 & 79.5 $\pm$ 4.0\\
 & F & Opp & NA & Mono & 5363 & 1.66 & 55.3 $\pm$ 0.4\\
 & F & Opp & NA & Poly & 705 & 7.01 & 86.3 $\pm$ 0.8\\
 & F & Same & NA & Mono & 740 & 1.05 & 25.0 $\pm$ 1.0\\
 & F & Same & NA & Poly & 41 & 4.41 & 52.9 $\pm$ 5.1\\
 & F & Bi & NA & Mono & 647 & 1.50 & 46.5 $\pm$ 1.2\\
 & F & Bi & NA & Poly & 97 & 5.84 & 84.8 $\pm$ 1.9\\
By S x O x A & M & Opp & 16-24 & Mono & 1312 & 4.90 & 73.6 $\pm$ 0.6\\
 & M & Opp & 16-24 & Poly & 189 & 21.10 & 98.7 $\pm$ 0.5\\
 & M & Opp & 25-34 & Mono & 951 & 2.63 & 73.0 $\pm$ 0.7\\
 & M & Opp & 25-34 & Poly & 119 & 11.93 & 98.6 $\pm$ 0.7\\
 & M & Opp & 35-44 & Mono & 981 & 2.19 & 63.7 $\pm$ 1.0\\
 & M & Opp & 35-44 & Poly & 110 & 9.41 & 95.7 $\pm$ 1.1\\
 & M & Opp & 45-54 & Mono & 1031 & 1.81 & 51.8 $\pm$ 0.9\\
 & M & Opp & 45-54 & Poly & 101 & 8.28 & 85.5 $\pm$ 1.9\\
 & M & Opp & 55-64 & Mono & 989 & 1.48 & 37.8 $\pm$ 1.1\\
 & M & Opp & 55-64 & Poly & 87 & 6.33 & 66.2 $\pm$ 2.9\\
 & M & Opp & 65-74 & Mono & 1016 & 1.10 & 18.1 $\pm$ 1.0\\
 & M & Opp & 65-74 & Poly & 86 & 4.21 & 33.1 $\pm$ 4.6\\
 & M & Same & 16-24 & Mono & 87 & 4.20 & 68.4 $\pm$ 3.3\\
 & M & Same & 16-24 & Poly & 15 & 8.40 & 89.1 $\pm$ 4.8\\
 & M & Same & 25-34 & Mono & 57 & 2.02 & 58.6 $\pm$ 3.8\\
 & M & Same & 25-34 & Poly & 7 & 6.29 & 98.8 $\pm$ 4.4\\
 & M & Same & 35-44 & Mono & 54 & 1.96 & 47.8 $\pm$ 4.4\\
 & M & Same & 35-44 & Poly & 6 & 5.33 & 61.6 $\pm$ 13.3\\
 & M & Same & 45-54 & Mono & 54 & 1.54 & 36.5 $\pm$ 4.6\\
 & M & Same & 45-54 & Poly & 11 & 5.27 & 75.4 $\pm$ 10.4\\
 & M & Same & 55-64 & Mono & 62 & 1.27 & 17.1 $\pm$ 3.9\\
 & M & Same & 55-64 & Poly & 4 & 3.75 & 11.8 $\pm$ 15.3\\
 & M & Same & 65-74 & Mono & 47 & 0.94 & 9.8 $\pm$ 4.1\\
 & M & Same & 65-74 & Poly & 4 & 2.75 & 12.5 $\pm$ 22.9\\
 & M & Bi & 16-24 & Mono & 81 & 5.06 & 74.8 $\pm$ 2.9\\
 & M & Bi & 16-24 & Poly & 8 & 23.12 & 100.0 $\pm$ 0.0\\
 & M & Bi & 25-34 & Mono & 45 & 2.47 & 67.3 $\pm$ 4.0\\
 & M & Bi & 25-34 & Poly & 5 & 10.60 & 89.6 $\pm$ 10.0\\
 & M & Bi & 35-44 & Mono & 64 & 2.33 & 65.2 $\pm$ 3.3\\
 & M & Bi & 35-44 & Poly & 3 & 10.33 & 100.0 $\pm$ 0.0\\
 & M & Bi & 45-54 & Mono & 51 & 2.00 & 57.0 $\pm$ 3.6\\
 & M & Bi & 45-54 & Poly & 9 & 7.11 & 84.4 $\pm$ 7.8\\
 & M & Bi & 55-64 & Mono & 50 & 1.58 & 22.6 $\pm$ 4.2\\
 & M & Bi & 55-64 & Poly & 5 & 7.80 & 61.5 $\pm$ 11.5\\
 & M & Bi & 65-74 & Mono & 63 & 1.22 & 16.5 $\pm$ 3.8\\
 & M & Bi & 65-74 & Poly & 6 & 3.50 & 34.0 $\pm$ 15.7\\
 & F & Opp & 16-24 & Mono & 1150 & 2.77 & 78.2 $\pm$ 0.6\\
 & F & Opp & 16-24 & Poly & 174 & 11.36 & 97.4 $\pm$ 0.6\\
 & F & Opp & 25-34 & Mono & 774 & 1.81 & 74.5 $\pm$ 0.9\\
 & F & Opp & 25-34 & Poly & 118 & 6.58 & 96.9 $\pm$ 0.8\\
 & F & Opp & 35-44 & Mono & 861 & 1.60 & 66.7 $\pm$ 0.9\\
 & F & Opp & 35-44 & Poly & 121 & 7.03 & 92.1 $\pm$ 1.2\\
 & F & Opp & 45-54 & Mono & 865 & 1.34 & 51.3 $\pm$ 1.0\\
 & F & Opp & 45-54 & Poly & 98 & 5.29 & 84.6 $\pm$ 1.6\\
 & F & Opp & 55-64 & Mono & 813 & 1.14 & 34.2 $\pm$ 1.3\\
 & F & Opp & 55-64 & Poly & 116 & 4.78 & 72.8 $\pm$ 2.2\\
 & F & Opp & 65-74 & Mono & 900 & 0.96 & 17.7 $\pm$ 1.2\\
 & F & Opp & 65-74 & Poly & 78 & 3.37 & 36.1 $\pm$ 6.7\\
 & F & Same & 16-24 & Mono & 156 & 1.91 & 30.9 $\pm$ 2.3\\
 & F & Same & 16-24 & Poly & 11 & 5.91 & 56.8 $\pm$ 9.7\\
 & F & Same & 25-34 & Mono & 99 & 1.23 & 38.2 $\pm$ 3.9\\
 & F & Same & 25-34 & Poly & 14 & 4.64 & 81.2 $\pm$ 7.6\\
 & F & Same & 35-44 & Mono & 110 & 1.11 & 34.6 $\pm$ 3.1\\
 & F & Same & 35-44 & Poly & 4 & 3.25 & 38.0 $\pm$ 14.6\\
 & F & Same & 45-54 & Mono & 108 & 0.77 & 16.8 $\pm$ 3.1\\
 & F & Same & 45-54 & Poly & 3 & 4.33 & 45.6 $\pm$ 9.5\\
 & F & Same & 55-64 & Mono & 135 & 0.61 & 16.7 $\pm$ 2.6\\
 & F & Same & 55-64 & Poly & 6 & 3.00 & 10.3 $\pm$ 11.8\\
 & F & Same & 65-74 & Mono & 132 & 0.53 & 11.9 $\pm$ 3.0\\
 & F & Bi & 16-24 & Mono & 135 & 2.56 & 63.7 $\pm$ 2.1\\
 & F & Bi & 16-24 & Poly & 23 & 8.09 & 87.1 $\pm$ 3.5\\
 & F & Bi & 25-34 & Mono & 91 & 1.71 & 66.1 $\pm$ 2.5\\
 & F & Bi & 25-34 & Poly & 17 & 6.35 & 91.0 $\pm$ 3.3\\
 & F & Bi & 35-44 & Mono & 98 & 1.26 & 51.1 $\pm$ 2.8\\
 & F & Bi & 35-44 & Poly & 20 & 4.05 & 100.0 $\pm$ 0.0\\
 & F & Bi & 45-54 & Mono & 108 & 1.19 & 48.7 $\pm$ 2.7\\
 & F & Bi & 45-54 & Poly & 15 & 5.33 & 78.0 $\pm$ 5.0\\
 & F & Bi & 55-64 & Mono & 104 & 1.13 & 29.0 $\pm$ 3.4\\
 & F & Bi & 55-64 & Poly & 13 & 4.46 & 77.8 $\pm$ 6.2\\
 & F & Bi & 65-74 & Mono & 111 & 0.89 & 17.5 $\pm$ 2.8\\
 & F & Bi & 65-74 & Poly & 9 & 5.89 & 17.0 $\pm$ 17.4\\
\bottomrule
\end{longtable}

\printbibliography
\end{refsection}

\clearpage
\beginsupplement
\setcounter{page}{1}
\renewcommand{\thepage}{S\arabic{page}}

\begin{refsection}

\clearpage
\begin{center}
{\LARGE \textbf{Supplementary Material}}\\[0.3cm]
    {\LARGE Modelling sexual partnership dynamics and population heterogeneities in agent-based dynamic network models}\\[0.5cm]
    {\normalsize Priyanka Nair-Turkich\textsuperscript{1}, Patricia T. Campbell\textsuperscript{2}, Nicholas Geard\textsuperscript{1} \\[0.2cm]
\textsuperscript{1}{School of Computing and Information Systems, The University of Melbourne, Parkville, Victoria, Australia} \\
\textsuperscript{2}{Department of Infectious Diseases, The University of Melbourne, at the Peter Doherty Institute for Infection and Immunity, Melbourne, VIC, Australia}
}
\end{center}
\vspace{0.5cm}

\beginsupplement
\section{Supplementary Material Summary}
This document is the supplementary material accompanying the main article and its appendices and includes: (1) the full Overview, Design Concepts and Details (ODD) protocol \citep{grimm2020odd} for the agent-based dynamic sexual partnership network model; (2) a brief description of modelling assumptions and design choices, and the rationale behind them; (3) annotated, minimal Python code blocks for the essential mechanisms (partnership formation, partnership dissolution, network construction, and the \textit{Susceptible-Infected-Susceptible} tra-nsmission model, intended to make the algorithmic logic reproducible. Parameter value tables (best-fit values, sampled ranges, and fixed simulation parameters) are omitted from this document, as they are reported in full in Appendix A of the main article. Equations that are already fully specified in the main text are stated here for ODD protocol completeness. 
\section{Overview, Design Concepts and Details (ODD) Protocol}
\label{sec:odd_protocol}
\subsection{Model overview}
\label{subsec:odd_overview}
The \textit{Partnersim-dynet} package simulates the influence of partnership dynamics such as formation, dissolution, and concurrency, and population heterogeneities on the network-based transmission of sexually transmitted infections (STIs). The model is designed to help understand how variations in partnership dynamics such as count, duration and concurrency, and population heterogeneities such as age, sex, and sexual orientation impact the spread of STIs in a population.

\subsection{Purpose and Patterns}
\label{subsec:odd_purpose}
This model was developed to simulate the formation and dissolution of sexual partnerships within a heterogeneous population, incorporating individual variation in sex, age, and sexual orientation, including bisexual individuals. The model is designed to reproduce the distribution of the partnership counts, stratified by age, sex, and sexual orientation, published in the third National Survey of Sexual Attitudes and Lifestyles (NATSAL-3) data \citep{Mercer2013SexualAttitudesNatsal, Clifton2023Natsal3RefTables} (Table 1 of the main article). The research questions and model calibration against NATSAL-3 data are described in full in Section 1 and Section 2.6 of the main article. Partnership formation and dissolution are governed by probabilistic, demographically-stratified mechanisms, including a duration-dependent hazard applied to dissolution and individual-level heterogeneity in sexual activity. The resulting dynamic networks provide a generalisable framework for simulating the transmission of bacterial STIs, illustrated in the main article using a \textit{Susceptible-Infected-Susceptible} (SIS) transmission model. 

\subsection{Entities, State Variables, and Scales}
\label{subsec:odd_entities}
Agents are characterised by both fixed attributes (sex and sexual orientation, including bisexual individuals) and time-varying attributes (age, sexually-active status, and current partnerships), with the distributions of sex and orientation calibrated to NATSAL-3 data. Table~\ref{tab:s-agents} lists the full state, variables, and scales of the agents.

\paragraph{Environment}
Every partnership generated during the simulation is recorded with the fields in Table~\ref{tab:s-env}. This is the full list of parameters recorded in the simulation, as referred to in Section 2.3.5 of the main text.

\paragraph{Temporal and Spatial Scale}
\label{subsubsec:odd_temporal}
\begin{itemize}
    \item \textbf{Time step:} 1 day
    \item \textbf{Duration of simulation:} 1,875 days (~5 years)
    \item \textbf{Population size:} $N = 15{,}000$, as used for the calibration of the partnership model and the outputs; disease simulations use \param{max\_steps} $=1{,}825$ (Appendix A, Table: disease transmission parameters, main article).
    \item \textbf{Population size:} $N = 15{,}000$, maintained approximately constant through the continuous replacement of agents leaving the sexually active pool at 75 years.
\end{itemize}

\begin{table}[H]
\centering
\caption{Agents and their state variables.}
\label{tab:s-agents}
\renewcommand{\arraystretch}{1}
\begin{tabular}{llp{3cm}>{\raggedright\arraybackslash}p{5.3cm}}
\toprule
\textbf{Variable} & \textbf{Type} & \textbf{Values} & \textbf{Description} \\
\midrule
\param{id} & int & $[0, N)$ & Unique agent identifier \\
\param{sex} ($s_i$) & categorical & \{Male, Female\} & Fixed biological sex, assigned at initialisation with equal probability \\
\param{orientation} ($o_i$) & categorical & \{Opposite-sex, Bisexual, Same-sex\} & Fixed sexual orientation, assigned conditional on sex \\
\param{age} ($a_i(t)$)  & int (years) & $[16, 74]$ & Age, incremented via daily birthday counter \\
\param{birthday\_offset} ($d_i(t)$) & int (days) & $[0, 364]$ & Days elapsed since last birthday \\
\param{sexually\_active} & bool & \{True, False\} & Determines eligibility for partnership formation \\
\param{concurrency\_eligible} & bool & \{True, False\} & Determines whether the agent may hold more than one active partnership \\
\param{$K_i$} & int & $\geq 2$ (concurrency-eligible only) & Individual maximum number of simultaneous partnerships \\
\param{partners} & set of ids & -- & Current set of active partnership identifiers \\
\param{$\eta^{\text{form}}_i$} & float & $\geq 1$ & Individual multiplier on partnership formation probability \\
\param{$\eta^{\text{diss}}_i$} & float & $\geq 1$ & Individual multiplier on partnership dissolution probability \\
\param{disease\_state} & categorical & \{S, I\} & Susceptible/Infectious state used by the SIS submodel (Section~\ref{subsubsec:odd_transmission}) \\
\bottomrule
\end{tabular}
\end{table}

\begin{table}[H]
\centering
\caption{Partnership record: fields logged for every partnership generated during the simulation.}
\label{tab:s-env}
\renewcommand{\arraystretch}{1.3}
\begin{tabular}{lp{7.5cm}}
\toprule
 \textbf{Field} & \textbf{Description} \\
\midrule
 \param{partner\_ids}           & Identifiers of the two agents in the partnership \\
 \param{partner\_demographics}  & Sex, orientation, age, and age group of both partners at the time of formation \\
 \param{start\_time}            & Time step (day) of partnership formation \\
 \param{end\_time}              & Time step (day) of partnership dissolution \\
 \param{duration}               & Observed duration from formation to dissolution (days) \\
 \param{censoring\_indicator}   & \texttt{False} if the partnership dissolved during the simulation; \texttt{True} if still active at the end of the simulation \\
 \param{external\_partnership\_indicator} & \texttt{True} if one partner was removed from the sexually active pool at age 75 and the partnership was retained for the surviving partner \\
\bottomrule
\end{tabular}
\end{table}

\subsection{Process Overview and Scheduling}
\label{subsec:odd_scheduling}

At each daily time step $t$, the following processes are executed in a fixed sequential order (schematically shown in Figure 3 of the main article), preventing logical inconsistencies such as an agent forming and dissolving a partnership on the same day or a removed agent being selected as a partner candidate:

\begin{enumerate}
    \item \textbf{Ageing and birthday updates:} all agents age by one day; agents whose birthday falls on the current time step have their age incremented by one year.
    \item \textbf{Sexual debut:} agents between the ages of 16--20 who reach a birthday are evaluated against age-specific debut probabilities in Table~\ref{tab:s-debut}; agents who debut become eligible for partnership formation from that time step onwards.
    \item \textbf{Agent removal:} agents whose age exceeds 74 are flagged and removed from the sexually active pool; the surviving partnerships of the removed agents are reclassified as external partnerships and remain at risk of dissolution under the same hazard mechanism (Section~\ref{subsubsec:odd_dissolution}).
    \item \textbf{Population replacement:} new 16-year-old agents are introduced to replace those removed, with sex and sexual orientation drawn from the assigned distributions and independent formation and dissolution heterogeneity multipliers assigned at initialisation (Section~\ref{subsubsec:odd_population_turnover}).
    \item \textbf{Partnership formation:} a pool of eligible agents is constructed from those who are sexually active and below their maximum allowed number of concurrent partnerships. Each eligible agent independently attempts to form a partnership; candidates are restricted to sex- and orientation-compatible pairs, and weighted by age similarity (Section~\ref{subsubsec:odd_formation}).
    \item \textbf{Partnership dissolution:} every active partnership is evaluated against a duration-dependent dissolution probability; partnerships formed during the current time step are protected from dissolution (Section~\ref{subsubsec:odd_dissolution}).
\end{enumerate}

\subsection{Design Concepts}
\label{subsec:odd_concepts}
\paragraph{Basic Principles:} The underlying principle of this model is the use of temporal sexual contact networks as an abstraction of population-level sexual activity, in which nodes represent individuals or agents, and edges represent sexual partnerships \citep{eames2002modeling}. Unlike static network representations, the model explicitly incorporates partnership formation, dissolution, and concurrency over time \citep{frieswijk2023time}. The model extends the agent-based modelling approaches of \citep{Tsoumanis_et_al} and \citep{azizi_using_2021} by incorporating a duration-dependent dissolution hazard \citep{Carroll2003WeibullModelSurvival} and by  representing bisexual individuals as a distinct demographic stratum capable of forming bridging partnerships between otherwise separate same-sex and opposite-sex sexual networks.

\paragraph{Emergence:}Network-level structural properties -- including the degree distribution, the prevalence and distribution of concurrent partnerships, age-assortative mixing patterns, and the extent of bridging between sub-networks of different orientations via bisexual individuals -- emerge from individual-level formation and dissolution rules rather than being directly imposed on the population.

\paragraph{Adaptation:}None. Agents do not adjust their partnership formation or dissolution behaviour in response to partnership history within a simulation run.

\paragraph{Objectives:}None. Partnership formation and dissolution are entirely governed by probabilistic mechanisms.

\paragraph{Learning:}None. Agents do not update their behaviour based on experience.

\paragraph{Prediction:}None. Agents do not forecast future states to guide current behaviour.

\paragraph{Sensing:}The agents perceive the sex, sexual orientation, age and current partnership count of the candidate partners; this information determines eligibility and selection weighting (Section \ref{subsubsec:odd_formation}).

\paragraph{Interaction:}Agents interact exclusively through the sexual partnership network. The compatibility of the combinations of sex and orientations is evaluated first; among compatible candidates, the selection of the partner is weighted by age similarity using a Gaussian kernel (Equation~\ref{eq:s-age_assortativity}).

\paragraph{Stochasticity:}The following processes are stochastic, detailed in Sections \ref{subsec:odd_init} and \ref{subsubsec:odd_transmission}). The assignment of sex and orientation; initial age and birthday offset; sexual debut; concurrency eligibility and individual cap $K_i$; individual heterogeneity multipliers $\eta^{\text{form}}_i, \eta^{\text{diss}}_i$; daily partnership-formation attempts; partner selection among compatible candidates; daily partnership-dissolution evaluation; disease transmission and recovery (Section~\ref{subsubsec:odd_transmission}).
The following processes are stochastic:

\begin{itemize}
    \item Assignment of biological sex: Bernoulli trial with probability 0.5.
    \item Assignment of sexual orientation: categorical draw conditional on sex 
    \item Age assigned at initialisation: discrete uniform draw over $[16, 74]$ from the respective age groups.
    \item Birthday offset: discrete uniform draw over $[0, 364]$.
    \item Sexual debut (ages 16--20): Bernoulli trial against age-specific cumulative debut probabilities.
    \item Concurrency eligibility: Bernoulli trial with probability $\pi_{\text{conc}}$.
    \item Individual concurrency cap $K_i$: Poisson draw with rate $\lambda$, floored at 2.
    \item Individual-level heterogeneity multipliers $\eta^{\text{form}}_i$, $\eta^{\text{diss}}_i$: independent negative binomial draws given in Eq ~\ref{eq:s-nb_multiplier}.
    \item Partnership formation: Bernoulli trial per eligible agent per day, with demographic stratum-specific and individually adjusted probability.
    \item Partner selection among compatible candidates: weighted random draw using the Gaussian age-assorta-tivity kernel (Equation~\ref{eq:s-age_assortativity}).
    \item Partnership dissolution: Bernoulli trial per active partnership per day, against the duration-dependent dissolution hazard (Equations~\ref{eq:s-dissolution_adj}--\ref{eq:s-hazard}).

\end{itemize}

\paragraph{Collectives:} None. Agents are stratified into demographic strata (sex $\times$ orientation $\times$ age group) to assign formation and dissolution probabilities, but these strata do not carry their own state variables or group-level behaviours.

\paragraph{Observation:} The partnership records given in (Table~\ref{tab:s-env}) are the primary observational data. Summary network statistics -- including degree distribution, partnership duration distribution, and concurrency prevalence -- are computed from these records and used as model calibration targets against NATSAL-3 (Table 1, main article).

\subsection{Initialisation}
\label{subsec:odd_init}

\begin{enumerate}
\item Agents $N = 15{,}000$ are created.
\item Each agent is assigned a biological sex $s_i$ with equal probability:
\begin{equation}
\Pr(s_i = \text{Male}) = \Pr(s_i = \text{Female}) = 0.5.
\end{equation}
\item Sexual orientation $o_i$ is assigned conditional on $s_i$. 
For males:
\begin{equation}
\label{eq:s-male_orientation}
\Pr(o_i \mid s_i = \text{Male}) =
\begin{cases}
0.90 & o_i = \text{Opposite-sex} \\
0.05 & o_i = \text{Same-sex} \\
0.05 & o_i = \text{Bisexual}
\end{cases}
\end{equation}
For females:
\begin{equation}
\label{eq:s-female_orientation}
\Pr(o_i \mid s_i = \text{Female}) =
\begin{cases}
0.80 & o_i = \text{Opposite-sex} \\
0.10 & o_i = \text{Same-sex} \\
0.10 & o_i = \text{Bisexual}
\end{cases}
\end{equation}
Since NATSAL-3 does not explicitly capture the sexual behaviour of bisexual individuals, this stratum was constructed from survey respondents who reported attraction: 
\begin{enumerate}
\item \textit{More often to (females/males) and at least once to (male/female)}
\item \textit{Approximately equally often to (females/males) and to (males/females)}
\item \textit{More often to (males/females) and at least once to (female/male)}
\end{enumerate}
\item The initial age of the agents is assigned by equal allocation, where the population is divided into equal blocks drawn across the six ten-year age groups (16--24, \ldots, 65--74), with any remaining agents from uneven division distributed between the age groups. The age of every agent is then drawn uniformly within its assigned age group range. Marginally, this reproduces an approximately uniform distribution of age of agents given by:
\begin{equation}
a_i(0) \sim \text{DiscreteUniform}(16, 74).
\end{equation}
A birthday offset is drawn independently per agent to avoid synchronised ageing events:
\begin{equation}
d_i(0) \sim \text{DiscreteUniform}(0, 364).
\end{equation}
Because initial offsets are uniformly distributed, birthdays are spread uniformly throughout the calendar year at the population level.
\item Agents aged 21 and over are treated as having already made their sexual debut. The model does not represent debut before the age of 16, which is applicable to a relatively smaller proportion of the population. For agents aged 16--20, debut is sampled against the cumulative debut probabilities in Table~\ref{tab:s-debut}, calibrated so that the modelled mean age at the first sexual contact is approximately 16 years, which is broadly consistent with the estimates from the NATSAL survey \citep{lewis2017heterosexual, wellings2001sexual}.
\begin{table}[H]
\centering
\caption{Cumulative probability of sexual debut by age, ages 16--20.}
\label{tab:s-debut}
\begin{tabular}{lc}
\toprule
Age & Cumulative probability of debut \\
\midrule
16 & 0.50 \\
17 & 0.70 \\
18 & 0.85 \\
19 & 0.95 \\
20 & 1.00 \\
\bottomrule
\end{tabular}
\end{table}
\item Each agent is designated concurrency-eligible with probability $\theta_{\text{conc}}$ (main text Section 2.3.4); and those agents are assigned an individual maximum:
\begin{equation}
\label{eq:s-concurrency_K}
K_i = \max\!\left(2,\ \text{Poisson}(\lambda)\right).
\end{equation}
where $K$ is the maximum number of concurrent partners per agent. Higher $\lambda$ indicates greater partnership concurrency in the population. Agents not designated concurrency-eligible are strictly monogamous ($K_i = 1$).
\item Individual-level heterogeneity multipliers
$\eta^{\text{form}}_i$ and $\eta^{\text{diss}}_i$ are drawn independently (Section~\ref{subsubsec:odd_heterogeneity}).
\end{enumerate}

\subsection{Input data}
\label{subsec:odd_input_data}
The model was calibrated to the data from the third National Survey of Sexual Attitudes and Lifestyles (NATSAL-3) \citep {Mercer2013SexualAttitudesNatsal, Clifton2023Natsal3RefTables}
\subsection{Submodels}
\label{subsec:odd_submodels}

\subsubsection{Partnership formation}
\label{subsubsec:odd_formation}

The daily probabilities of the formation and dissolution of partnerships are stratified by sex, sexual orientation, and age group ($6$ sex--orientation combinations $\times$ $6$ age groups $= 36$ strata), with the reference stratum being $p^{\text{form}}_{\text{base}}$ (opposite-sex females, 25--34). Probabilities for other strata follow the multiplicative structure defined in the main text (Eq.~9):
\begin{equation}
\label{eq:s-formation_strat}
p^{\text{form}}_{s,o,g} = p^{\text{form}}_{\text{base}} \cdot \mu^{\text{ori}}_{s,o} \cdot \mu^{\text{age}}_{g},
\end{equation}
with the age multiplier given by:
\begin{equation}
\label{eq:s-age_mult}
\mu^{\text{age}}_{g} =
\begin{cases}
\beta & g = 1 \ (\text{ages } 16\text{--}24) \\
1 & g = 2 \ (\text{ages } 25\text{--}34, \text{ reference}) \\
\exp\!\left(-\kappa \cdot (g - 2)\right) & g = 3, 4, \ldots, 6,
\end{cases}
\end{equation}
where $\beta$ (\textit{youth boost}) and $\kappa$ (\textit{age decay}) are calibrated parameters (Appendix A, main article). The same stratification and multiplicative structure apply to the baseline dissolution probability $p^{\text{diss}}_{\text{base},s,o,g}$ (Section~\ref{subsubsec:odd_dissolution}).

For example, if $p^{\text{form}}_{\text{base}} = 0.0025$, an opposite-sex male aged 16--24 with $\mu^{\text{ori}}_{M,\text{Opposite-sex}} = 1.3$ and $\beta = 1.8$ has daily formation probability
\[
0.0025 \times 1.3 \times 1.8 = 0.00585.
\]
\textbf{Partnership formation steps:} At each time step, a pool of agents eligible to initiate partnership formation is constructed. An agent is eligible if they are active, sexually active, and their current number of partnerships is below their allowed maximum (one for monogamous agents; $K_i \geq 2$ for concurrency-eligible agents). Eligible agents are processed in random order. Each agent independently attempts to initiate a partnership with a daily formation probability specific to their demographic stratum (after individual-level adjustment, Section~\ref{subsubsec:odd_heterogeneity}). When an agent initiates, a partner is drawn from candidates that satisfy the following conditions:
\begin{enumerate}
\item \textbf{Self-exclusion:} an agent cannot partner with themselves.
\item \textbf{Availability:} candidates must be sexually active and below their allowed partnership maximum.
\item \textbf{ Sex and orientation compatibility:}
\begin{itemize}
    \item Opposite-sex agents seek opposite-sex partners whose orientation is opposite-sex or bisexual.
    \item Same-sex agents seek same-sex partners whose orientation is same-sex or bisexual.
    \item Bisexual agents may seek partners of either sex: a bisexual male is compatible with females who are opposite-sex-oriented and with males that are same-sex-oriented; a bisexual female is compatible with males that are opposite-sex-oriented and with females that are same-sex-oriented.
\end{itemize}
\item \textbf{Single formation per time step:} agents who have already formed a partnership in the current time step are excluded from further selection.
\end{enumerate}

Among the resulting compatible candidate set, partner selection uses Gaussian weighting on age difference:
\begin{equation}
\label{eq:s-age_assortativity}
w_{ij} \propto \exp\!\left(-\frac{(a_j - a_i)^2}{2\sigma^2}\right), \qquad \sigma = 4.0,
\end{equation}
so that partners closer in age are more likely to be selected.   
Age-assortativity is represented through a Gaussian kernel with fixed bandwidth $\sigma = 4$ years; this is a simplification of more complex empirical patterns of partner-age preference, which differ by sex and age. $\sigma = 4$ years gives an age difference distribution centred at 0 and most values being typically within $\pm 10$ years, such that agents are most likely to form partnerships with others of similar age. An agent whose initiation attempt fails to find a partner remains in the pool only as a passive candidate during that time step but does not attempt initiation again. Once a partnership is formed, both partners' counts are incremented, and both are removed from further formation consideration for the remainder of the time step. 
\subsubsection{Partnership dissolution}
\label{subsubsec:odd_dissolution}
The daily baseline dissolution probabilities $p^{\text{diss}}_{\text{base}, s, o, g}$ are stored in a three-dimensional matrix stratified by sex, sexual orientation, and age group, with the same reference structure, multipliers applied as a product of age, sex, and orientation, as the partnership formation process described in and Equation~\eqref{eq:s-formation_strat}. To represent the empirical observation that the risk of dissolution is highest near the inception of the partnership and decreases with the increase in partnership longevity, we model the risk of partnership dissolution using a Weibull-like hazard function in which the daily probability of partnership dissolution decays smoothly with the age of the partnership \citep{Carroll2003WeibullModelSurvival}. 
\begin{equation}
\label{eq:s-dissolution_adj}
p^{\text{diss}}_{\text{adj}}(d) = p^{\text{diss}}_{\text{base}, s, o, g} \cdot f(d),
\end{equation}
with
\begin{equation}
\label{eq:s-hazard}
f(d) = \left(1 + \frac{d}{\alpha}\right)^{-\gamma}, \qquad \alpha = 1500, \ \gamma = 2,
\end{equation}
where $d$ is the duration of the partnership in days, $\alpha$ is a scale parameter, and $\gamma$ is a shape parameter. The same parameter values are applied to internal partnerships (both partners are active) and external partnerships (one partner removed at age 75). Partnerships are protected from dissolution on the day of formation; the dissolution evaluation applies only when $d > 1$. At each time step, for every active partnership, the model:
\begin{enumerate}
\item computes the current partnership duration $d$;
\item evaluates $p^{\text{diss}}_{\text{adj}}(d)$ from Equations~\eqref{eq:s-dissolution_adj} and~\eqref{eq:s-hazard};
\item draws a uniform random number $u \sim \text{Uniform}(0,1)$;
\item dissolves the partnership if $u < p^{\text{diss}}_{\text{adj}}(d)$.
\end{enumerate}
We assume that the dissolution risk depends only on the demographic stratum of the participating agents and the duration of the partnership, and that the hazard function is shared across all demographic strata. The Weibull-like hazard function implies that very long-duration partnerships approach a near-zero daily dissolution rate, consistent with the persistence of stable long-term relationships. 
\subsubsection{Individual-level heterogeneity}
\label{subsubsec:odd_heterogeneity}
To incorporate heterogeneity in sexual activity within agents of same attributes such as age, sex, and orientation, each agent is assigned independent multiplicative effects on partnership formation and dissolution at initialisation. Each agent draws an integer $X_i$ from a negative binomial distribution with parameters $(r, p)$:
\begin{equation}
X_i \sim \text{NegBin}(r, p),
\end{equation}
which has mean $\mathbb{E}[X_i] = r(1-p)/p$. The integer draw is converted into a multiplier
\begin{equation}
\label{eq:s-nb_multiplier}
\eta_i = 1 + \frac{X_i}{\mathbb{E}[X_i]} \geq 1,
\end{equation}
which is applied to the agent's demographic stratum-specific baseline probability:
\begin{equation}
p^{\text{form}}_{i} = \eta^{\text{form}}_{i} \cdot p^{\text{form}}_{s_i, o_i, g_i(t)}, \qquad p^{\text{diss}}_{i} = \eta^{\text{diss}}_{i} \cdot p^{\text{diss}}_{\text{base}, s_i, o_i, g_i(t)}.
\end{equation}
Independent multipliers $\eta^{\text{form}}_{i}$ and $\eta^{\text{diss}}_{i}$ are drawn separately and remain fixed throughout the simulation. The resulting probabilities are bounded below by a small lower limit ($10^{-4}$) and above by 0.99 to avoid degenerate values. This mechanism yields a right-skewed, heavy-tailed distribution of activity levels, allowing a small proportion of agents to account for disproportionately high sexual activity.

\begin{figure}[H]
\centering
\includegraphics[width=\linewidth,height=9cm]{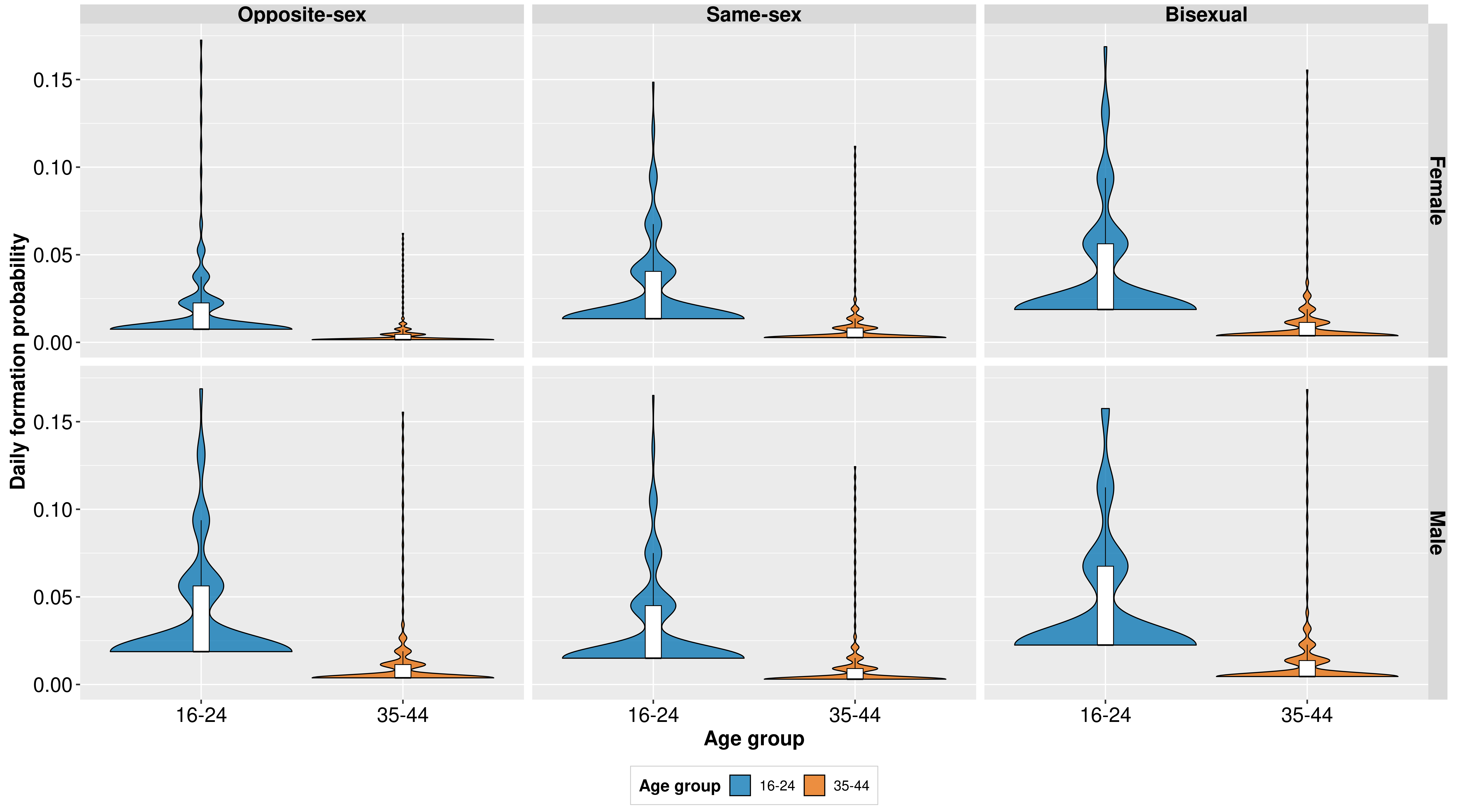}
\caption{Distribution of formation and dissolution probabilities within the age groups 16--24 and 35--44 for males and females. Violin width represents probability distribution of agent-specific formation probabilities.}
\label{fig:nb_distribution}
\end{figure}

Given $r = 0.5$ and $p = 0.5$, for a population of $N = 15{,}000$ agents, we use the following example to illustrate the heterogeneity in formation and dissolution probabilities, which includes some of the demographic strata multipliers presented in Eq \ref{eq:s-formation_strat} and \ref{eq:s-age_mult} to see the resulting heterogeneity in the probability of partnership formation in Figure~\ref{fig:nb_distribution}:
\begin{itemize}
    \item $p_{\text{base}}$ (Female, Opposite-sex, 25--34) = 0.0025
    \item Sex multipliers ($\mu_s$): Female = 1.0, Male = 1.0
    \item Orientation multipliers ($\mu_o$):
    \begin{itemize}
        \item Female: Opposite-sex = 1.0, Same-sex = 1.8, Bisexual = 2.5
        \item Male: Opposite-sex = 2.5, Same-sex = 2.0, Bisexual = 3.0
    \end{itemize}
    \item Age multiplier ($\mu_g$): 
    \begin{itemize}
        \item 16--24: 3.0 (youth boost)
        \item 35--44: $e^{-0.5} \approx 0.606$ (age decay)
    \end{itemize}
\end{itemize}

Because $\eta_i \geq 1$ by construction, the multiplier only inflates the probabilities of the baseline values, which means that agents do not have multipliers that reduce them below the baseline. The values of $\eta^{\text{form}}_{i}$ and $\eta^{\text{diss}}_{i}$ are independent, which means that there is no correlation between the agent and the propensity to form and dissolve partnerships.
\subsubsection{Population turnover}
\label{subsubsec:odd_population_turnover}
Agents whose age exceeds 74 years are flagged for removal from the sexually active pool. This occurs through the continuous ageing process described in the previous sub-section by using birthday counters. The model removes an agent when they reach the age of 75. To maintain a constant population size, the model adds new 16-year-old agents to replace all agents over the age of 75 that are removed from the pool of active agents. Each replacement agent is assigned:
\begin{enumerate}
\item a unique agent identifier, one greater than the previous maximum;
\item an initial age of 16 years;
\item sex and orientation, sampled from equations~\eqref{eq:s-male_orientation} and~\eqref{eq:s-female_orientation};
\item an empty partnership set;
\item a uniformly distributed birthday offset $d_i \sim \text{DiscreteUniform}(0, 364)$;
\item independent formation and dissolution heterogeneity multipliers (see Section~\ref{subsubsec:odd_heterogeneity}).
\end{enumerate}
When an agent over the age of 74 is removed, their partners who remain in the active pool retain a record of the partnership. Such partnerships are reclassified as \textit{external partnerships} and remain at risk of dissolution under the same hazard mechanism as internal partnerships (Section~\ref{subsubsec:odd_dissolution}). We assume that population size and the distribution of ages remain stable over time, with agent removals at age 75 balanced by inflows of new 16-year-old agents. The retention of partnerships across the age threshold (as external partnerships) reflects the assumption that the older partner's exit from the pool of sexually active agents does not end the relationship from the perspective of the surviving partner.

\subsubsection{Dynamic network construction}
\label{subsubsec:odd_network_construction}
Partnership formation and dissolution (Sections~\ref{subsubsec:odd_formation} and~\ref{subsubsec:odd_dissolution}) generate a flat timestamped record of the start and end time of every partnership. The sexual contact network at a given time step $t$ is a graph whose nodes are the agents active at $t$ and whose edges are the partnerships active at $t$.

An agent is active at the time step $t$ if
\begin{equation}
\label{eq:s-active_interval}
\texttt{EntryTimestep} \leq t < \texttt{ExitTimestep},
\end{equation}
Agents who are still active at the end of the simulation are treated as active at every $t$ up to and including
the final time step. The network graph $G(t)$ is constructed as follows:
\begin{enumerate}
\item \textbf{Nodes:} every agent active at $t$, including agents with zero
current partnerships, who appear as isolated nodes.
\item \textbf{Edges:} every partnership satisfying the active-partnership
condition above, restricted to pairs where \emph{both} agents are active at
$t$.
\end{enumerate}

\subsubsection{Transmission}
\label{subsubsec:odd_transmission}

\textit{partnersim-dynet-sti} package models disease transmission using a discrete-time \textit{Susceptible-Infectious-Susceptible} (SIS) framework layered on top of the dynamic partnership network described in Section~\ref{subsubsec:odd_formation}--\ref{subsubsec:odd_dissolution} \citep{keeling_rohani}, as introduced in main text Section 2.7. Each agent occupies one of two mutually exclusive disease states at any time step: \textbf{$S$: Susceptible}, denoting an agent who is currently uninfected (including agents who have previously cleared an infection), and \textbf{$I$: Infectious}, denoting an agent who is currently infected and capable of transmitting the pathogen to a susceptible partner. Unlike the partnership formation and dissolution processes (Sections~\ref{subsubsec:odd_formation} and~\ref{subsubsec:odd_dissolution}), which depend on demographic stratum and individual-level heterogeneity multipliers, transmission risk is assumed to be homogeneous across all agents and partnership types, and depends only on the current disease state of both partners in an active partnership. Agents are seeded at $t=51$ time step with initial infectious proportion, drawn randomly independent of sex, orientation, age or concurrency status. Each active partnership has an independent with per-partnership, per-time-step probability $\beta_{\text{trans}}$. External partnerships are not included in the disease transmission simulation as they are not active partnerships.  

\subsubsection{Recovery}
\label{subsubsec:odd_recovery}

Every infectious agent recovers independently of their partnership status or partner disease state, with probability $\gamma_{\text{recov}}$ per time step, and returns immediately to $S$ state without temporary or long-term immunity. This mechanism is consistent with the natural history of many bacterial STIs, in which reinfection can occur immediately following clearance \citep{kretzschmar2017pair}. Transmission and recovery mechanisms are evaluated as independent Bernoulli processes at each time step. Disease parameter values $\beta_{\text{trans}}$, $\gamma_{\text{recov}}$, \param{infection\_start}, \param{max\_steps} and \param{initial\_infected} are given in Appendix A of the main article.     

\subsubsection{Model calibration}
\label{subsubsec:odd_model_calibration}

\begin{figure}[H]
\centering
\begin{subfigure}{\linewidth}
\includegraphics[width=1\linewidth, height=16cm, keepaspectratio]{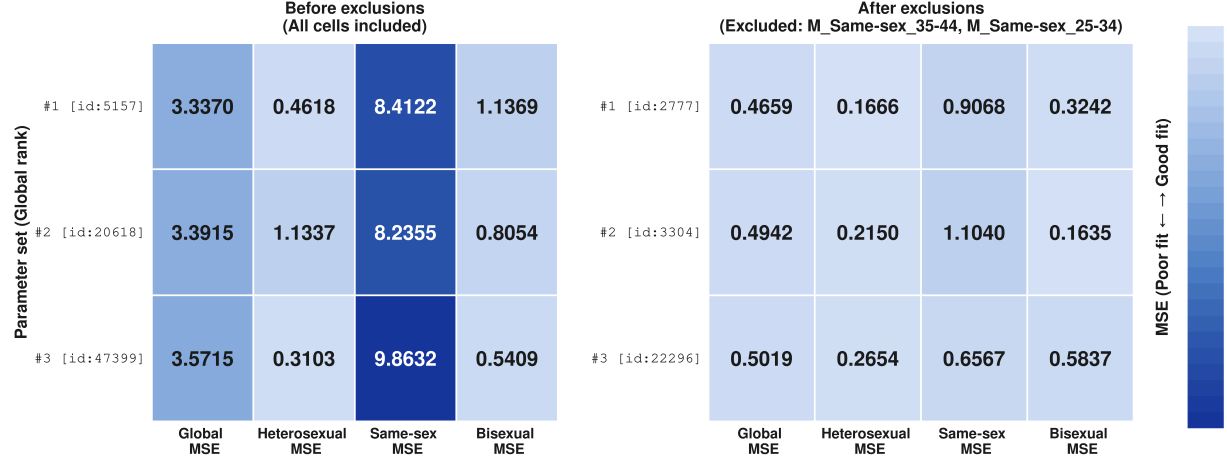}
\caption{No concurrency ($\theta_{\mathrm{conc}} = 0$), best-fit parameter set id:2777}
\end{subfigure}
\vspace{12pt}
\begin{subfigure}{\linewidth}
\includegraphics[width=1\linewidth, height=18cm, keepaspectratio]{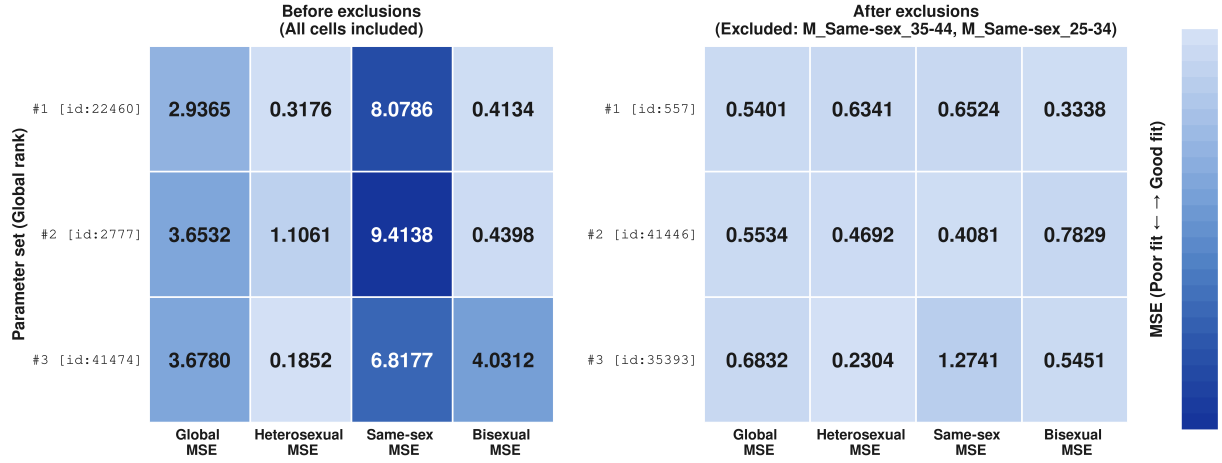}
\caption{15\% concurrency ($\theta_{\mathrm{conc}} = 0.15$), best-fit parameter set id:557}
\end{subfigure}
\caption{Ranked model fit (global and orientation-specific MSE) before and 
after exclusions of the same-sex male 25--34 and 35--44 outlier targets, for the top three parameter sets by global rank for no-concurrency and with concurrency scenarios. Full best-fit parameter values are given in Appendix A.}
\label{fig:ranked_model_fit}
\end{figure}

Latin Hypercube Sampling (LHS) \citep{McKay1979ComparisonThreeMethods} was implemented in \textit{partnersim-dynet-lhs} package to explore 60,000 parameter combinations per concurrency scenario, ranked by Mean Squared Error (MSE) against NATSAL-3 targets for calibration. The ranges of sampled parameters and best-fit values are reported in Appendix A of the main article. Following the exclusion of mean partnership count for same-sex males of age groups 25--34 and 35--44 (main text Section 2.6.3), the calibration performance improved substantially. Prior to the exclusion, these subgroups contributed disproportionately to the global MSE across the top-ranked parameter sets, with same-sex MSE values consistently being high despite comparatively lower MSE in the heterosexual and bisexual subgroups, resulting in elevated global MSE values. Figure~\ref{fig:ranked_model_fit} compares the ranked calibration performance of the top three parameter sets before and after this exclusion. Following exclusion, same-sex MSE improved and the global MSE decreased substantially, indicating an improved overall fit of the model. 

\section{Model Design}
\label{sec:odd_design_choices}

\begin{description}[leftmargin=0pt, style=nextline, font=\bfseries]
\item[Binary sex classification and bisexual stratum]
The binary sex classification reflects the structure of available empirical data from sexual behaviour surveys such as NATSAL-3, which collect data using male/female categories. The NATSAL-3 survey does not explicitly capture the sexual behaviour of bisexual individuals; therefore, an additional demographic stratum of bisexual individuals was constructed based on the responses given to the question related to sexual attraction in the survey. 

\item[Concurrency]
Concurrency eligibility and the individual cap $K_i$ are assigned once, at initialisation (or at replacement). This keeps the effect of the mechanism identifiable and isolates the structural effect of concurrency from any assumption about within-individual behaviour change. Figure \ref{fig:concurrency_distributions} shows the impact of different values of $\lambda$ on the maximum number of concurrent partners per agent. For our model, we have chosen $\lambda$ = 2.
\begin{figure}[H]
\centering
\includegraphics[width=1\linewidth, height=7cm]{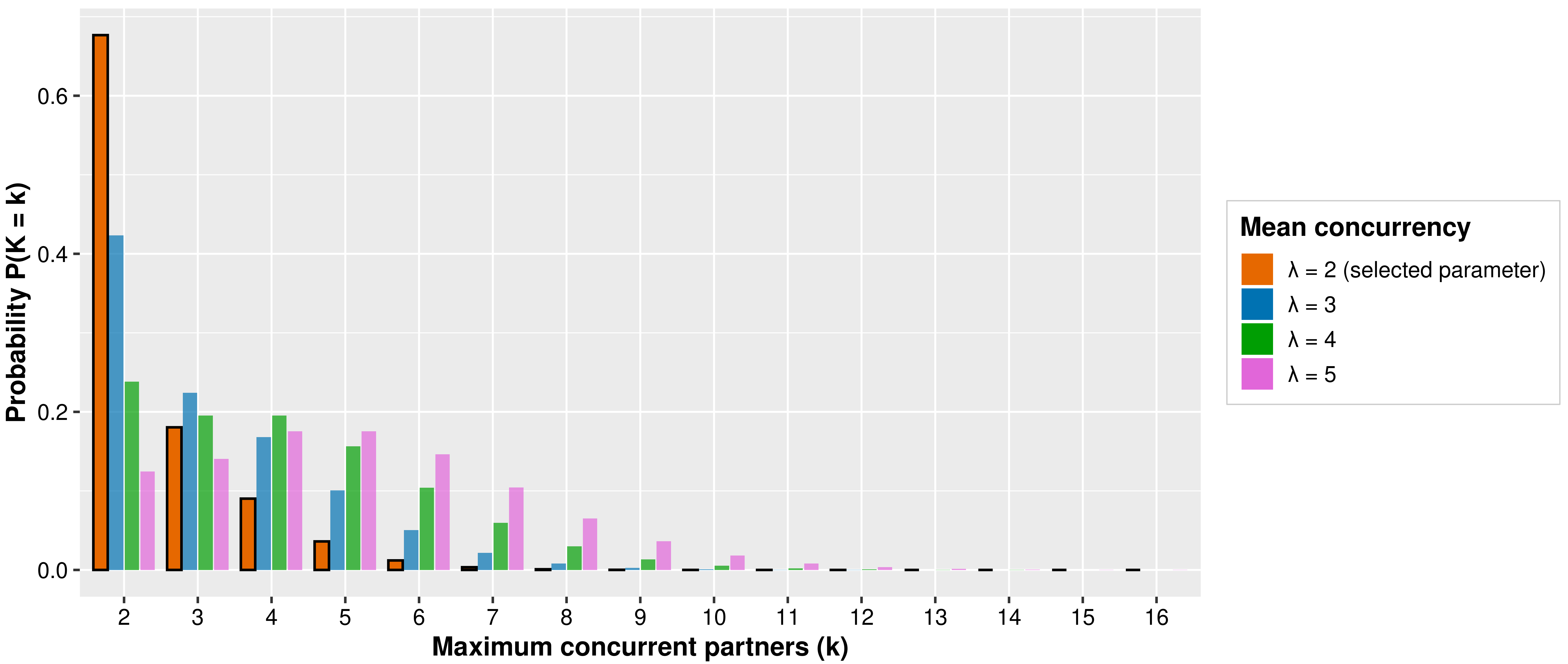}
\caption{Distribution of maximum concurrent partners. Bars: probability of \textit{k} concurrent partnerships under \textit{K = max(2, Poisson($\lambda$))}. Higher ($\lambda$) indicates greater partnership concurrency in the population. \textit{K} is bounded below by 2 (minimum concurrent partnerships). }
\label{fig:concurrency_distributions}
\end{figure}

\item[Weibull-like, duration-dependent dissolution hazard]
A constant daily dissolution probability would imply an exponential (memory-less) duration distribution, which is inconsistent with the empirically higher risk of dissolution shortly after partnership formation \citep{nelson2010age}. We have modelled a hazard of dissolution that decays smoothly with duration. Figure \ref{fig:partnership_dissolution_hazard} shows the impact of the different values of $\alpha$, the scale parameter, and $\gamma$ the shape parameter. The values of $\alpha$ = 1500 and $\gamma$ = 2 result in a slower decay in the risk of partnership dissolution. 
\begin{figure}[H]
\centering
\includegraphics[width=\textwidth, height=18cm]{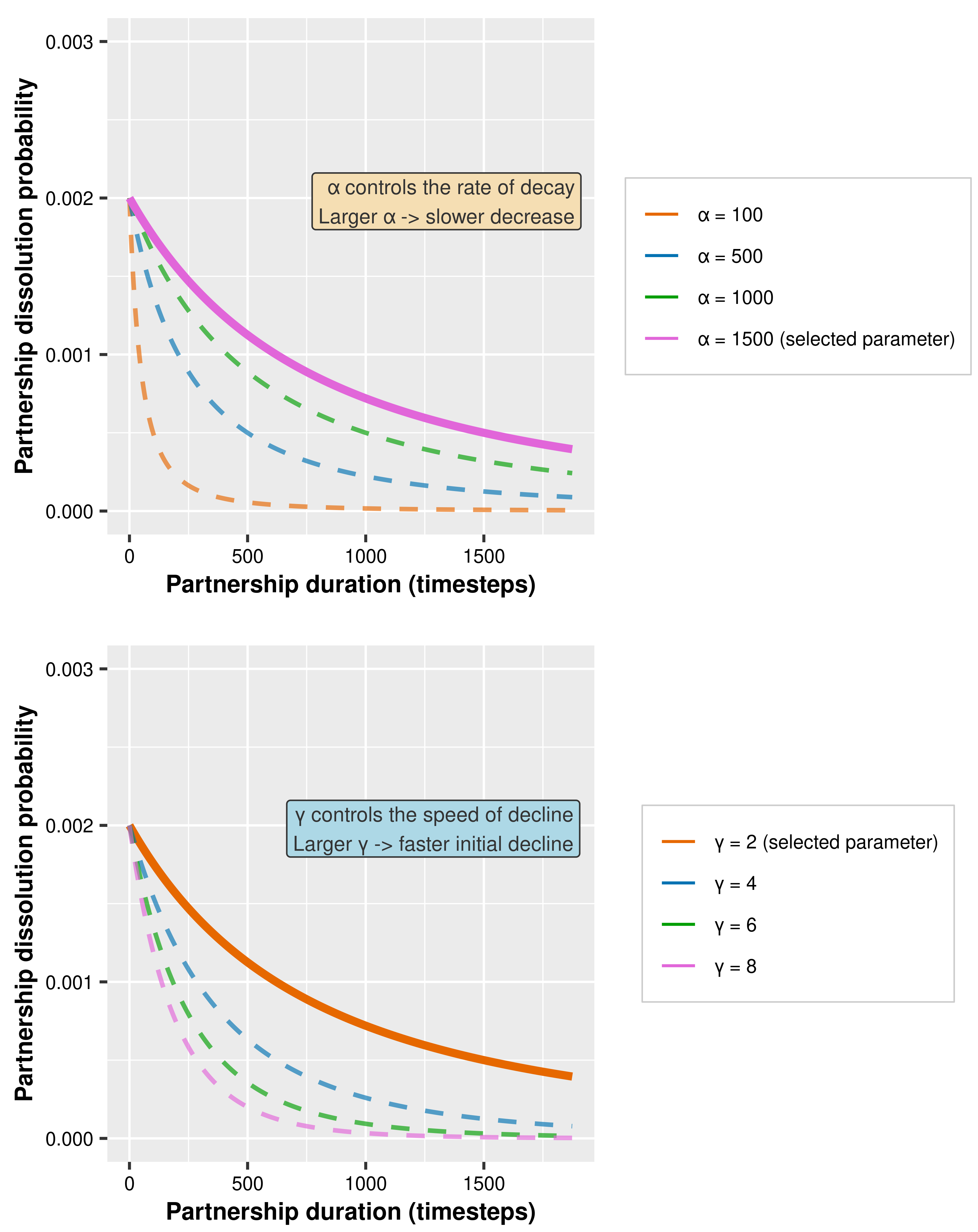}
\caption{Partnership dissolution hazard model: for a given baseline dissolution probability of 0.002, the figure shows the impact of the hazard function on the probability of partnership dissolution over time. The top plot shows the effect of scale parameter (selected $\gamma = 2$) and the bottom plot shows the effect of shape parameter (selected $\gamma = 1500$)}
\label{fig:partnership_dissolution_hazard}
\end{figure}

\item[Gaussian age-assortative weighting]
This mechanism was chosen to reproduce the empirically observed tendency of partners being similar in age. Figure \ref{fig:age_assortative_weighting} shows that setting $\sigma = 4$ years gives an age difference distribution centred at 0 and most values being within $\pm 10$ years, such that agents are most likely to form partnerships with others of similar age. If no compatible candidate is available, no partnership is formed. 
\begin{figure}[htbp]
\centering
\includegraphics[width=\linewidth]{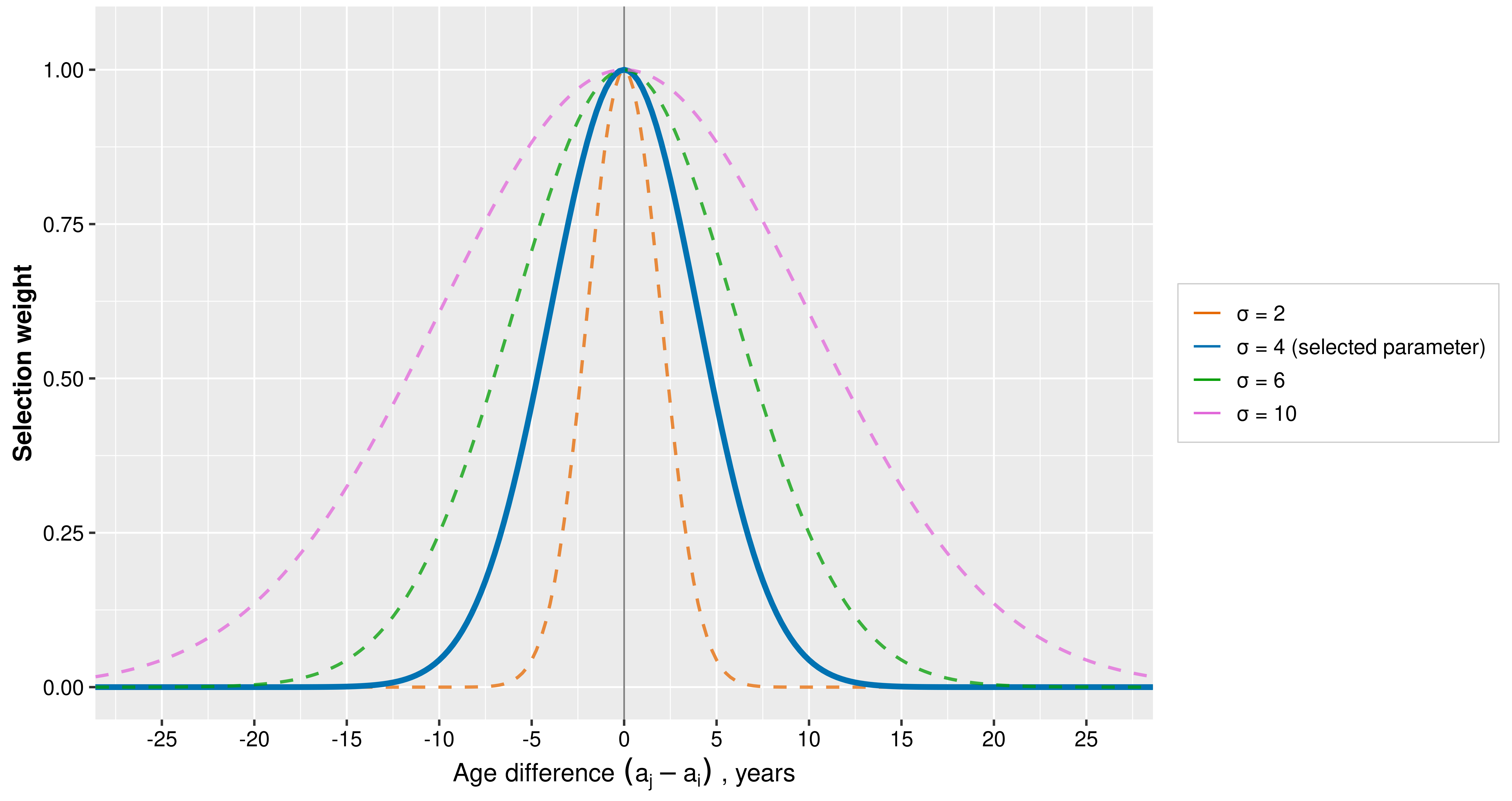}
\caption{Age-assortative Partner Selection: Gaussian weighting applied for age differences in partnerships }
\label{fig:age_assortative_weighting}
\end{figure}

\item[Negative-binomial multiplier]
Individual-level sexual heterogeneity has been modelled with a multiplier assigned to individuals. This is included to represent the diversity in sexual activity among agents sharing the same age, sex, and orientation. Figure \ref{fig:nb_multiplier} shows the range of sampled values $X_i$ from a negative binomial distribution with varying parameters $(r, p)$. 

\begin{figure}[H]
\centering
\includegraphics[width=\linewidth, height=12cm]{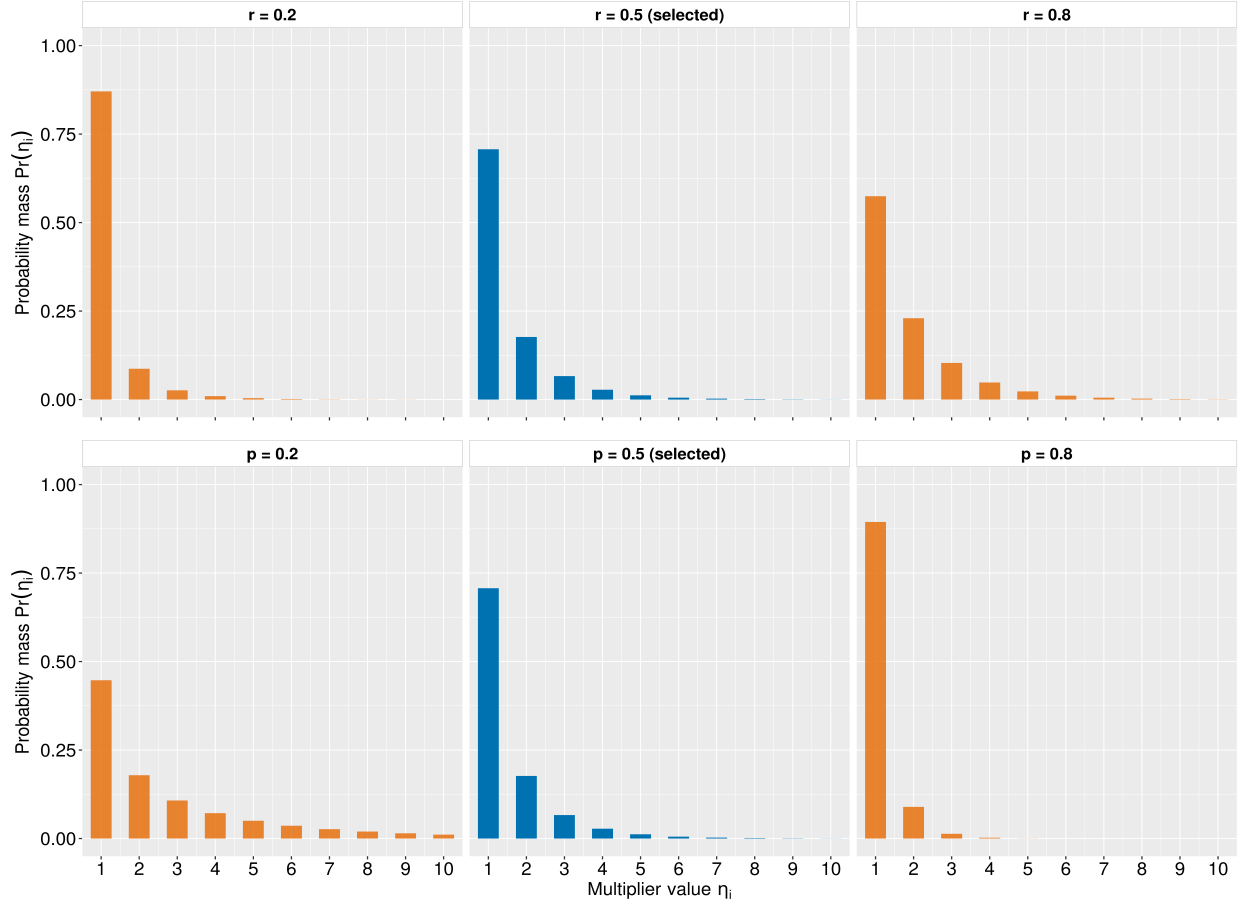}
\caption{Distribution of multipliers from the negative binomial draws.}
\label{fig:nb_multiplier}
\end{figure}
\item[\textit{Youth boost} and exponential \textit{age decay}]
Motivated by the empirical NATSAL-3 pattern that the youngest age band reports the highest mean partnership count, declining smoothly with age thereafter \citep{Clifton2023Natsal3RefTables}. Parameterising this as a single multiplicative boost for the 16--24 band plus an exponential decay for older bands keeps the computational cost of simulating the age-specific heterogeneity low. 

\item[Population turnover via age-75 replacement]
Ageing agents out at 75 and replacing them with 16-year-olds keeps population size and age structure stable over a 5-year simulation. The partnership- and network-focused research questions of this study of sexual partnerships does not strictly require the inclusion of births and deaths. 
\end{description}

\section{Python Implementation}
\label{sec:code}

\subsection{Architecture}
\label{sec:code_arch}
Agents are stored as a structure-of-arrays (\texttt{sex\_arr}, \texttt{ori\_arr}, \texttt{age\_arr}, \dots) indexed by an internal \emph{slot}, with an \texttt{idx2id}/\texttt{id2idx} mapping between slots and stable agent IDs; this allows a removed agent's slot to be reused immediately by a new agent without renumbering the population. Partnerships are stored as a sparse dict-of-dicts keyed by slot (\texttt{self.partnerships[i][j] = start\_time}); external partnerships (Section~\ref{subsubsec:odd_population_turnover}) are stored separately in \texttt{self.external\_partner}, keyed by the removed partner's stable ID rather than by slot, since that slot may already have been reassigned. Population size (\texttt{capacity}) is held fixed at $N$; all random draws go through a single seeded generator \texttt{self.\_rng = np.random.default\_rng(seed)}, so that a given \texttt{(config, seed)} pair is fully reproducible.

\subsection{Partnership formation}
\begin{lstlisting}[caption={Daily partnership-formation step, trimmed (Section~\ref{subsubsec:odd_formation}). \texttt{compat\_idx} encodes the sex/orientation compatibility rules; \texttt{fast\_normal\_pdf} implements the Gaussian age-assortative kernel (Equation~\eqref{eq:s-age_assortativity}).}]
# Initiator pool: currently-single agents, plus concurrency-eligible agents
# below their personal cap (internal + external partners both count).
initiators = set(self.single_agents)
for aid in self.concurrent_agents_ids:
    idx = self.id2idx.get(aid)
    if idx is None or not self.active[idx] or not self.sexually_active_arr[idx]:
        continue
    cap = self.concurrent_agent_max_partners.get(aid, self.cfg.concurrency_min_partner_cap)
    total = int(self.partner_count_arr[idx]) + len(self.external_partner[idx])
    if total < cap:
        initiators.add(aid)

for aid in list(initiators):
    idx = self.id2idx[aid]
    if formed_bool[idx]:
        continue
    sc, oc = int(self.sex_arr[idx]), int(self.ori_arr[idx])
    age_group = _AGE_GROUP_LABELS_FOR_NUMBA[age_group_codes_arr[idx]]
    formation_prob = self._formation_prob(sc, oc, str(age_group), idx=idx)

    if self._rng.random() >= formation_prob:
        continue  # attempt formation with probability formation_prob,

    candidate_idx = np.intersect1d(current_candidate_idx, compat_idx[(sc, oc)], assume_unique=True)
    if candidate_idx.size == 0:
        continue

    age_diffs = self.age_arr[candidate_idx] - int(self.age_arr[idx])
    weights = fast_normal_pdf(age_diffs.astype(np.float64), loc=0.0, scale=self.cfg.age_difference_scale)
    wsum = weights.sum()
    probs = weights / wsum if (wsum > 0 and not np.isnan(wsum)) else None
    partner_idx = int(self._rng.choice(candidate_idx, p=probs))

    self.partnerships[idx][partner_idx] = t
    self.partnerships[partner_idx][idx] = t
    self.partner_count_arr[idx] += 1
    self.partner_count_arr[partner_idx] += 1
    formed_bool[idx] = formed_bool[partner_idx] = True
    self.single_agents.discard(aid)
    self.single_agents.discard(int(self.idx2id[partner_idx]))
\end{lstlisting}

\subsection{Partnership dissolution}
\begin{lstlisting}[caption={Duration-dependent dissolution, trimmed (Section~\ref{subsubsec:odd_dissolution}). Internal (both partners active) and external (one partner removed) partnerships are collected into flat arrays and evaluated in a single vectorised call.}]
def _collect_internal_partnerships(self, t, active_idx, breakage_probs_arr):
    """One row per undirected internal partnership (order i < j)."""
    pairs_a, pairs_b, durations, base_probs = [], [], [], []
    for i in active_idx:
        for j, start_time in self.partnerships[i].items():
            if i >= j:
                continue
            pairs_a.append(int(i)); pairs_b.append(int(j))
            durations.append(t - int(start_time))
            base_probs.append(float(breakage_probs_arr[i]))
    return (np.asarray(pairs_a), np.asarray(pairs_b),
            np.asarray(durations), np.asarray(base_probs))
# --- per-timestep usage ---
pairs_a, pairs_b, durations, base_probs = self._collect_internal_partnerships(
    t, agent_idx, breakage_probs_arr
)
uniforms = self._rng.random(len(pairs_a))
dissolves = compute_breakage_events(
    durations=durations, base_breakage_probs=base_probs,
    alpha=self.cfg.dissolution_alpha, gamma=self.cfg.dissolution_gamma,
    uniforms=uniforms,
) 
for k in np.where(dissolves)[0]:
    self._record_and_dissolve_internal(int(pairs_a[k]), int(pairs_b[k]), t - int(durations[k]), t, partnership_data)
# External partnerships (one partner already removed) follow the same
# collect -> hazard applied -> dissolve pattern via
# _collect_external_partnerships() / _record_and_dissolve_external().
\end{lstlisting}

\subsection{Network construction}
\begin{lstlisting}[caption={Active-interval lookup and snapshot graph
construction, trimmed (Section~\ref{subsubsec:odd_network_construction}).
\texttt{ActiveIntervals} is built once per analysis from the agent log;
\texttt{build\_graph\_at} is then called once per snapshot.}]
def active_at(self, t: int) -> set[int]:
    """Agent IDs active at t: entry_t <= t < exit_t """
    mask = (self.entry_t <= t) & (t < self.exit_t)
    return set(self.agent_ids[mask].tolist())

def build_graph_at(t, partnerships, active):
    G = nx.Graph()

    # Nodes: every agent active at t, including isolated ones
    active_nodes = active.active_at_array(t)
    G.add_nodes_from(active_nodes.tolist())

    # Edges: partnerships active at t (start <= t < end)
    active_mask = (partnerships.start <= t) & (t < partnerships.end)
    if not active_mask.any():
        return G

    a = partnerships.agent[active_mask]
    b = partnerships.partner[active_mask]

    # Filter: both endpoints must be active at t
    active_set = set(active_nodes.tolist())
    edges = [
        (int(ai), int(bi)) for ai, bi in zip(a, b, strict=False)
        if int(ai) in active_set and int(bi) in active_set and ai != bi
    ]
    G.add_edges_from(edges)  # re-partnered pairs collapse to one edge
    return G
\end{lstlisting}

\subsection{Transmission and recovery}
\begin{lstlisting}[caption={Per-step SIS transmission and recovery, trimmed
(Section~\ref{subsubsec:odd_transmission}). \texttt{transmission} is called
on every infectious agent each time step and iterates over the agent's
current neighbours in the snapshot graph $G(t)$ returned by
\texttt{build\_graph\_at} (Section~\ref{subsubsec:odd_network_construction}). Evaluated independently of \texttt{transmission} each time step; an agent infected during the current step is not eligible to recover during that same step, since \texttt{recovery} only acts on agents whose \texttt{condition} was already \texttt{"I"} at the start of the step. }]
def transmission(self) -> None:
    """Attempt to infect each S neighbour with probability infection_prob."""
    if self.condition != "I" or self.network is None or self.id not in self.network:
        return

    p_infect = self.model.p.infection_prob
    for neighbour in self.network.neighbors(self.id):
        nb = self.network.nodes[neighbour].get("agent")
        if nb is None or nb.condition != "S":
            continue
        if np.random.rand() < p_infect:
            nb._receive_infection(infector=self)  # Eq. s-transmission

def recovery(self) -> None:
    """Recover to S with probability recovery_prob."""
    if self.condition == "I" and np.random.rand() < self.model.p.recovery_prob:
        self.condition = "S"

# --- per-timestep model update (trimmed) ---
graph = build_graph_at(t, self._partnerships, self._active)
active_ids = self._active.active_at(t)
for agent in self.agents:
    agent.network = graph if agent.id in active_ids else None

# --- per-timestep model update (trimmed) ---
# transmission() and recovery() are both called once per active agent,
# per time step.}).
if t >= self.p.infection_start_step:
    for agent in self.agents:
        agent.transmission()
    for agent in self.agents:
        agent.recovery()
\end{lstlisting}

\section{Code Availability}

The full source code is available at: \url{https://github.com/pnt-id-models/partnersim-dynet-lhs} (Partnership model), \url{https://github.com/pnt-id-models/partnersim-dynet-lhs} (LHS Analysis), and \url{https://github.com/pnt-id-models/partnersim-dynet-sti} (\textit{SIS} transmission model)

\printbibliography
\end{refsection}

\end{document}